\documentclass[twocolappendix]{emulateapj}

\pdfoutput=1
\usepackage{graphicx}
\usepackage{array}
\usepackage{hyperref}
\usepackage{longtable}

\begin{document}

\newcommand{\rotse}    {\sc{ROTSE}}
\newcommand{\rar}       {\rightarrow}
\newcommand{\eps}       {\epsilon}

%
%

\title{Extensive Spectroscopy and Photometry of the Type IIP SUPERNOVA 2013\MakeLowercase{ej}}

\author{G. Dhungana\altaffilmark{1,*}}
\author{R. Kehoe\altaffilmark{1}}
\author{J. Vinko\altaffilmark{2,3}}
\author{J. M. Silverman\altaffilmark{2}}
\author{J. C. Wheeler\altaffilmark{2}}
\author{W. Zheng\altaffilmark{4}}
\author{G. H. Marion\altaffilmark{2}}
\author{O. D. Fox\altaffilmark{4}}
\author{C. Akerlof\altaffilmark{5}}
\author{B. I. Biro\altaffilmark{6}}
\author{T. Borkovits\altaffilmark{6}}
\author{S. B. Cenko\altaffilmark{7,8}}
\author{K. I. Clubb\altaffilmark{4}}
\author{A. V. Filippenko\altaffilmark{4}}
\author{F. V. Ferrante\altaffilmark{1}}
\author{C. A. Gibson\altaffilmark{9}}
\author{M. L. Graham\altaffilmark{4}}
\author{T. Hegedus\altaffilmark{6}}
\author{P. Kelly\altaffilmark{4}}
\author{J. Kelemen\altaffilmark{10}}
\author{W. H. Lee\altaffilmark{12}}
\author{G. Marschalko\altaffilmark{10}}
\author{L. Moln\'ar\altaffilmark{10}}
\author{A. P. Nagy\altaffilmark{3}}
\author{A. Ordasi\altaffilmark{3}}
\author{A. Pal\altaffilmark{10}}
\author{K. Sarneczky\altaffilmark{10}}
\author{I. Shivvers\altaffilmark{4}}
\author{R. Szakats\altaffilmark{10}}
\author{T. Szalai\altaffilmark{3}}
\author{E. Szegedi-Elek\altaffilmark{10}}
\author{P. Sz\'ekely\altaffilmark{11}}
\author{A. Szing\altaffilmark{6}}
\author{K. Tak\'ats\altaffilmark{13,14}}
\author{K. Vida\altaffilmark{10}}

\altaffiltext{1}{Department of Physics, Southern Methodist University, Dallas, TX, USA}
\altaffiltext{2}{University of Texas at Austin, TX, USA}
\altaffiltext{3}{Department of Optics and Quantum Electronics, University of Szeged, Szeged, Hungary}
\altaffiltext{4}{Department of Astronomy, University of California, Berkeley, California, USA}
\altaffiltext{5}{University of Michigan, Ann Arbor, Michigan, USA}
\altaffiltext{6}{Baja Observatory of University of Szeged, Baja, Hungary}
\altaffiltext{7}{Astrophysics Science Division, NASA Goddard Space Flight Center, Mail Code 661, Greenbelt, MD 20771, USA}
\altaffiltext{8}{Joint Space-Science Institute, University of Maryland, College Park, MD 20742, USA}
\altaffiltext{9}{McDonald Observatory, University of Texas at Austin, TX, USA}
\altaffiltext{10}{Konkoly Observatory of the Hungarian Academy of Sciences, Budapest, Hungary}
\altaffiltext{11}{Department of Experimental Physics, University of Szeged, Szeged, Hungary}
\altaffiltext{12}{Instituto de Astronom\'ia, Universidad Nacional Aut\'onoma de M\'exico, Apartado Postal 70-264, 04510 Ms\'exico D.F., Mexico}
\altaffiltext{13}{Millennium Institute of Astrophysics, Vicuna Mackenna 4860, 7820436 Macul, Santiago, Chile}
\altaffiltext{14}{Departamento de Ciencias Fisicas, Universidad Andres Bello, Avda. Republica 252, 32349 Santiago, Chile}
\altaffiltext{*}{gdhungana@smu.edu}

\shorttitle{SN~2013ej}
\shortauthors{Dhungana et al.}

\begin{abstract}

We present extensive optical ($UBVRI$, $g'r'i'z'$, and open CCD) and
near-infrared ($ZYJH$) photometry for the very nearby Type IIP SN
~2013ej extending from +1 to +461 days after shock breakout, 
estimated to be MJD $56496.9\pm0.3$. Substantial time series
ultraviolet  and optical spectroscopy obtained from +8 to +135 days are
also presented. Considering well-observed SNe IIP from the
literature, we derive  $UBVRIJHK$ bolometric   calibrations from
$UBVRI$ and unfiltered  measurements that potentially reach 2\%
precision with a $B-V$  color-dependent correction. We observe
moderately strong Si II  $\lambda6355$ as early as +8 days. The
photospheric velocity ($v_{\rm  ph}$) is determined by modeling the
spectra in the vicinity of Fe II  $\lambda5169$ whenever observed, and
interpolating at photometric epochs based on a semianalytic method. This
gives $v_{\rm ph} = 4500\pm500$ km s$^{-1}$ at +50 days. We also observe
spectral homogeneity of  ultraviolet spectra at +10--12 days for SNe IIP,
while variations are  evident a week after  explosion. Using the expanding
photosphere  method, from combined analysis of SN 2013ej and SN 2002ap, we
estimate  the distance to the host galaxy to be $9.0_{-0.6}^{+0.4}$ Mpc,
consistent  with distance estimates from other methods. Photometric and
spectroscopic analysis during the plateau phase, which we estimated to be
$94\pm7$ days long, yields an explosion energy of  $0.9\pm0.3\times10^{51}$ ergs, a final pre-explosion 
progenitor mass of 
$15.2\pm4.2$~M$_\odot$ and a radius of $250\pm70$~R$_\odot$.  We
observe a broken exponential profile beyond +120 days, with a break point
at +$183\pm16$ days. Measurements beyond this break time yield a
$^{56}$Ni mass of $0.013\pm0.001$~M$_\odot$.

\end{abstract}

\maketitle

\newpage
\section{Introduction} \label{intro}

Supernovae (SNe) exhibiting substantial hydrogen in their spectra 
are classified as Type II \citep{Filippenko97}.  These events are 
considered to result from the sudden core collapse (CC) 
of massive stars that still retain substantial hydrogen
envelopes. Early-time spectra are basically blue continua with P Cygni lines of hydrogen.  
SNe~II manifest in a variety of subtypes,
with Type SNe~IIP yielding distinctive plateaus of bright optical 
emission lasting roughly 100 days.  The plateau phase is 
believed to arise from a particularly extended hydrogen outer layer that 
sustains optical emission through recombination as the photosphere
recedes and the outer envelope cools over time. After the plateau phase ends, 
subsequent evolution is powered by radioactive decay. This behavior yields 
direct measurement of radioactive material produced from the explosion. 
While some variation is observed in the late-time properties among SNe~IIP, 
variation is more evident in the properties during early times and the photospheric phase,
such as rise time, absolute peak magnitude, 
plateau length and slope (e.g. \citet{Anderson14}). 
Unlike thermonuclear SNe~Ia, which 
are thought to come mostly 
from near-Chandrasekhar-mass white dwarf thermonuclear explosions, 
SNe~IIP are believed to arise from massive progenitors 
\citep{Heger03, UtrobinChugai09} ranging from 8 to 25 M$_\odot$. 
Using pre-SN imaging data, \citet{smartt09b} obtained a Zero Age Main Sequence ($ZAMS$) mass range of 8 to 17 M$_\odot$ for these events.  
Nevertheless, their characteristics have lent themselves to use 
as cosmic distance indicators and possible independent probes of dark 
energy \citep{hamuy01, hamuy02, nugent06, poznanski10}.

SNe~IIP present the opportunity to measure a wealth of physical parameters 
from the explosion, and the extensive data available for nearby events are 
crucial to pinning down the mechanisms involved.  This, in turn, is important 
to any use as cosmological probes from the most frequently occuring 
SN types (e.g., \cite{Li11}).  

On 2013 July 25 (UT dates are used throughout this paper), 
discovery with the 0.76-m Katzman Automatic Imaging Telescope 
(KAIT) at Lick Observatory of a new SN~IIP in 
M74 was announced \citep {CBET3606}. This made SN~2013ej one of the 
closest SNe ever discovered. Prediscovery photometry was obtained with the 
Lulin telescope \citep{ATel5466} and the ROTSE-IIIb telescope at 
McDonald Observatory \citep{CBET3609}, making this also one of the 
best-observed young SNe~IIP. Follow-up spectroscopy was performed using the 
Hobby Eberly Telescope (HET), and the Kast spectrograph at Lick Observatory, 
providing a classification and a redshift. \cite{V14} performed an 
analysis of the first month of photometry and spectroscopy, yielding constraints
on the object and indicating it to be one of the more slowly evolving SNe~IIP at early times.
They identified a moderately strong Si~II feature, blueward of H$\alpha$ in the first month.
Pre-explosion images obtained with the {\it Hubble Space Telescope (HST)} were 
analyzed by \cite{Fraser14}, from which they proposed two possible progenitors, with the redder 
source being the more likely candidate. Using an M-type 
supergiant bolometric correction, they estimated the mass of the progenitor 
to be 8--15.5 M$_\odot$. More recently, from hydrodynamic simulations, 
\cite{Huang15} found the progenitor to be a red supergiant with a derived mass 
of 12--13 M$_\odot$ prior to explosion. We also note that, from an 
independent dataset, \cite{bose15} have favored SN~2013ej to be a Type IIL event, 
accounting for the observed steep plateau and the systematically high velocity of strong H~I lines.

We present an extensive analysis of unfiltered CCD and broadband photometry from
the ultraviolet (UV) through the infrared (IR), 
and a time series of UV and optical spectroscopy, for SN~2013ej.  
We consider all the measurements relative to 
2013 July 23.9 (MJD 56496.9) unless otherwise explicitly stated. 
Section \ref{data} presents the data, while
Section \ref{data_reduc} describes the photometric and spectroscopic reductions.
Utilizing open-CCD and broadband photometry, we analyze the early-time photometry to 
derive the time of shock breakout in Section \ref{photo}. This section also presents bolometric 
calibration of unfiltered and broadband photometry, as well as a derivation of photometric 
observables such as color and temperature. 
Analysis of UV and optical spectroscopic features from +8~d to +135~d
is discussed in Section \ref{spectro}. In Section \ref{dist} we derive the photospheric velocity at photometric epochs, 
from which we utilize the expanding photosphere method (EPM) to estimate the distance to 
SN 2013ej. 
Kinematics of the explosion, properties of the progenitor, Ni mass yield, and other physical 
properties are derived in Section \ref{explo}. The discussion and our conclusions are presented in 
Section \ref{discuss}.

\section{Observations} \label{data}
\subsection{Photometry}
\begin{figure}
\begin{center}
\includegraphics[width=85mm,height=80mm,keepaspectratio]{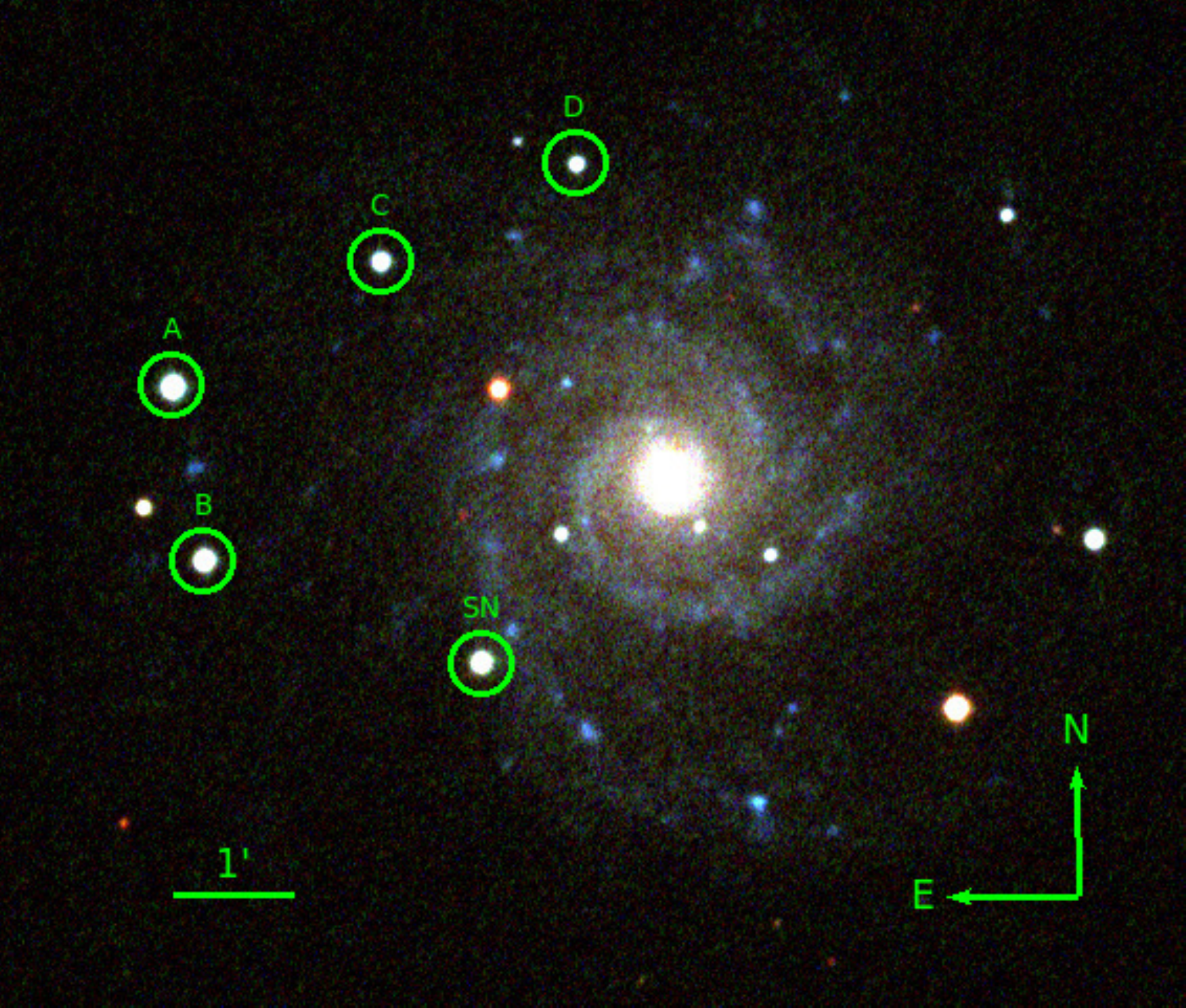}
\caption{The field around SN~2013ej on a color-combined ($BVI$) CCD frame taken with the 
0.6~m Schmidt telescope at Konkoly Observatory, Piszkesteto, Hungary.  
Konkoly photometry reference stars in the vicinity of SN~2013ej are shown.}
\label{fig:sn13ej_konkoly}
\end{center}
\end{figure}

SN~2013ej was discovered by the Lick Observatory Supernova Search (LOSS;
Filippenko et al. 2001) on 
2013 July 25.45 \citep{CBET3606}, using unfiltered data taken with KAIT. A color combined frame 
of the SN~2013ej and the host galaxy M74 is shown in Fig. \ref{fig:sn13ej_konkoly}.
The 0.45~m ROTSE-IIIb telescope also observed 
SN~2013ej in automated sky patrol mode, first on 2013 July 31.36.  
ROTSE-IIIb is operated with an unfiltered CCD with broad
wavelength transmission over the range 3000--10,600 \AA.  Precursor ROTSE images from July 14.42 
rule out any emission at a limiting magnitude of 16.8. Careful analysis of additional 
ROTSE-IIIb observations reveals the earliest detection at July 25.38, about 100 minutes prior 
to the discovery epoch \citep{CBET3609}. 
Following discovery, we scheduled 
follow-up observations with the goal of obtaining well-sampled 
photometry of this bright, nearby SN.  Unfortunately, weather conditions were not optimal for 
the following 5 days for ROTSE-IIIb when the SN was nearing its peak. We then continued 
observations for 200 days (see Fig. \ref{fig:broadband}).

We obtained broadband photometry with the 60/90~cm Schmidt telescope of the Konkoly Observatory at
Piszkesteto Mountain Station, Hungary, through Bessell {\it BVRI} filters. This Konkoly 
dataset spans from +8~d to +130~d. Photometric observations were also performed at 
Baja Observatory, Hungary, with the 50~cm BART telescope equipped with an Apogee-Ultra CCD and 
Sloan $g'r'i'z'$ filters.

Photometry was also obtained with the multi-channel Reionization And Transients InfraRed camera \citep[RATIR;][]{butler12} mounted on the 1.5~m Johnson telescope at the Mexican Observatorio Astrono\'mico Nacional on Sierra San Pedro M\'artir in Baja California, M\'exico \citep{watson12}. Typical observations include a series of 80~s exposures in the $ri$~bands and 60~s exposures in the $ZYJH$~bands, with dithering between exposures. Near-IR data from RATIR span from +3~d to +125~d.

\begin{table*} 
\begin{center}
\caption{Tertiary Konkoly {\it BVRI} measurements of the standard stars in the vicinity of SN~2013\MakeLowercase{ej} used for Konkoly photometry\tablenotemark{a}}
\label{tab:konkolystd}
\begin{tabular}{lcccccc}
\hline
\hline
Star & $\alpha$ & $\delta$ & $B$ & $V$ & $R$ & $I$ \\
 ~     & (J2000) & (J2000) & (mag) & (mag) & (mag) & (mag) \\
\hline
2MASS\tablenotemark{b} J01365863+1547463 (A) & 01:36:58.60 & +15:47:47.32 & 13.19 (0.02) & 12.60 (.01) & 12.32 (0.02) & 11.90 (0.02)\\
2MASS J01365760+1546218 (B) & 01:36:57.56 & +15:46:22.04 & 13.95 (0.02) & 13.16 (.02) & 12.77 (0.02) & 12.30 (0.02)\\
2MASS J01365154+1548473 (C) & 01:36:51.51 & +15:48:48.04 & 14.62 (0.03) & 13.99 (.02) & 13.69 (0.02) & 13.26 (0.02)\\
2MASS J01364487+1549344 (D) & 01:36:44.88 & +15:49:35.88 & 15.74 (0.03) & 14.93 (.02) & 14.59 (0.02) & 14.10 (0.02)\\
\hline
\hline

\end{tabular}
\end{center}
\tablenotetext{1}{~Photometric uncertainties are given inside parentheses.}
\tablenotetext{2}{~Two Micron All-Sky Survey}

\end{table*}

In addition to the unfiltered data taken in discovery mode, scheduled follow-up Bessell 
$BVRI$ photometry was obtained with KAIT and the Nickel 1-meter telescope located at Lick Observatory. 
Starting on June 30, a thorough sample of both unfiltered and $BVRI$ 
measurements was obtained until late in the nebular phase. Unfiltered KAIT data extend 
to +213~d, while $BVRI$ data span from +7~d to +461~d 
(see Fig. ~\ref{fig:broadband} and Fig. ~\ref{fig:Lbolexp}). 

SN~2013ej was also monitored with the {\it UVOT} instrument onboard the NASA {\it Swift} 
space telescope through the {\it uvw2, uvm2, uvw1, u, b, v} filters. These frames were 
collected from the {\it Swift} archive\footnote{http://heasarc.gsfc.nasa.gov/cgi-bin/W3Browse/swift.pl}. 
{\it Swift} data range from +7~d to +138~d. The {\it Swift} dataset has been published by \cite{V14}, \cite{Huang15}, \cite{bose15}. Our reduction of {\it Swift} frames is consistent with these works in the $u$, $b$ and $v$ filters in the plateau, and until +30~d after the explosion in the $uvw2$, $uvm2$ and $uvw1$  filters. However, we obtain significantly brighter magnitudes than \cite{Huang15} beyond +30~d in the later three filters. Given the fact that $uvw2$, and $uvw1$ filters have extended red tails \citep[e.g.][]{ergon14} and using our photometry, all three of these have a marginal contribution to the total flux after +30~d, we ignore the flux from $uvw2$, $uvm2$ and $uvw1$ bands beyond this epoch (also see Section~\ref{bolo}).

We also note that the detection of SN 2013ej at its youngest observed phase was 
announced by~\cite{ATel5466} in the $BVR$ bands, 
on July 24.8, which is 15~hr earlier than the first ROTSE-IIIb detection. 
A nonphotometric prediscovery detection on images 
taken on July 24.125 was also reported by C. Feliciano on the Bright Supernovae 
website\footnote{http://www.rochesterastronomy.org/supernova.html}. 
No emission on 2013 July 23.54 at $V = 16.7$ mag was seen by ASAS-SN \citep{ATel5237}.

\subsection{Optical and Ultraviolet Spectra}

A total of 17 low-resolution optical spectra of SN 2013ej were obtained using the
Marcario Low-Resolution Spectrograph (LRS; \cite{Hill98}) on the
9.2~m Hobby-Eberly Telescope (HET) at McDonald Observatory, the
dual-arm Kast spectrograph \citep{MillerStone93} on the Lick 3~m Shane
telescope, and the DEep Imaging Multi-Object Spectrograph (DEIMOS;
\cite{Faber03}) on the Keck-II 10~m telescope. The Kast and DEIMOS
observations were aligned along the parallactic angle to reduce
differential light losses \citep{Filippenko82}. 
These optical spectra span from +8~d to +135~d.

Near-UV spectra of SN~2013ej were taken with UVOT/UGRISM onboard $Swift$, covering the 
wavelength range 2000--5000 \AA\ and spanning +8--16~d.
 
\section{Data Reduction} \label{data_reduc}
\subsection{Photometry}
ROTSE data were reduced online using an image-reduction pipeline \citep{Yuan08}, 
followed by a DAOPHOT-based point-spread-function (PSF) photometry technique \citep{Stetson87}.  
Because of significant photometric artifacts and reduced efficiency of image differencing, 
we performed aperture photometry of SN~2013ej (e.g., \citet{Howie15}). An aperture size of 
1 full width at half-maximum intensity (FWHM) of the median PSF on each image was considered, and we chose a background-sky annulus having inner and outer 
radii of 2 and 4.5 times the FWHM.
Additionally, a reference template image was smeared to reflect the PSF at each epoch, and the 
underlying host-galaxy contribution inside the aperture was subtracted.  
The typical FWHM of the PSF during the observation timescale was 3--4$''$.  
We calibrated the derived relative flux to the $R$ band from the USNO B1.0 catalog. 
The instrumental calibration and comparison to other data for analysis are presented  
in Section ~\ref{photo}.

Filtered data from Konkoly were reduced with standard {\it IRAF}\footnote{IRAF is 
distributed by the National Optical Astronomy Observatories, which are operated by the 
Association of Universities for Research in Astronomy, Inc., 
under cooperative agreement with the National Science Foundation (NSF).} routines
to get the SN magnitudes. The instrumental magnitudes were transformed to the standard 
Johnson-Cousins system via local tertiary standards tied to \citet{landolt92} standards on a 
photometric night (see Table~\ref{tab:konkolystd} and Fig.~\ref{fig:sn13ej_konkoly}).
The {\it g'r'i'z'} data from Baja Observatory 
were standardized using $\sim 100$ stars within the $\sim 40 \times 40$ arcmin$^2$ 
field of view around the SN, taken from the Sloan Digital Sky Survey (SDSS) Data Release 12 catalog. In order to avoid selecting 
saturated stars from the SDSS catalog, a magnitude cut $14 < r' < 18$ was 
applied during the photometric calibration.

PSF photometry was performed on KAIT and Nickel reduced data \citep{Ganeshalingam10} using DAOPHOT.
Several nearby stars were chosen from the APASS\footnote{http://www.aavso.org/apass} 
catalog, and the magnitudes were first 
transformed to the Landolt 
system\footnote{http://www.sdss.org/dr7/algorithms/sdssUBVRITransform.html\newline \#Lupton2005} 
before calibrating KAIT data. We used APASS $R$ band magnitudes to calibrate the KAIT unfiltered photometry.
Image subtraction was not performed for KAIT data, as the object was 
extremely bright and far from the galaxy core.

For RATIR data reduction, no off-target sky frames were 
obtained on the optical CCDs, but the small galaxy size and 
sufficient dithering allowed for a sky frame to be created 
from a median stack of all the images in each filter.  
Flat-field frames consist of evening sky exposures. Given the 
lack of a cold shutter in RATIR's design, IR dark frames are not available.  
Laboratory testing, however, confirms that the dark current is 
negligible in both IR detectors \citep{fox12}. RATIR data were reduced, 
coadded, and analyzed using standard CCD and IR processing techniques 
in IDL and Python, utilizing online astrometry programs {\tt SExtractor} 
and {\tt SWarp}\footnote{SExtractor and SWarp can be accessed 
from http://www.astromatic.net/software.}. Calibration was performed 
using field stars with reported fluxes in both 2MASS \citep{skrutskie06} 
and the SDSS Data Release 9 Catalogue \citep{ahn12}.

\begin{figure}
\begin{center}
\includegraphics[width=85mm,height=90mm]{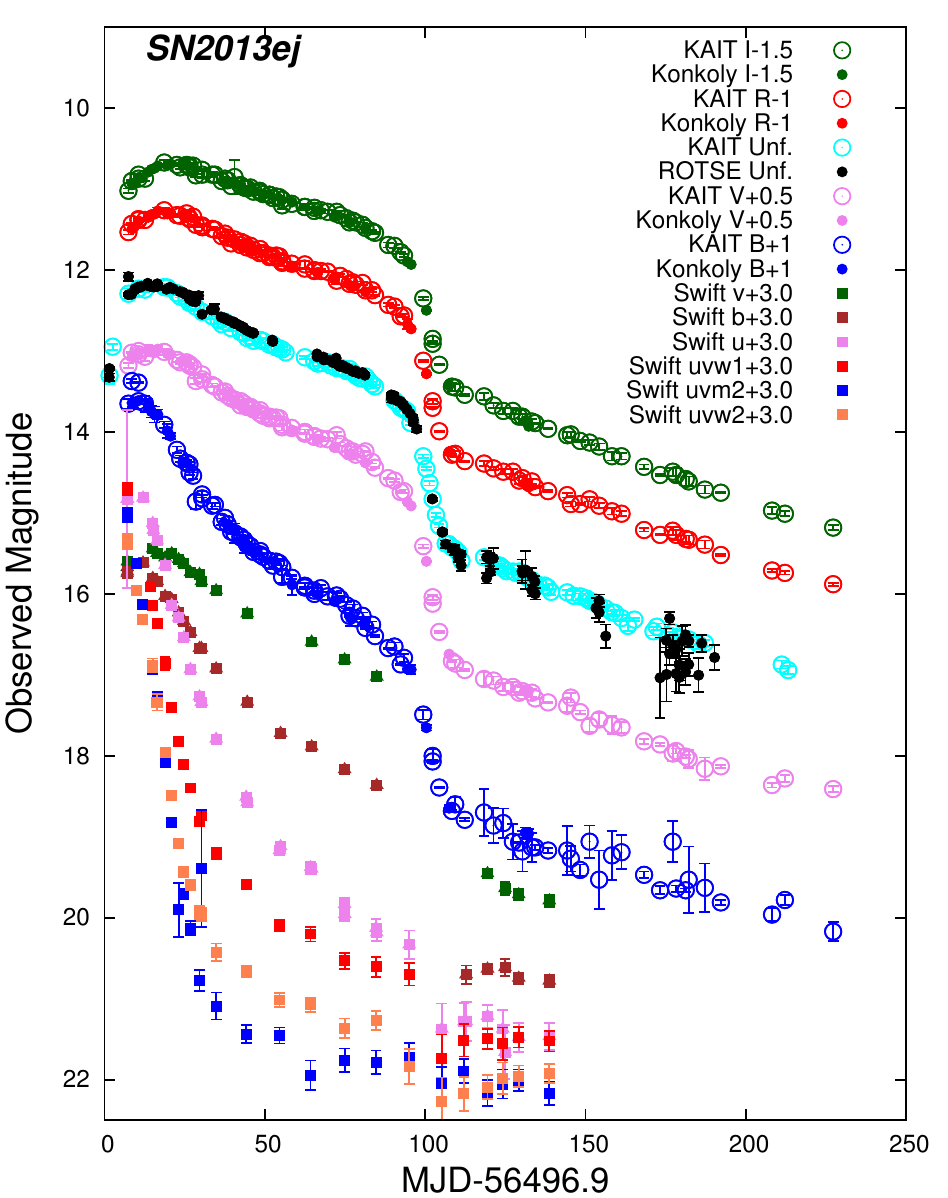}
\caption{Open CCD and multiband photometry of SN 2013ej. KAIT $BVRI$ and unfiltered data 
points are shown with empty circles, while Konkoly $BVRI$ and ROTSE points are solid circles. {\it Swift} photometry is represented with filled square symbols.}
\label{fig:broadband}
\end{center}
\end{figure}

\begin{figure}
\begin{center}
\includegraphics[width=85mm,height=90mm]{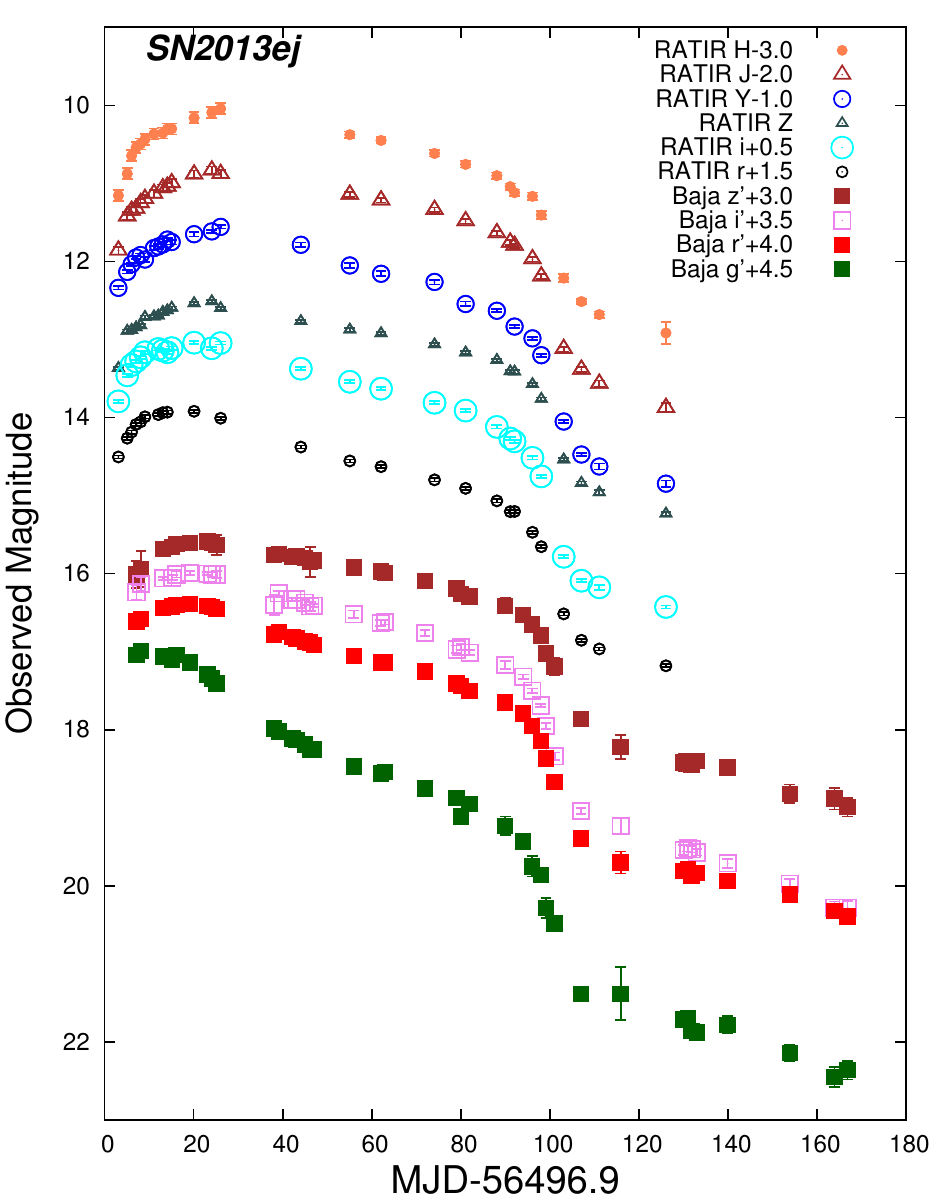}
\caption{SDSS {\it g'r'i'z'} photometry from Baja Observatory, as well as RATIR optical 
and near-IR photometry, of SN 2013ej.}
\label{fig:ratirbaja}
\end{center}
\end{figure}

Figure \ref{fig:broadband} shows the final calibrated SN 2013ej light curves in the ROTSE and 
KAIT unfiltered bands, the KAIT and Konkoly $BVRI$ bands, and the $Swift$ UVOT bands. 
Comparison of the data from various sources revealed that they are generally consistent 
within $\pm 0.1$ mag in all optical bands.
Figure \ref{fig:ratirbaja} illustrates {\it g'r'i'z'} Baja photometry and $riZYJH$ 
RATIR photometry. The Appendix 
tabulates ROTSE, Konkoly, Baja, KAIT and Nickel, and RATIR photometry in Tables \ref{tab:rotselc}, 
\ref{tab:konkolyphot}, \ref{tab:bajaphot}, \ref{tab:kaitlc}, and \ref{tab:ratirphot}, respectively.

\subsection{Spectroscopy}
\begin{figure}
\begin{center}
\includegraphics[width=90mm,height=90mm]{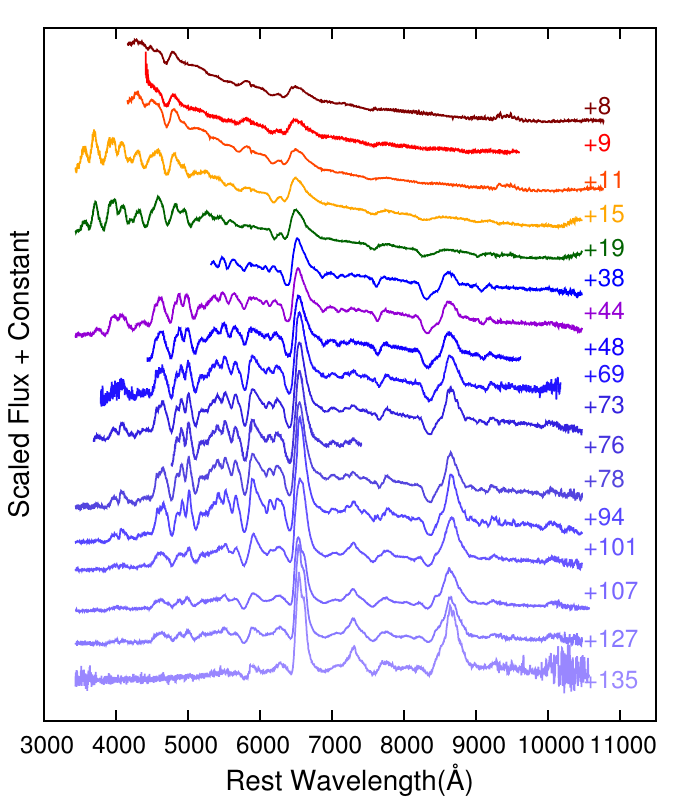}
\caption{Time series of optical spectra of SN~2013ej. Phases are in days since 
explosion (MJD 56496.9). The log of observations is 
given in Table \ref{tab:obslog}.}
\label{fig:sn13ej_spec}
\end{center}
\end{figure}

All of our optical spectra were reduced 
using standard techniques (e.g., \citet{Silverman2012}). Routine CCD processing 
and spectrum extraction were completed with IRAF, and the data were extracted with
the optimal algorithm of \cite{Horne86}. We obtained the wavelength scale
from low-order polynomial fits to calibration-lamp spectra. Small wavelength shifts 
were then applied to the data after cross-correlating a template sky to the 
night-sky lines that were extracted with the SN. Using our own IDL routines, 
we fit spectrophotometric standard-star spectra to the data in order to flux calibrate 
our spectra and to remove telluric lines \citep{WH88,Matheson00}. 
A log of observed optical spectra is given in Table~\ref{tab:obslog} and 
plotted in Figure ~\ref{fig:sn13ej_spec}. HET spectra are 
archived on WISEREP\footnote{http://wiserep.weizmann.ac.il} \citep{wise12}, and all 
of our spectra will be made 
publicly available from the database.

\begin{figure}
\begin{center}
\includegraphics[scale=0.6,keepaspectratio]{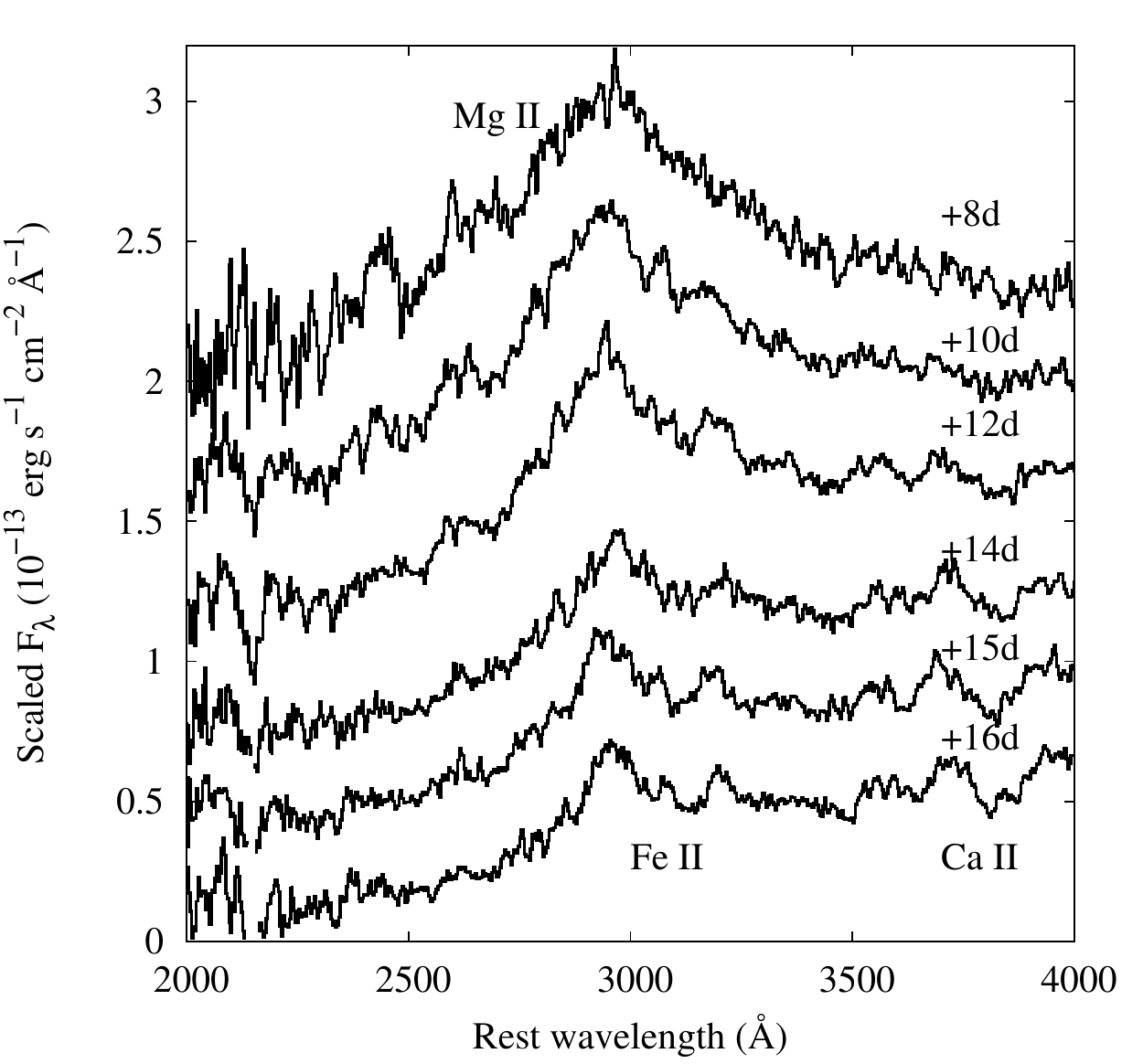}
\caption{The spectral evolution of SN~2013ej in the UV based on $Swift$/UGRISM
observations. Labels next to each spectrum indicate the days since explosion (MJD 56496.9). 
Feature identifications are based on \citet{bufano09}.}
\label{fig:uvsp}
\end{center}
\end{figure}

UV spectra were collected from the $Swift$ archive, and were reduced using the $uvotimgrism$ 
task in $HEAsoft$\footnote{http://heasarc.nasa.gov/lheasoft/}.  
The log of the UGRISM spectral observations is given in Table~\ref{tab:ugrism} and the 
spectra are plotted in Figure \ref{fig:uvsp}.

\begin{table}
\caption{Observing log of SN~2013\MakeLowercase{ej} optical spectra.}
\label{tab:obslog}
\begin{center}
\begin{tabular}{lccc}
\hline
\hline
UT Date & MJD & Epoch\tablenotemark{a} (days)  & Instrument \\
\hline
2013 Aug 1.41 & 56505.41 &  +8 & HET \\
2013 Aug 2.46 & 56506.46 &  +9 & DEIMOS \\
2013 Aug 4.38 & 56508.38 &  +11 & HET  \\
2013 Aug 4.51 & 56508.51 &  +11 & Kast \\
2013 Aug 8.52 & 56512.52 &  +15 & Kast \\
2013 Aug 12.50 & 56516.50 & +19 & Kast \\ 
2013 Aug 30.50 & 56534.50 & +37 & Kast \\
2013 Sept 6.41 & 56541.41 & +44 & DEIMOS \\
2013 Sept 10.60 & 56545.60 & +48 & Kast \\
2013 Oct 1.54 & 56566.54 & +69 & Kast \\
2013 Oct 5.33 & 56570.33 & +73 & Kast \\
2013 Oct 8.48 & 56573.48 & +76 & DEIMOS \\
2013 Oct 10.29 & 56575.29 & +78 & Kast \\
2013 Oct 26.26 & 56591.26 & +94 & Kast \\
2013 Nov 2.34 & 56598.34 & +101 & Kast \\
2013 Nov 8.31 & 56604.31 & +107 & Kast \\
2013 Nov 28.37 & 56624.37 & +127 & Kast \\
2013 Dec 6.39 & 56632.39 & +135 & Kast \\
\hline
\end{tabular} 
\end{center}
\tablenotetext{1}{~Epochs are rounded to days since explosion.}
\end{table}

\begin{table}
\begin{center}
\caption{Observing log of $Swift$ UVOT/UGRISM spectra}
\label{tab:ugrism}
\begin{tabular}{lcccc}
\hline
\hline
UT Date & MJD & Epoch & Exposure & S/N \\
     &     & (days) &  (s)  &  \\ 
\hline
2013 Jul 31.8 & 56504.8 & +8 & 4142 & 30 \\
2013 Aug 3.1 & 56507.1 & +10 & 4946 & 41 \\
2013 Aug 4.8 & 56508.8 & +12 & 4896 & 36 \\
2013 Aug 7.4 & 56511.4 & +14 & 3861 & 25 \\
2013 Aug 8.2 & 56512.2 & +15 & 4449 & 21 \\
2013 Aug 9.2 & 56513.2 & +16 & 4449 & 25 \\
\hline
\end{tabular}
\end{center}
\end{table}

\section{Photometric Analysis} \label{photo}
From the lack of narrow Na~I~D lines from the host galaxy, 
\cite{V14} showed that the reddening from M74 in the direction 
toward SN~2013ej is negligible. No evidence of Na~I~D lines from the host was seen in spectra of \cite{bose15} and in our own sample.
Thus, we do not consider any host extinction. We adopt the Milky Way 
reddening value of $E(B-V) = 0.061$ mag \citep{sf11} in our data sample. 
We note that \cite{Huang15} used $E(B-V)_{\rm tot}=0.12$ mag for SN 2013ej 
based on the $V-I$ color information, 
while \cite{bose15} adopted $E(B-V)_{\rm tot}=0.06$ mag.
\subsection{Early-Time Photometry} \label{early}

After explosion, the gravitational waves and neutrinos soon escape while the 
electromagnetic signal is initially trapped in the envelope. 
Only when the hydrodynamic front reaches the photosphere, 
which takes hours to days, is the 
rise in intensity from the star observed. This epoch indicates 
the time of first light and marks the beginning of the shock-breakout 
phase. Precise knowledge of 
the epoch of shock breakout is crucial to constrain explosion 
parameters and progenitor properties. It is also 
instrumental for distance estimates using methods such 
as EPM (see Section \ref{dist}).

Several efforts have been carried 
out to model the shock breakout of the 
compact progenitor of SN 1987A (e.g., \citet{Hoflich1991, Eastman1994}). 
Notably, it has been shown that the breakout peak depends upon 
envelope mass and density structure 
\citep{FalkArnett1977}, so the very early light curve may provide 
clues on the envelope structure of massive stars. Recently, 
substantial theoretical studies have been performed with the goal of 
understanding the shock breakout of SNe~II 
through a variety of processes in several progenitor 
scenarios (e.g., \citet{nakarsari10, svirskinakar14}).
\cite{couch15} argue that strong aspherical shocks can lead to 
breakout at different times along the periphery
of the star compared to a spherical shock from a spherical star. 

To estimate the time of shock breakout for SN~2013ej,  
we consider several datasets during the first few days. 
To study the rise behavior, we combine ROTSE and KAIT unfiltered data, 
calibrated to $R$ magnitudes, with the earliest 
prediscovery $R$ band detection from Lulin Observatory. Because of 
the lack of sufficient data points in any of the independent datasets, 
we calibrate the Lulin $R$ magnitude to the ROTSE magnitude in the following way. 
We saw a systematic variation of KAIT and Konkoly $R$ band photometry of SN 2013ej. Allowing 
a similar offset to exist between Lulin $R$ and KAIT or Konkoly $R$, we calculate 
the average of differences of KAIT $R$ and Konkoly $R$ magnitudes with ROTSE unfiltered magnitudes and add this 
as a correction to the Lulin data point to bring it to the ROTSE system. 
For this, we limit the observations to between +30~d and +90~d in the plateau, 
where they are densely sampled in both sets and the spectral energy distribution 
(SED) is smooth compared to the rapid evolution during early times (see Section \ref{color}). 
Furthermore, we add an 
additional systematic uncertainty to the Lulin point from the root-mean square (RMS) of KAIT $R$ and Konkoly $R$ 
magnitude differences. We note that the Lulin observation already has an uncertainty higher than $0.2$ mag. Any systematic uncertainty that arises from translating the calibration from plateau to early rise time is likely to be smaller than this. Specifically, \cite{butler06} found a correction of around of $0.1$ mag between KAIT unfiltered and Lulin $R$  band magnitudes for Gamma Ray Burst study. ROTSE unfiltered and KAIT unfiltered data, both of which closely track the $R$  magnitudes to early times, have been independently cross-calibrated to the same unfiltered source for early time studies of both SN Ia and SN IIP \citep [e.g.][]{Quimby2007, zheng13}. On the first night of detection with ROTSE-IIIb, we had better 
time granularity of about 2~hr between coadds of two sets of images. As the SN was still 
young ($\sim 1$~d after explosion), a significant rise even in only 2~hr is detectable.

A functional form of the SN~IIP rise behavior has not been well established. 
A simple power law, specifically a $t^2$ rise law, has been tested in the context of SNe~Ia many times, 
while recent studies \citep[e.g.][]{zheng13, Howie15} have shown departure from $t^2$ rise at very early times. 
In SNe~Ia, the heat loss due to cooling of the ejecta can be thought to be compensated by the
radioactive heating, thus maintaining the steady temperature,
while in SNe~IIP, the adiabatic cooling of the shock-heated envelope is 
expected to result in a steep drop of temperature. 
\cite{Quimby2007} fitted the early rise of SN IIP 2006bp with a $t^{2}$ law at very early times. 
While $t^2$ may be a valid approximation until a few days after explosion in SNe IIP, it is clearly not 
valid as long as it seems to hold in SNe~Ia.

\begin{figure}
\begin{center}
\includegraphics[width=90mm,height=100mm]{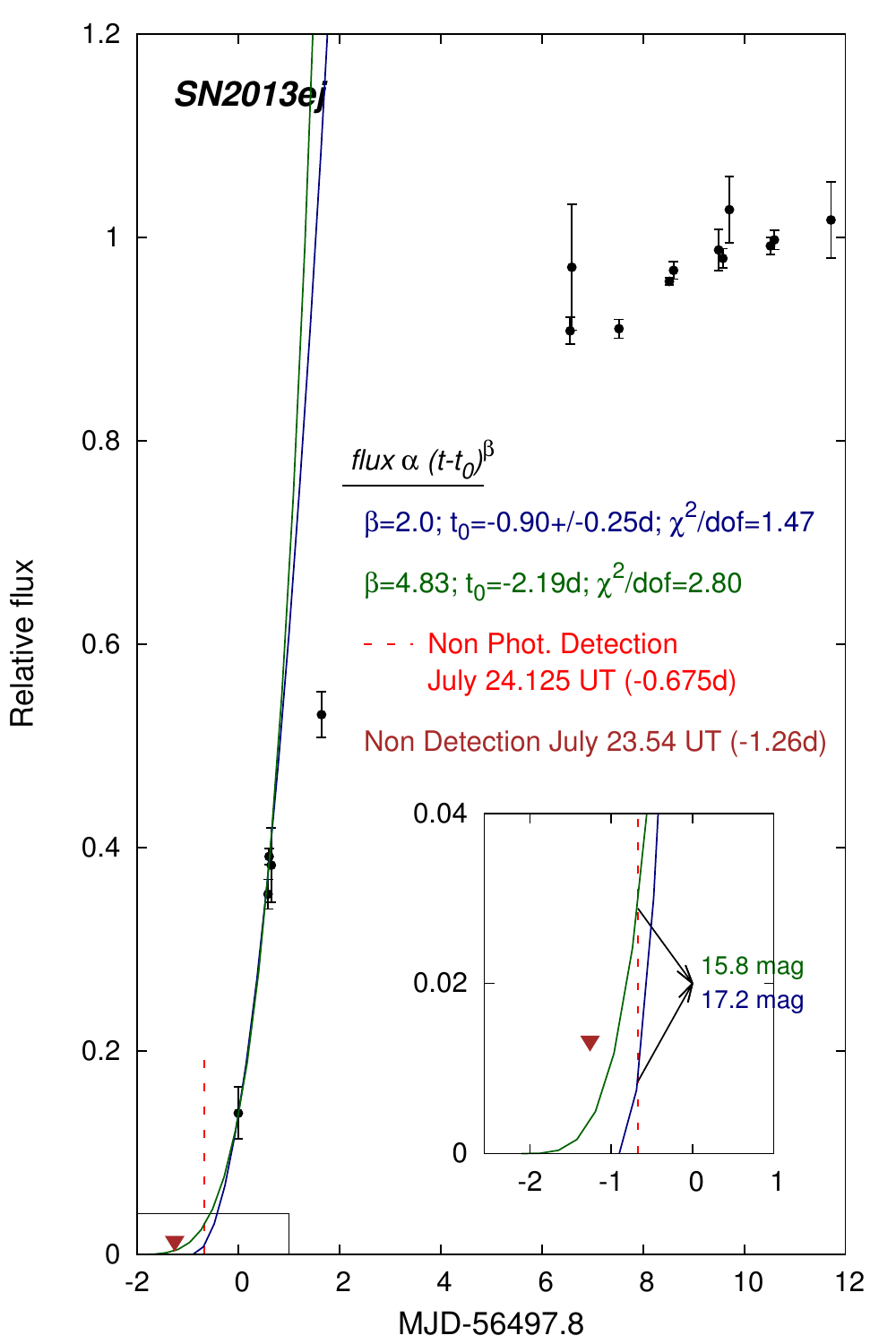}
\caption{Early rise behavior of SN~2013ej. Multiple datasets calibrated to ROTSE 
magnitudes (see text) are shown. Solid lines are power-law 
fits to data obtained before +2~d since explosion. 
The triangle point is a nondetection limit on July 23.54 at $V \approx 16.7$ mag, shown 
here for reference. The dashed line indicates the detection on July 
24.125 with no photometry available. The inset illustrates the projection of where 
the emission would be for floating index (green) and fixed index $\beta=2$ (blue).}
\label{fig:earlyrise}
\end{center}
\end{figure}

Keeping this uncertainty in mind, we perform a least-squares 
fit of the rising light curve of SN~2013ej to a single power law, given by
\begin{equation}
F(t) ~=~ {A~(t-t_{0})^{\beta}},
\end{equation}
\noindent  where $A$ is a constant, $t_{0}$ is the time of shock breakout, and $\beta$ is the 
power-law index. Only data points earlier than +2~d since explosion are considered for 
fitting, effectively including the first 4 points. None of the fits including data beyond +2~d 
were consistent with the detection and nondetections. We perform the $t_{0}$ 
estimation relative to the Lulin observation point on July 24.8 (MJD 56497.8).
As SN~2013ej is a SN~IIP with particularly early photometry, we first test the power-law hypothesis by 
letting $\beta$ float. This yields $t_{0}=-2.19$ days and $\beta=4.83$ with $\chi^2/dof = 2.80$ 
(see Fig.~\ref {fig:earlyrise}). Keeping $\beta=2$ fixed, we obtain $t_{0} = -0.90\pm 0.25$ days, 
which corresponds to July $23.9\pm 0.25$.  The reasonable $\chi^2/dof$ value of $1.47$ 
indicates consistency of the early evolution with the 
$t^2$ model until $\sim 2$~d after explosion. Deviation of $\beta$ from 2 might be indicative of asymmetry in the 
explosion itself and is still an open question. 
This is an important question for SNe IIP and requires further observation of very early times. For SN~2013ej, while the sparse data do not rule out the the power index of 4.83, the $t^{2}$ model yields better $\chi^2/dof$ and is consistent with all reported early 
detections and nondetections. Noting this, we take MJD $56496.9\pm0.3$ as the epoch of shock breakout.

\subsection{Unfiltered and Broadband Photometry} \label{broad}

SN~IIP light curves have a unique signature. After the shock breakout, 
the hot ejecta are believed to expand violently. The photon diffusion 
timescale being much longer than the expansion timescale, very 
little photon energy gets diffused.  The ejecta would follow a 
homologous adiabatic expansion, cooling quickly from the outside. 
Soon after  the ejecta cool to $\sim 6000$ K, hydrogen ions start 
to recombine, the opacity plummets, and diffusion cooling 
becomes dominant. This will result in an ionization front that 
recedes inward as a recombination wave, giving a characterstic, 
slowly declining, almost linear, plateau phase that lasts for approximately 
100 days. This plateau is observed as a result of decreasing opacity 
because of less scattering due to declining electron density. 
The photosphere remains contiguous with the receding ionisation front.  
After all the hydrogen recombines, the photosphere recedes into the inner,
heavy element core and the light curve 
transitions to the radioactive tail phase. This tail is expected to 
decline by the $^{56}{\rm Co} \rightarrow ^{56}{\rm Fe}$ decay at a rate of 
0.98 mag per 100 days if all the gamma rays and positrons are trapped.

   Figure \ref{fig:broadband} shows the apparent magnitude light curves of SN 2013ej 
with unfiltered and $BVRI$ broadband observations. Each $BVRI$ set consists 
of data that starts from around the peak, extends through a characteristic plateau 
phase lasting about 100 days, and proceeds to the well-observed radioactive tail phase. 
In the ROTSE light curve, the peak for SN~2013ej occurs at +18~d, where the 
absolute magnitude reaches $-17.5$. This peak is consistent with the KAIT 
unfiltered data, both in phase and magnitude. 
On both the KAIT and Konkoly $BVRI$ light curves, the peak occurs 
on +12.5~d in $B$, +15.5~d in $V$, +19.5~d in $R$, and +20~d in $I$. 
We do not observe any obvious secondary peak like that seen by \cite{Bose13} 
in SN 2012aw at about +50~d in the $V$ band, or an obvious minimum around +42~d in $V$, 
which would be indicative of the end of free adiabatic cooling.  It is thus more 
challenging to ascertain the advent of the plateau phase in the photometry. 
We will estimate the plateau length in Section \ref{explo}. 
From their respective peaks, the light curves 
decline by 0.038, 0.021, 0.016, and 0.012 mag per day in $B$, $V$, $R$, 
and $I$ (respectively) until +90~d. The Konkoly and KAIT data are consistent 
in this decline behavior. Our $V$ band slope is steeper compared 
to the 0.017 mag day$^{-1}$ given by \cite{bose15}. This could possibly be a sampling issue, 
because their photometry during the plateau is sparsely sampled and 
the peak is less well constrained. 

\begin{figure}
\begin{center}
\includegraphics[width=85mm,height=80mm,keepaspectratio]{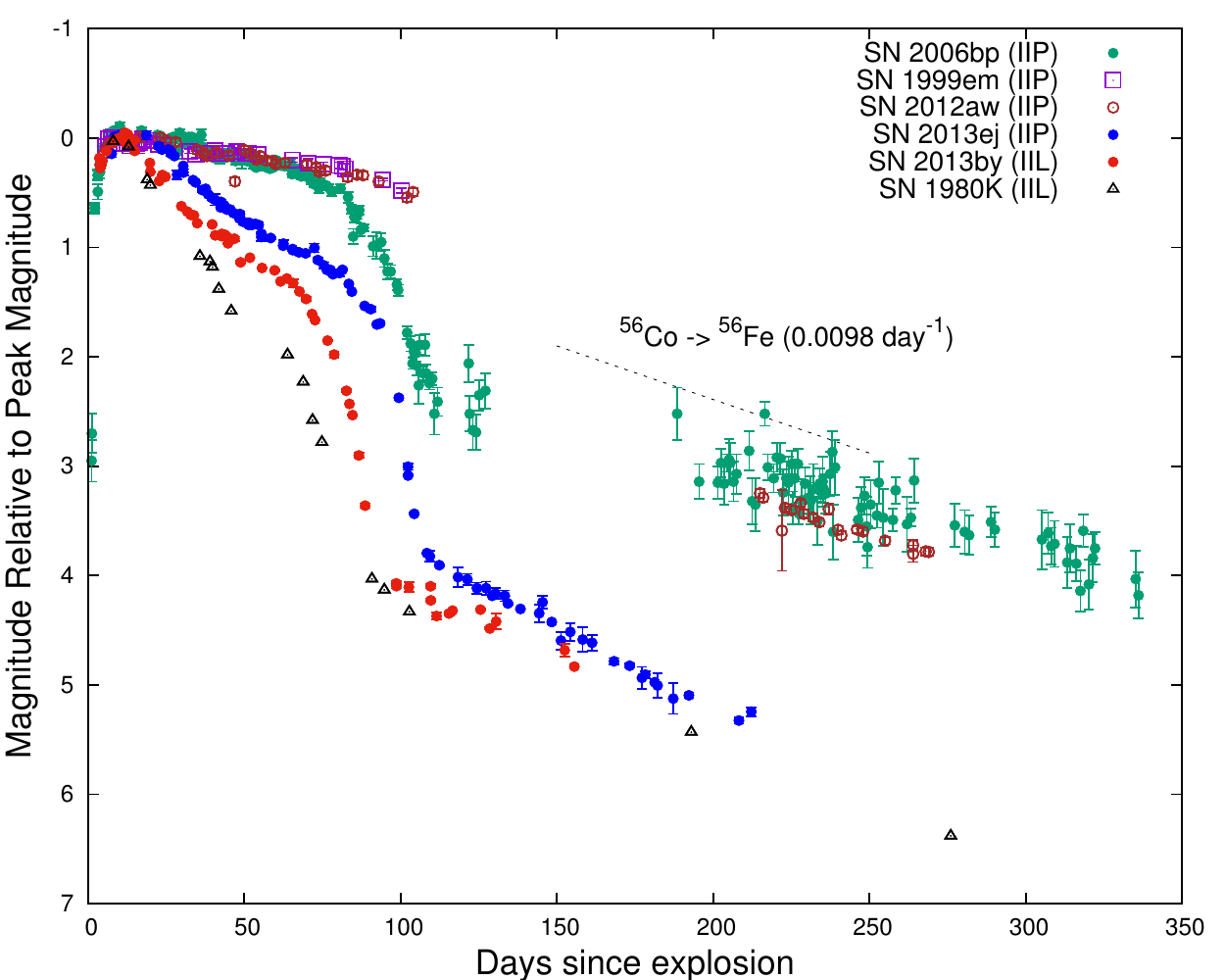}
\caption{Comparison of some SN~IIP and SN~IIL light curves in the $V$ band, except SN~2006bp (ROTSE). 
SN~2013ej has a systematically steeper plateau, with a sharp drop at the end of the plateau. 
SNe~IIL 2013by and 1980K have steep linear evolution after peak, but the 
drop at the end of the plateau is not as sharp. SN~2013ej also exhibits a steeper tail decline than the events shown here.}
\label{fig:comparelc}
\end{center}
\end{figure}

SN~2013ej has one of the steepest plateaus among SNe~IIP 
(see Fig. \ref{fig:comparelc} for a comparison with other normal SNe~IIP).
We note that the $B$ band decline for SN 2012aw was 1.74 mag until +104~d, while the $R$ band 
showed no change in brightness over the plateau \citep{Bose13}. Classic SN~IIP 1999em 
also evolved similarly \citep{leo02}, while the more energetic SN~2004et had a faster 
decline of $\sim 2.2$ mag from the $B$ band peak until +100~d \citep{Bose13}. The decline rate for 
SN 2013ej is consistently higher in all bands. 
Example SN~IIL light curves of the recent SN 2013by \citep{valenti15} and the archetype 
SN~1980K \citep{barbon82} are also shown. The magnitude fall of SN~2013ej in the $V$ band 
from peak to +50~d is about 0.75 mag. 
This puts SN~2013ej within the SN~IIL category of \cite{Faran14}, where they use a cut of 0.5 mag for a SN~IIL event. 
\cite{valenti15} observed a fall of $1.46\pm0.06$ mag in $V$ for SN~2013by, and also pointed out the SN~IIP-like behavior 
in its light-curve drop to the radioactive phase. 
They show a handful of objects for which the difference of the $V$ band peak and +50~d magnitude 
is more than 0.5 mag, which would be in the SN~IIL class of \cite{Faran14}.  

SN~2013ej, in spite of having such a steep plateau, also exhibits 
a drop at the end of the plateau that is significantly sharper than the decline in the plateau, 
which is also a characterstic feature of SNe~IIP. 
A steep plateau for a SN~IIP object may also indicate 
an inefficient thermalization of the ejecta.  
Additionally, with such a steep plateau, very little nickel yield might be expected. \cite{Bersten11} showed from 
hydrodynamic modelling that extensive mixing from $^{56}$Ni is required to reproduce 
flat plateaus. The higher the Ni yield is, the sooner the plateau starts to flatten, and this will also 
affect the extension of the plateau duration because of radioactive heating. 

    When the plateau ends at about 100~d after the explosion, the light curve suddenly 
transitions into the radioactive tail. 
From the luminosity derived from radiactive decay of synthesized materials, 
in Section~\ref{explo} we will estimate the mass of nickel produced. 
We represent the decline behavior by separate linear fits to the data 
from +120~d to +183~d and from +183~d to +461~d. The epoch +120~d was chosen 
to ensure the late-time decay phase, and +183 was chosen as the break time of the 
late time behavior (see Section \ref{nimass}).  
Table ~\ref{tab:tailbehav} lists the decay rate along 
with $\chi^2$ per degree of freedom of the respective fits. 
It is clear that the light-curve decline in the tail is 
much steeper in all bands and unfiltered photometry before +183 d. 
Only $B$ band has a slope shallower than $^{56}{\rm Co} \rightarrow~ ^{56}{\rm Fe}$ after +183d.
While SN~2006bp had a tail phase decline of $0.73\pm 0.04$ mag per 100~d \citep{Quimby2007}, 
which is less steep than full trapping of gamma rays from radioactive decay, 
SN~2013ej exhibits the opposite behavior. 
\begin{center}
\begin{table}
\caption{Radioactive Tail Decline Behavior. $\chi^{2}/{\rm dof}$ are given inside paranthesis.}
\label{tab:tailbehav}
\begin{tabular}{lccccc}
\hline
\hline
Band & +120~d to +183~d       &  +183~d to +461~d \\

  & (mag/100 days) & (mag/100 days)\\
\hline
KAIT $B$ & $1.15 \pm 0.08$ (0.37) & $0.75 \pm 0.02$ (1.75)\\
KAIT $V$ & $1.46 \pm 0.04$  (1.01) & $1.10 \pm 0.02$ (2.69)\\
KAIT $R$ & $1.54 \pm 0.04$  (2.66) & $1.30 \pm 0.01$ (3.92)\\
KAIT $I$ & $1.63 \pm 0.03$  (0.78) & $1.18 \pm 0.02$ (2.96)\\
KAIT unfiltered & $1.51 \pm 0.02$  (0.88) & \\
ROTSE unfiltered & $1.64 \pm 0.07$ (1.13) & \\
KAIT $BVRI$ & $1.36 \pm 0.09$ (0.48) & $1.06 \pm 0.02$ (2.15) \\
\hline 
\hline
\end{tabular}
\end{table}
\end{center}

\subsection{Color and SED Evolution} \label{color}

\begin{figure}
\begin{center}
\includegraphics{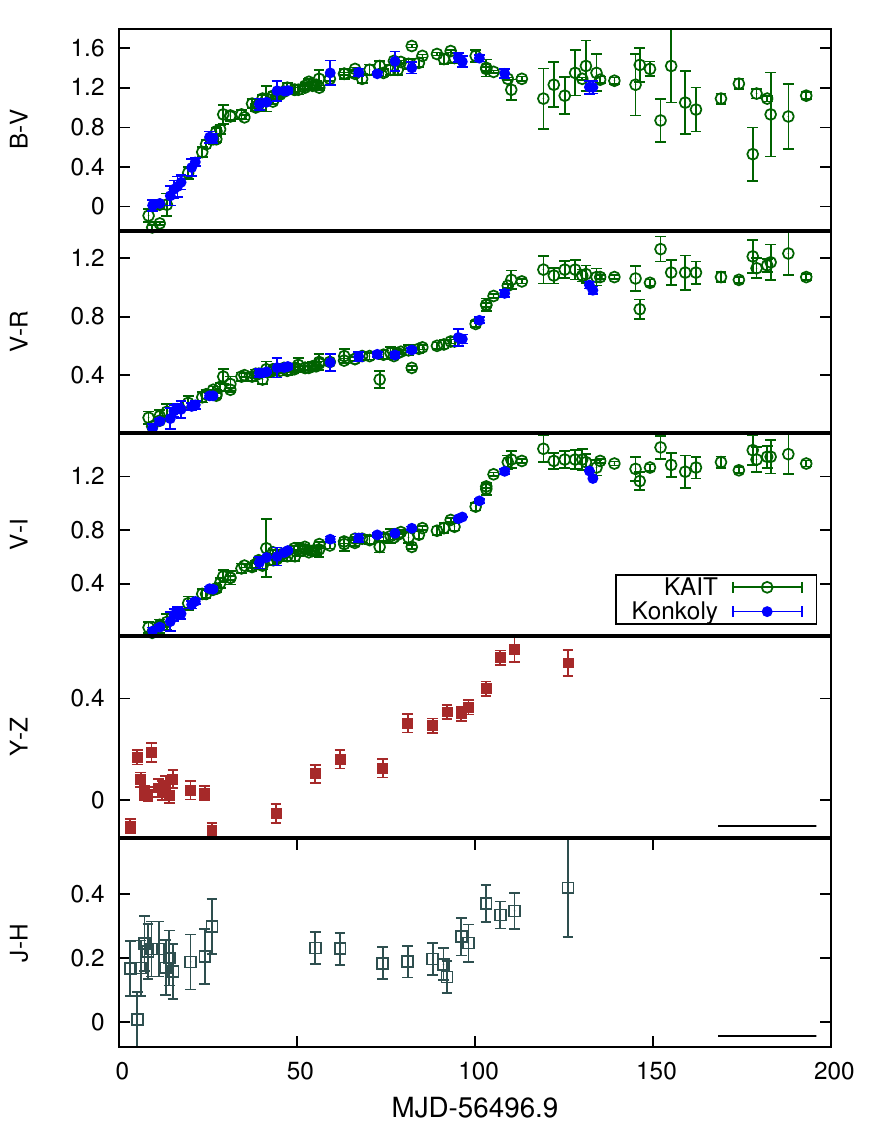}
\caption{Optical and near-IR color evolution of SN~2013ej. Here, the $J-H$ and $Y-Z$ colors from RATIR are shown in the Vega system for consistency.}
\label{fig:colorevolve}
\end{center}
\end{figure}

The color evolution of SN~2013ej in the optical exhibits a rapid change in 
the first 30 days, as shown in Figure \ref{fig:colorevolve}.
This is due to the fact that the $U$ and $B$ fluxes {\bf decline} rapidly at early phases. 
While the evolution of $B-V$ is more rapid in the first 30 days, the $V-R$ and $V-I$ colors are 
smooth and slowly rising.  Soon after, when the temperature 
has fallen to around 6000~K (see Section \ref{temp}), the trends are more alike, 
and the SED is more uniform. 
Later, as the light curve approaches the tail, 
both the $V-R$ and $V-I$ colors show a rapid rise, as an effect from a greater decline of flux in $V$ 
relative to the $I$ band. The transition from plateau to the tail is evident in both optical 
and near-IR colors.

\begin{figure}
\begin{center}
\includegraphics[width=90mm,height=80mm,keepaspectratio]{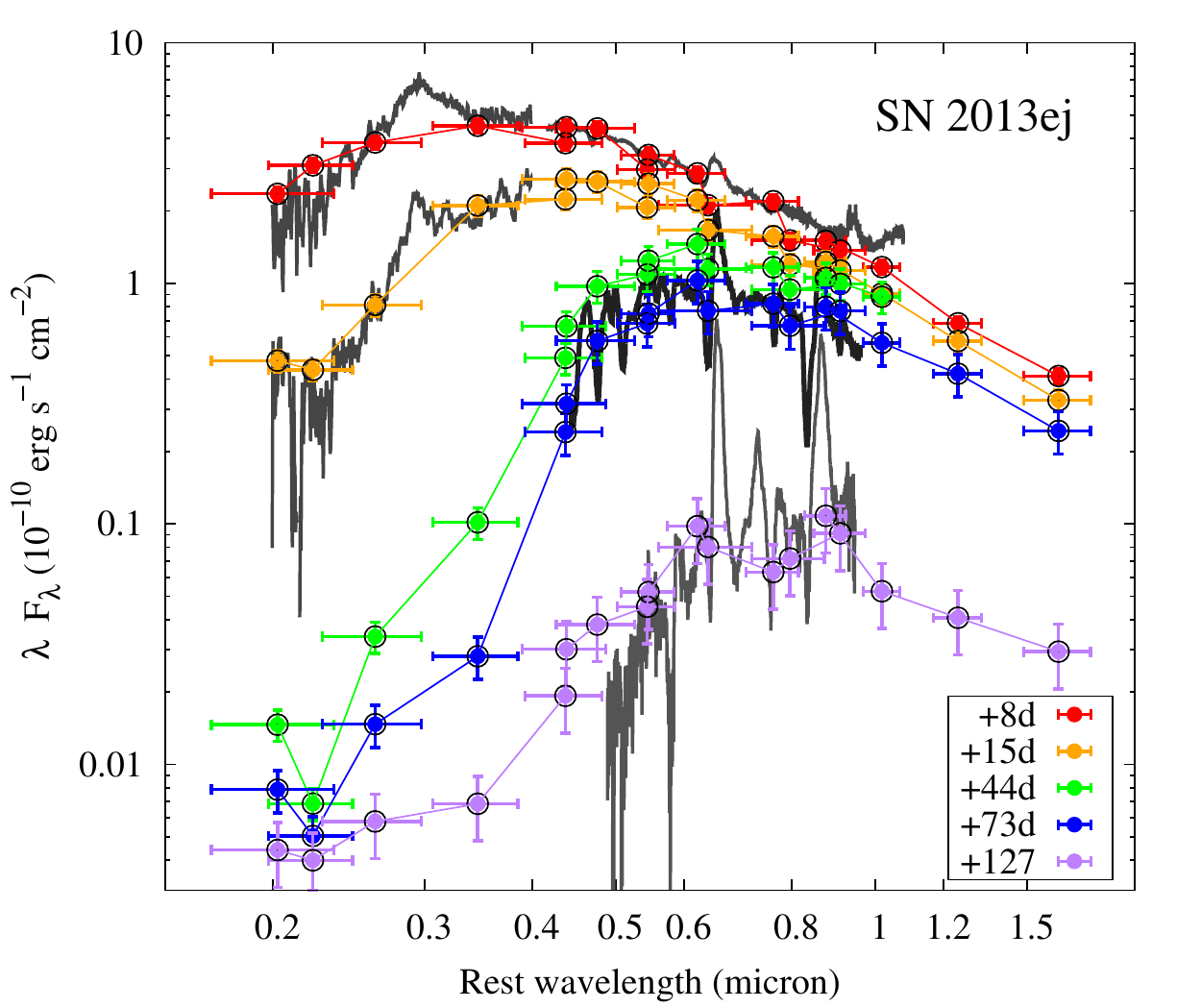}
\caption{The evolution of the spectral energy distribution (SED) of SN~2013ej. Fluxes from 
broadband photometry from the UV, optical, and near-IR are plotted with filled circles, and the horizontal
bars represent the FWHM of each filter. Phases are coded by colors and indicated
in the legends. The optical and UV spectra at certain epochs (where available) are
also overplotted for comparison. Dereddening with $E(B-V)=0.061$ mag has 
been applied.}
\label{fig:sedevolve}
\end{center}
\end{figure}

In Figure~\ref{fig:sedevolve}, the evolution of the SED ($\lambda F_{\lambda}$)
is shown, together with some of the contemporaneous UV and optical spectra (see Section \ref{spectro}).
This observed SED evolution is in agreement with the general characteristics of SNe~IIP: a strong decline of the UV flux accompanied by a monotonic decrease of the continuum
slope in the optical during the plateau phase, in accord with the continuously reddening 
optical colors seen in Figure~\ref{fig:colorevolve}. 

\subsection{Photospheric Temperature} \label{temp}

\begin{figure}
\begin{center}
\includegraphics[keepaspectratio]{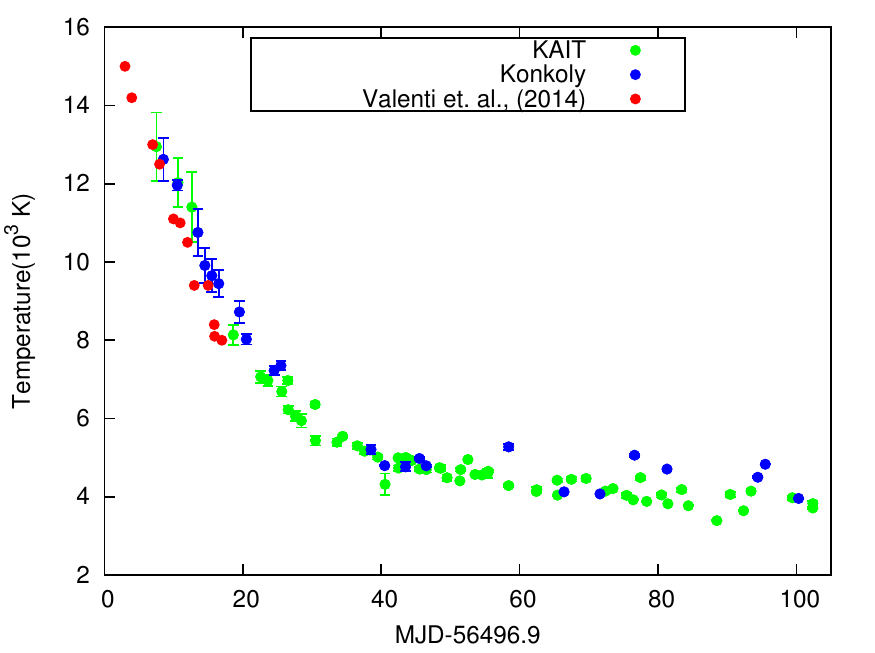}
\caption{Evolution of the photospheric temperature using the KAIT (green points) and Konkoly (blue points) $BVI$ datasets. Temperature estimates from \cite{V14} are shown with red points.}
\label{fig:temp}
\end{center}
\end{figure}

 We determine the temporal evolution of the photospheric temperature by fitting the KAIT and Konkoly $BVI$ fluxes ($R$ fluxes are omitted to avoid contamination from the strong H${\alpha}$ feature) to a  
Planck function at each epoch until $\sim$+100 d.
 Beyond this, the light curve enters the 
radioactive phase, and the energy mostly comes out in strong nebular lines. 
The $BVI$ set covers the wavelength range 
3935--8750~{\AA}. No UV flux is considered, as
this would heavily bias the blackbody fits because of the many metallic
blends occurring at shorter wavelengths. The temperature drops
from 12,500~K at +8~d to 6400~K at +24~d, and it declines very slowly to 4000~K at +100~d as shown in Figure \ref{fig:temp}. An independent 
estimate of the temperature by \cite{V14} is also shown, and it exhibits reasonable agreement with our result. 
The rapid temperature drop in the first few weeks encapsulates quick adiabatic cooling, while later in the plateau phase the smooth slow decline signifies the cooling from photon energy diffusion during recombination at nearly constant temperature dictated by atomic physics.

\subsection{Bolometry} \label{bolo}
\begin{figure}
\begin{center}
\includegraphics[width=100mm,height=80mm,keepaspectratio]{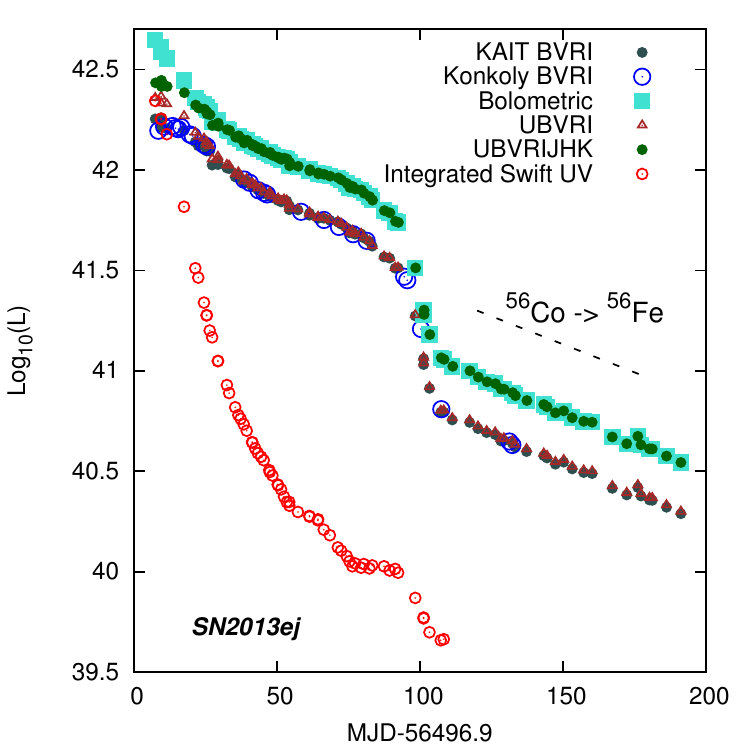}
\includegraphics[scale=0.45,keepaspectratio]{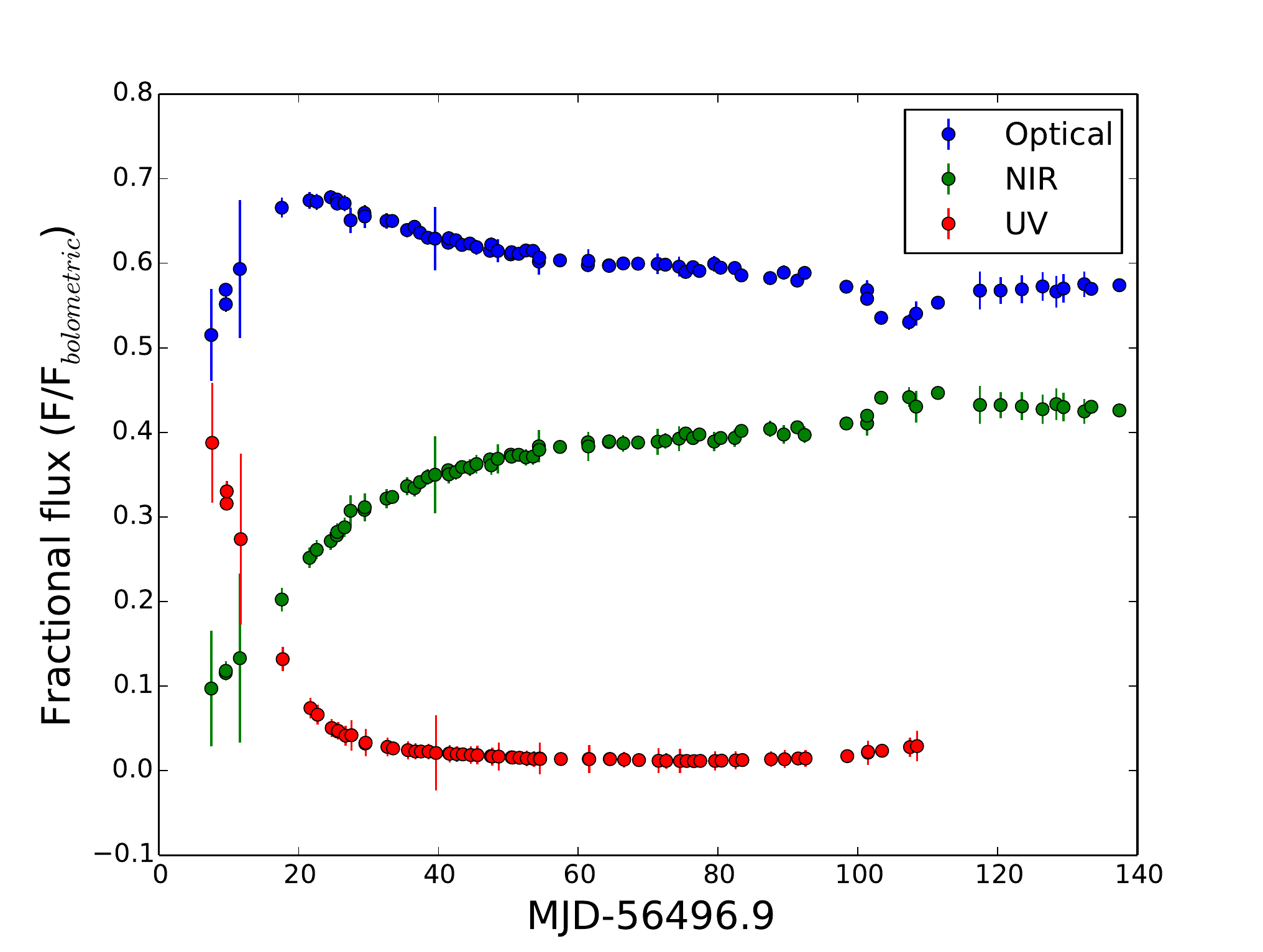}
\caption{{\it Top Panel}: Integrated $BVRI$, $UBVRI$, $UBVRIJHK$, and 
bolometric light curves of SN 2013ej. $UBVRI$ and $UBVRIJHK$ shown here 
are derived using $Swift$ $u$, KAIT $BVRI$, RATIR $JH$, and estimated flux 
from the $K$ 
band. The bolometric light curve includes an additional contribution 
from UV flux below $\sim3200$~\AA ~before +30d. Beyond +30d, $UBVRIJHK$ closely resembles the bolometric light 
curve in the plateau. {\it Bottom Panel}: UV, optical, and near-IR 
fractional flux.
The UV covers flux below the $U$ band, the near-IR (NIR) covers flux above 
the $I$ band, while optical covers in between ($UBVRI$). The UV fraction is shown curtailed at the calculated end of plateau, because of the marginal contribution and possible contamination from red tails (see text).}
\label{fig:bolometric}
\end{center}
\end{figure}

  Bolometric photometry permits determination of several 
explosion parameters, including the direct estimate of the amount 
of Ni synthesized during the explosion.  UV flux  in SNe~Ia 
and SNe~Ib/c is a small fraction of the total flux, since the high-energy photons are 
nearly entirely absorbed by transition lines of ionized heavy elements.  
In SNe~II, however, the UV flux dominates at early times. 
Bright UV and X-ray emission flashes are expected as the shock breaks out, 
followed by a post-break UV plateau lasting a few days.  
The spectral index evolves rapidly in the first few weeks, with 
a large portion of the flux yield in UV. Later evolution transitions 
to dominant emission in the $R$, $I$, and near-IR bands.  Without data in all 
wavebands, it is generally impossible to obtain an exact bolometric flux. 
The availability of extensive sets of data spanning from UV to IR wavelengths 
for SN~2013ej gives an ideal opportunity to obtain the most accurate 
estimate of the bolometric flux for SN IIP. Additionally, this helps to derive 
bolometric calibration relations for a broadband sample limited in 
wavelength and unfiltered sets such as from ROTSE or KAIT. 
We adopt the distance to the SN to be $9.0_{-0.6}^{+0.4}$ Mpc (see Section \ref{dist}) 
to estimate the integrated luminosity. Pseudo-bolometric and bolometric light curves
of SN~2013ej are shown in the top panel of Figure \ref{fig:bolometric}. 
To estimate the bolometric flux in the late time when we do not have UV and NIR observations, we multiply the late time $BVRI$ flux by a scale factor, which we derive by dividing the bolometric flux by the $BVRI$ flux between +120~d and +137~d. We also note that the flux from $uvw2$, $uvm2$ and $uvw1$ bands contribute a total of 1\% or less to the bolometric flux beyond +30d according to our reduction (see Fig. \ref{fig:bolometric}). Photometry by \cite{Huang15} would contribute much less. Given the marginal contribution from these three bands after +30~d and potential complication of red leaks for $uvw2$ and $uvw1$ \citep[e.g.][]{ergon14}, these three filters were omitted beyond +30~d in the bolometric flux calculation.

\begin{figure}
\begin{center}
\includegraphics[width=80mm,height=80mm]{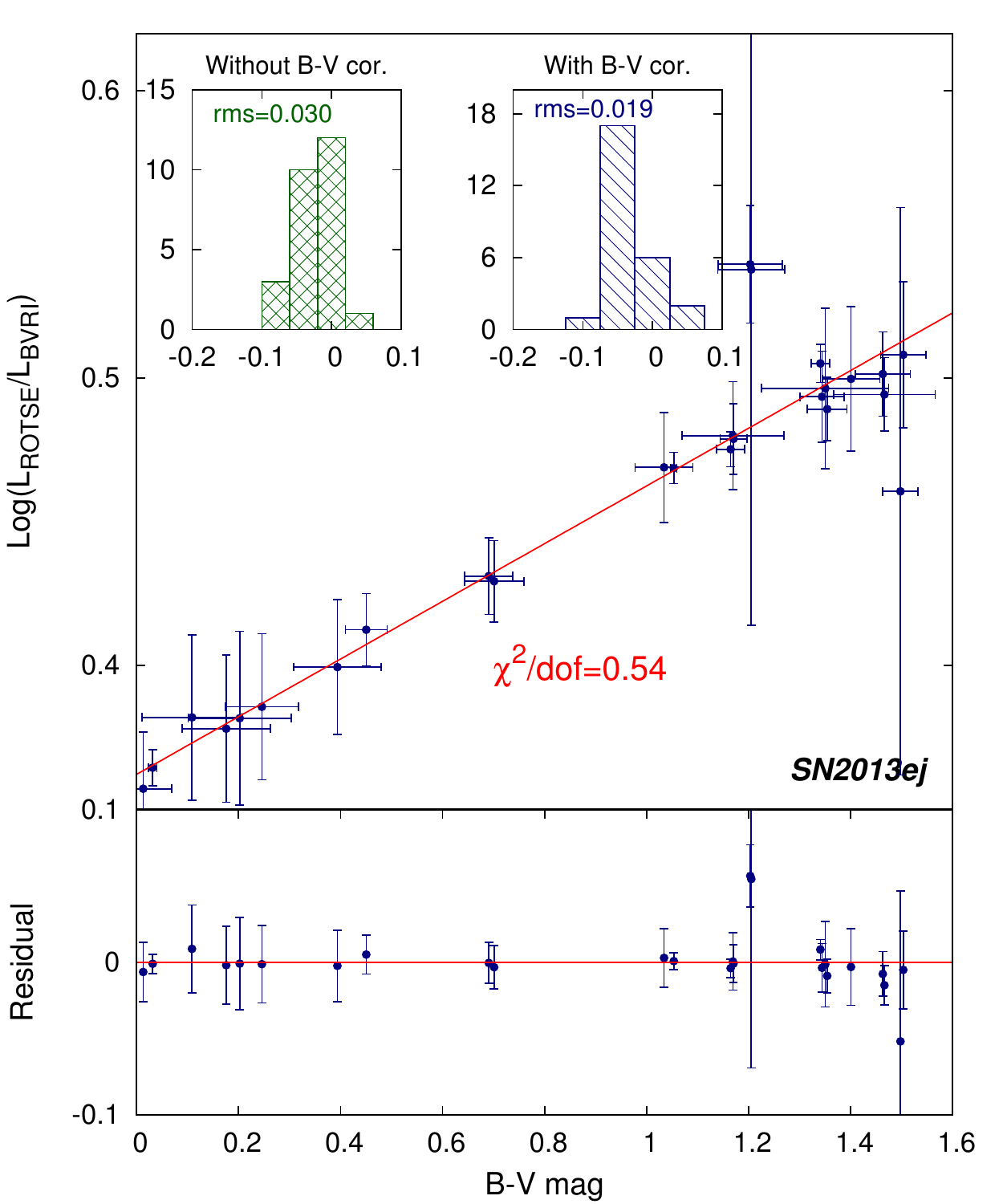}
\includegraphics[width=80mm,height=80mm]{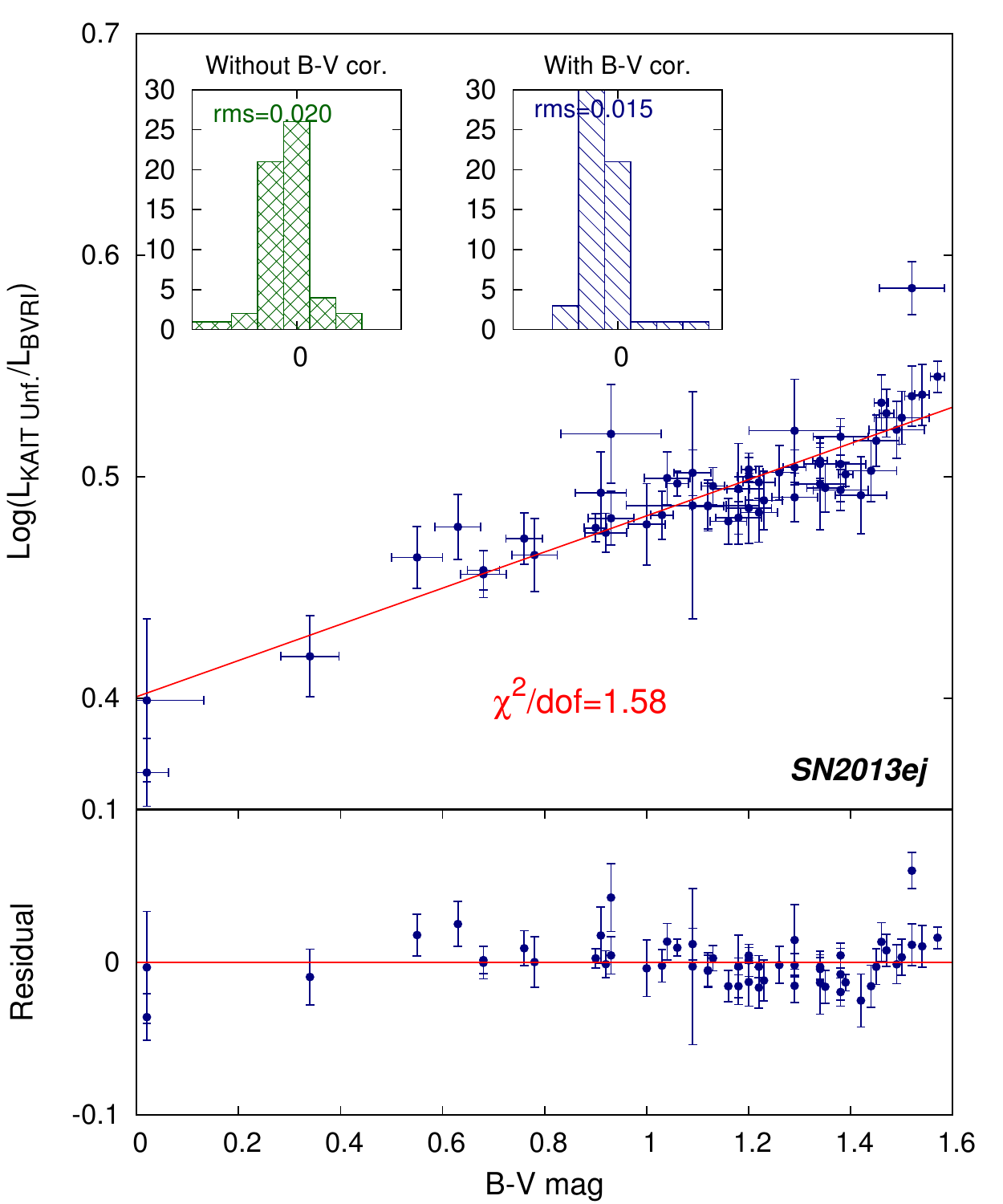}
\caption{{\it Top Panel}: Pseudo-bolometric $BVRI$ calibration of SN 2013ej 
from ROTSE unfiltered photometry compared to Konkoly $BVRI$ photometry. Residual from a $B-V$ color-dependent correction 
(histogram in blue) shows less than 5\% uncertainty. The histogram 
shown on the left (green) is obtained from the residuals by comparing the 
two fluxes without any color correction.
{\it Bottom panel}: Same as in top panel, but for KAIT unfiltered to 
KAIT $BVRI$ calibration. We obtain 6\% residuals by direct comparison 
and 4\% residuals when applying a $B-V$ dependence. Fit equations for $B-V$ dependence 
are given in Table \ref{tab:bvrifit}.}
\label{fig:rotsebvri}
\end{center}
\end{figure}

As a first step, from the fact that the open CCD transmission is broad, 
we establish a calibration relation for the ROTSE and KAIT unfiltered flux with integrated 
$BVRI$ flux as follows. Both $BVRI$ datasets are converted to 
absolute flux using the relations given by \cite{Bessel1998} corresponding 
to an A0 star (see Table A4 of their paper). The value of 
${\rm log}_{10}(L_{\rm ROTSE}/L_{BVRI})$ 
shows a direct relation with the $B-V$ color. A linear fit of 
${\rm log}_{10}(L_{\rm ROTSE}/L_{BVRI})$ versus $B-V$ is shown in the top panel 
of Figure \ref{fig:rotsebvri}. From the RMS of the residuals, we obtain a 
calibration precision of better than 5\%, while about 8\% precision is 
obtained from the residual without accounting for the $B-V$ dependence. 
A similar analysis for KAIT unfiltered and KAIT $BVRI$ datasets yields
 4\% and 6\% precision, respectively, as shown in the bottom panel of 
Figure \ref{fig:rotsebvri}. A summary of the fits is 
given in Table \ref{tab:bvrifit}.

\begin{table}
\begin{center}
\caption{$B-V$ Dependent Pseudo-Bolometric $BVRI$ and $UBVRI$ 
Calibration of ROTSE and KAIT Unfiltered Data}
\label{tab:bvrifit}
\begin{tabular}{lccc}
\hline
\hline
Calibration  & Intercept  &  Slope & $\chi^2/{\rm dof}$ \\
\hline
ROTSE Unf. \\
~~~~~~--~Konkoly $BVRI$  & $0.362\pm{0.004}$ & $0.1004\pm{0.004}$ & 0.54 \\
KAIT Unf. \\
~~~~~~--~KAIT $BVRI$ &   $0.401\pm{0.007}$ & $0.082\pm{0.006}$ & 1.58 \\
\hline
ROTSE Unf. \\
~~~~~~--~$UBVRI$  & $0.299\pm0.009$  & $0.161\pm0.007$  & 1.05 \\
KAIT Unf. \\
~~~~~~--~$UBVRI$  &  $0.363\pm0.006$  & $0.112\pm0.006$ & 2.71  \\
\hline
\end{tabular}
\end{center}
\end{table}

\begin{figure}
\begin{center}
\includegraphics[width=80mm,height=80mm]{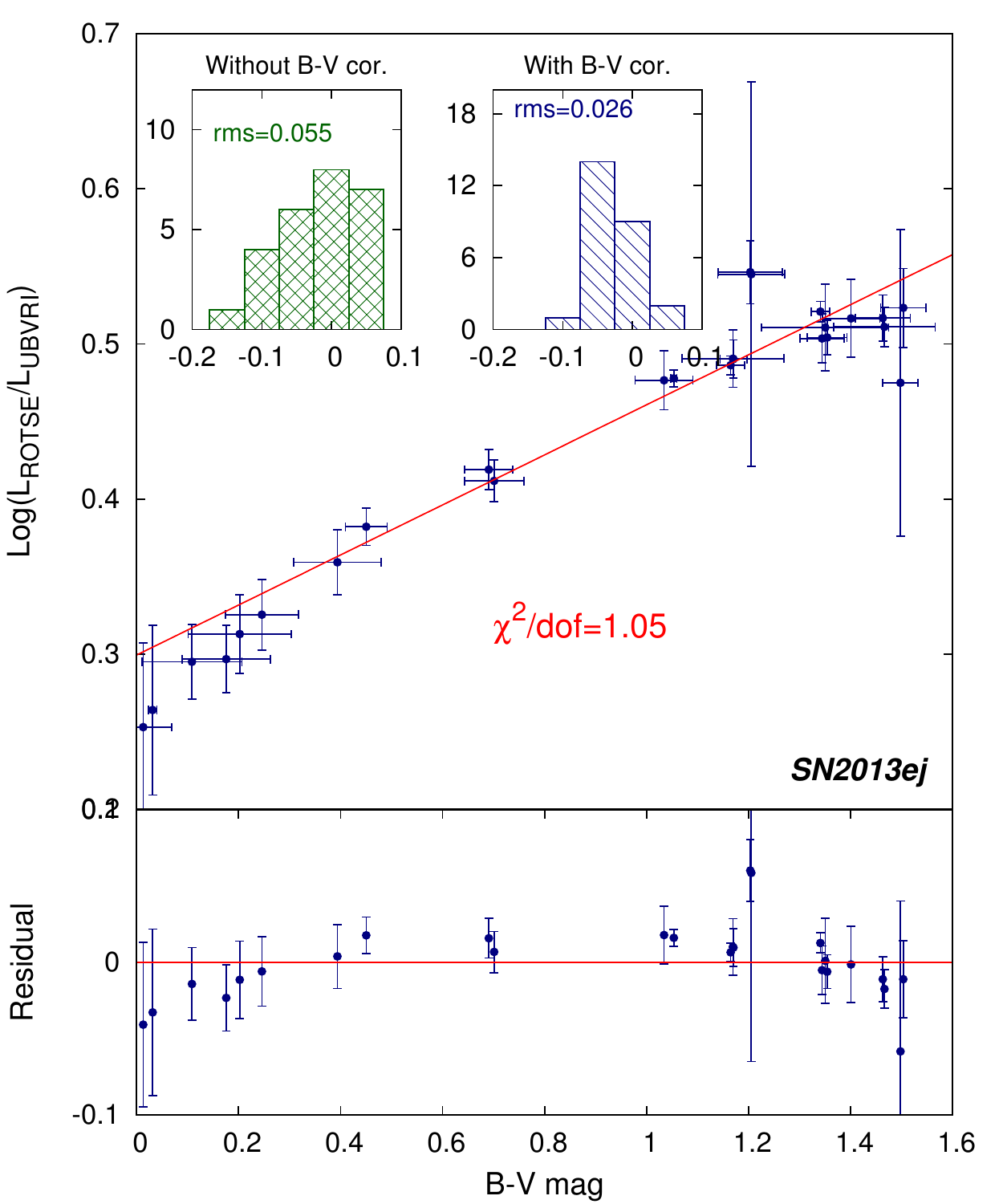}
\caption{Pseudo-bolometric $UBVRI$ calibration of SN 2013ej from the 
ROTSE luminosity. The $B-V$ color-dependent correction (histogram in blue) 
improves the measurement by about a factor of 2 compared to 
direct comparison (histogram in green).}
\label{fig:rotseubvri}
\end{center}
\end{figure}

In the second step, since ROTSE and KAIT unfiltered are also sensitive to the near-UV, we look at 
the behavior by integrating UV data from $Swift$. As pseudo-bolometric 
flux based on Johnson-Cousins $UBVRI$ filters is commonly derived, 
we first calibrate $Swift$ $u$ to the Johnson-Cousins $U$ band following 
\cite{Poole08} and integrate with the $BVRI$ dataset. We limit the integration 
to the wavelength range 3285--8750~\AA, where we have extended the lower 
and upper bounds by the half width at half-maximum intensity (HWHM) 
in the $U$ and $I$ bands. 
As before, from observations of the evolution 
of ${\rm log}_{10}(L_{\rm ROTSE}/L_{UBVRI})$, the RMS of the residuals after 
subtracting the mean reveals 13\% precision. Calibration with $B-V$ dependence improves 
the precision to 6\% as shown in Figure \ref{fig:rotseubvri}, where 
a $\chi^2/{\rm dof}$ of 1.05 is obtained for the fit. Analogously, KAIT unfiltered to $UBVRI$ (using $Swift~u$ and KAIT $BVRI$) 
yields about 5\% precision after a color-dependent correction, but with larger $\chi^2/{\rm dof}$.

\begin{table*}
\begin{center}
\caption{SN~IIP Bolometric Calibration Sample based on Well-Sampled 
Photometry from $U$ through $K$}
\label{tab:calib}
\begin{tabular}{lcccccc}

\hline
\hline
Object	& Host	& Distance (Mpc) & Total $E(B-V)$ (mag) & $V$ Plateau Slope (mag/100 days) & Feature & References	\\
\hline
SN 1999em & NGC 1637 & $11.7\pm{1.0}$  & 0.10 & $0.31\pm{0.05}$ & Normal & 1,2,3,4 \\
SN 2004et  & NGC 6946 & $5.6\pm{0.3}$ & 0.41 & $0.72\pm{0.03}$ & Over Luminous & 5,6 \\
SN 2005cs  & M51    & $8.4\pm{0.7}$ & 0.05 & $-0.10\pm{0.05}$ & Subluminous & 7,8 \\
SN 2013ej  &  M74     &  $9.0_{-0.6}^{+0.4}$  & 0.06 & $1.95\pm{0.06}$ & Normal & {\bf This paper} \\
\hline
\end{tabular}
\hspace{0.4in}
\\
(1) \cite{elmhamdi03};~~(2)~\cite{leo02};~~(3)~\cite{krisciunas09};~~(4)~\cite{leo03}\\
(5) \cite{sahu06};~~(6)~\cite{maguire10};~~(7)~\cite{pastorello09};~~(8)~\cite{vinko12} 
\end{center}
\label{tab:bolometriccalibration}
\end{table*}

It is very unusual to obtain a consistently complete set of data 
for a single object in all bands and still have minimal systematic effects. 
Various calibration and correction methods have been developed to better 
estimate the bolometric flux, but all are limited in one way or another. 
Here we revisit this problem based on the most extensive sets of data from 
the literature. The calibration sample is provided in Table \ref{tab:calib}. 
This set includes extensive photometry from the UV to the IR. 
To avoid any confusion, 
we dub the values obtained by integrating fluxes from data as ``$UBVRIJHK$," 
while we label an equivalent flux derived from our calibration as ``$UtoK$." 
The same convention also holds in other cases. By ``bolometric" flux, 
we mean integration from $Swift~uvw2$ at blue end to $H$ band in the red end, added with 
contribution from $K$ band as estimated below. We note that the IR flux past $K$ band 
will be significant as the SN cools over time. 
We have not accounted for any correction 
from beyond $K$ band in this procedure. 

The light curves shown in the top panel of Figure \ref{fig:bolometric} are 
derived by integrating data in the $UBVRIJHK$ bands, the wavelength range
3285--23,850~\AA. To obtain the $UBVRIJHK$ flux of SN~2013ej, we integrate the observed 
flux in the $u$ band from $Swift$ (after calibrating to Bessell $U$), $BVRI$ 
from the Konkoly or KAIT data, and near-IR $JH$ data from RATIR, where they are 
linearly extrapolated in the tail phase. We add an additional 
contribution from the $K$ band by estimating the average fractional flux in $K$ with respect to the $UBVRIJH$ flux using the calibration sample given in 
Table \ref{tab:calib}. We find that the $K$ band contributes $\sim 2$\% at +10~d, rising to 5--6\% at +80~d \cite{Huang15}. have published $K$ band data but their data is rather sparse. Comparing their $K$ band measurmemt with our estimated flux at matching epochs yielded an offset of less than 1\% of the total bolometric flux in the plateau while they both agreed in the tail.

From Figure \ref{fig:bolometric}, it is clear that the pseudo-bolometric 
$UBVRI$ flux is significantly lower compared to the bolometric flux, and 
the difference monotonically grows over time as the source cools.  
The bolometric luminosity declines very fast, by 0.4 dex in the 
first 30 days, and relatively slower by another 0.4 dex in the next 50 days. 
The $UBVRIJHK$ luminosity is significantly different from the bolometric 
luminosity only before about +20~d; 
otherwise, $UBVRIJHK$ closely resembles the bolometric flux. 
The bottom panel of Figure \ref{fig:bolometric} shows the fractional 
contribution from each of the UV, optical, and near-IR regions to the 
bolometric flux. The UV portion of the total flux drops from 
about 38\% at +8~d to below 10\% by +22~d. After this, the optical 
contribution drops very slowly and remains above 60\% until the end of 
the plateau, dropping slightly during the tail phase. The near- IR 
flux contributes 
about 40\% during the plateau phase, and remains almost constant in the tail.

\begin{figure}
\begin{center}
\includegraphics[scale=0.7,keepaspectratio]{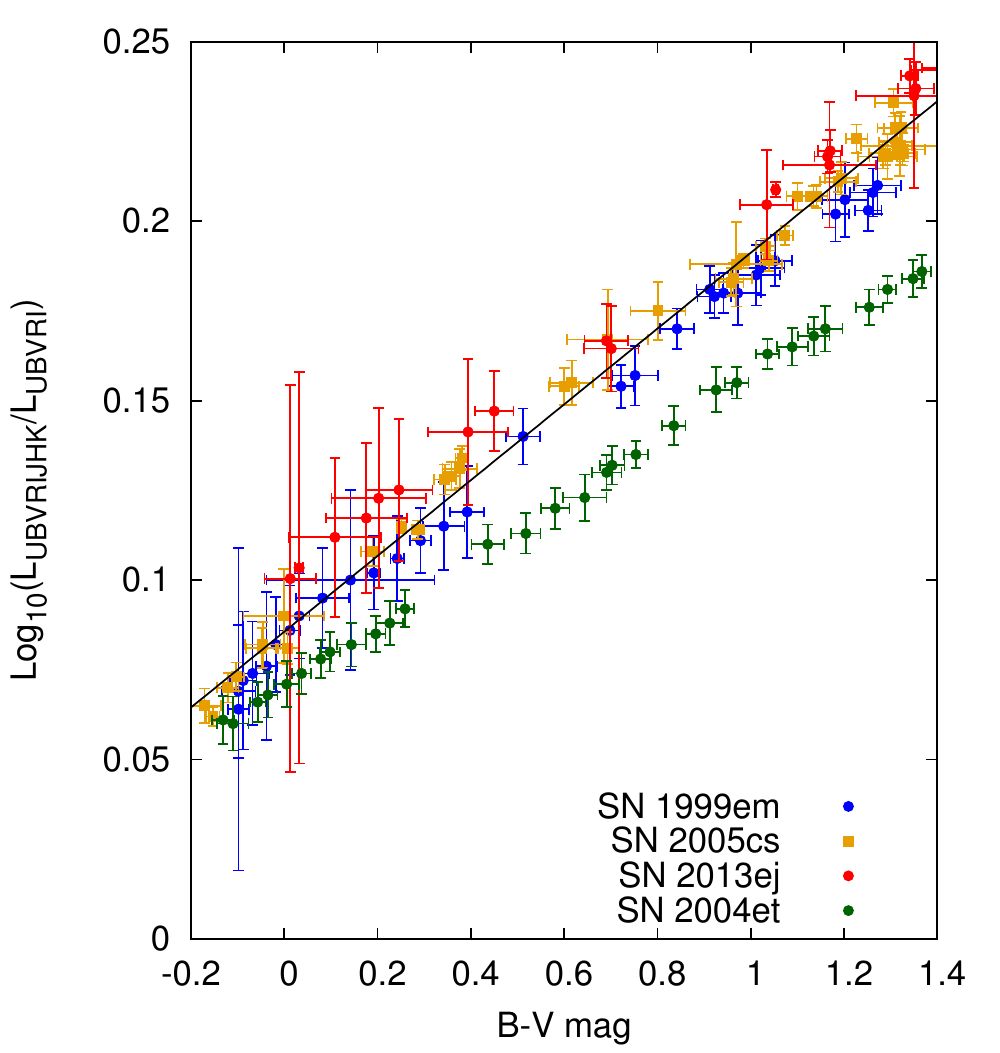}
\caption{Linear behavior of the log of the ratio of flux with $B-V$. 
A tight correlation of ${\rm log}_{10}(UBVRIJHK/UBVRI)$ with $B-V$ is seen for 
SNe 1999em, 2005cs, and 2013ej. The fit has $\chi^2/{\rm dof} = 1.51$. 
The atypical SN~IIP 2004et is shown and not included in the fit. 
The derived fit is given by Eq. \ref{eq:calib}.}
\label{fig:bol1}
\end{center}
\end{figure}

The log of the ratio of $UBVRIJHK$ luminosity to $UBVRI$ luminosity in 
the calibration sample (Table \ref{tab:calib}) shows a tight correlation 
with $B-V$ color. We fit ${\rm log}_{10}(L_{UBVRIJHK}/L_{UBVRI})$ versus 
$B-V$ color with a straight line using three of the four objects 
in this sample (Fig.\ref{fig:bol1}). SN~2004et is an energetic, 
atypical SN~IIP with largely uncertain $E(B-V)$; it is clearly an outlier, 
so we do not include it in the final fit given by Eq. \ref{eq:calib} below:

\begin{eqnarray}
&{\rm log}_{10}(\frac{L_{UtoK}}{L_{UBVRI}})~=~ (0.0856\pm{0.0012})\nonumber\\
&~~~~~~~~~~+(0.1056\pm{0.0012})\times(B-V).
\label{eq:calib}
\end{eqnarray}

\begin{figure}
\begin{center}
\includegraphics[scale=0.55,keepaspectratio]{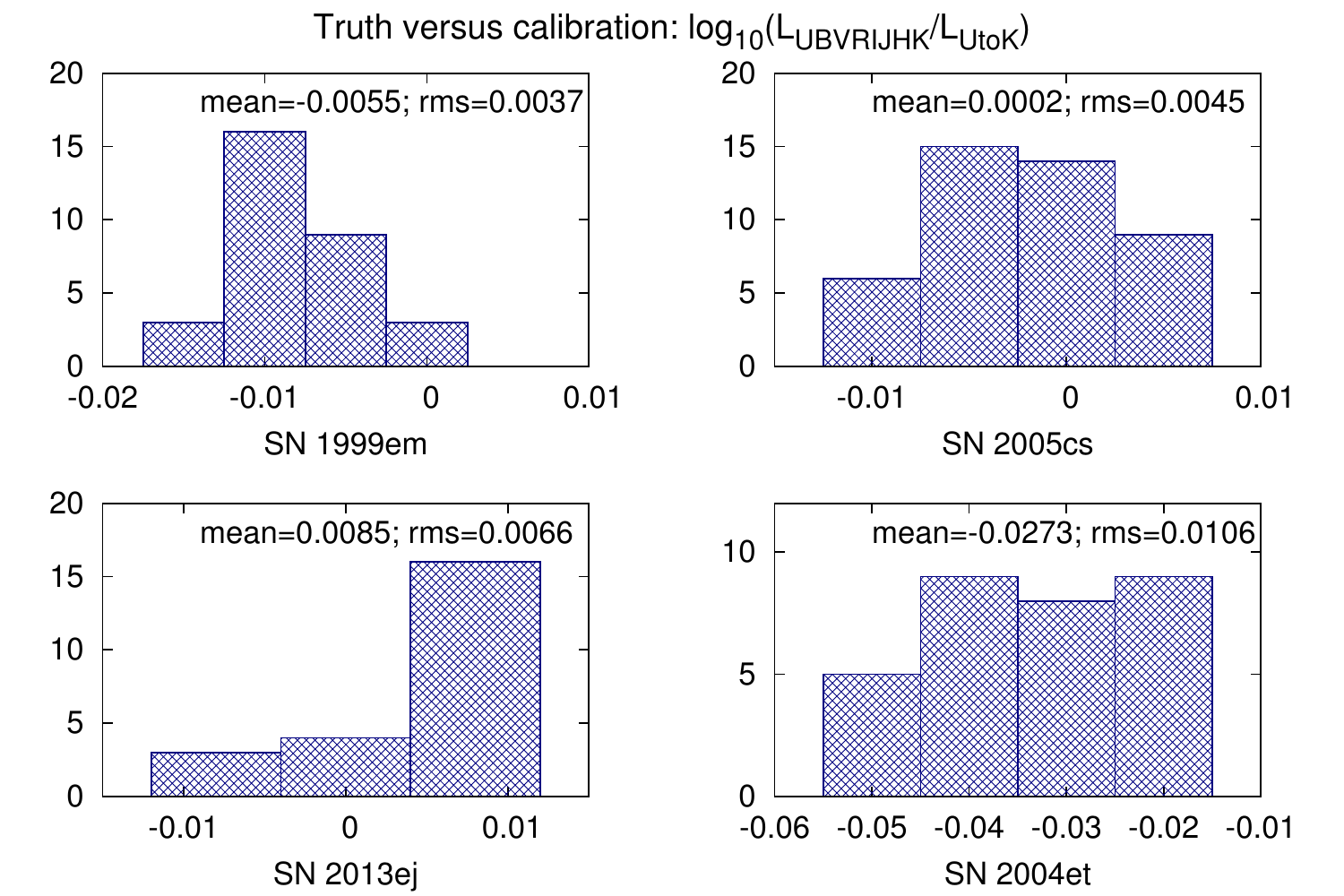}
\caption{Histograms obtained after subtracting ${\rm log}_{10}(L_{UtoK}/L_{UBVRI})$
from ${\rm log}_{10}(L_{UBVRIJHK}/L_{UBVRI})$. The mean and the RMS of the 
difference confirm 1--2\% precision for SN 2013ej, SN 1999em, 
and SN 2005cs. Applying the same calibration to SN 2004et 
(not included in the fit) yields $\sim 6$\% precision. 
Note that the values given in the figures are in ${\rm log}_{10}$.}
\label{fig:bolresidual}
\end{center}
\end{figure}

\begin{figure}
\begin{center}
\includegraphics[scale=0.65,keepaspectratio]{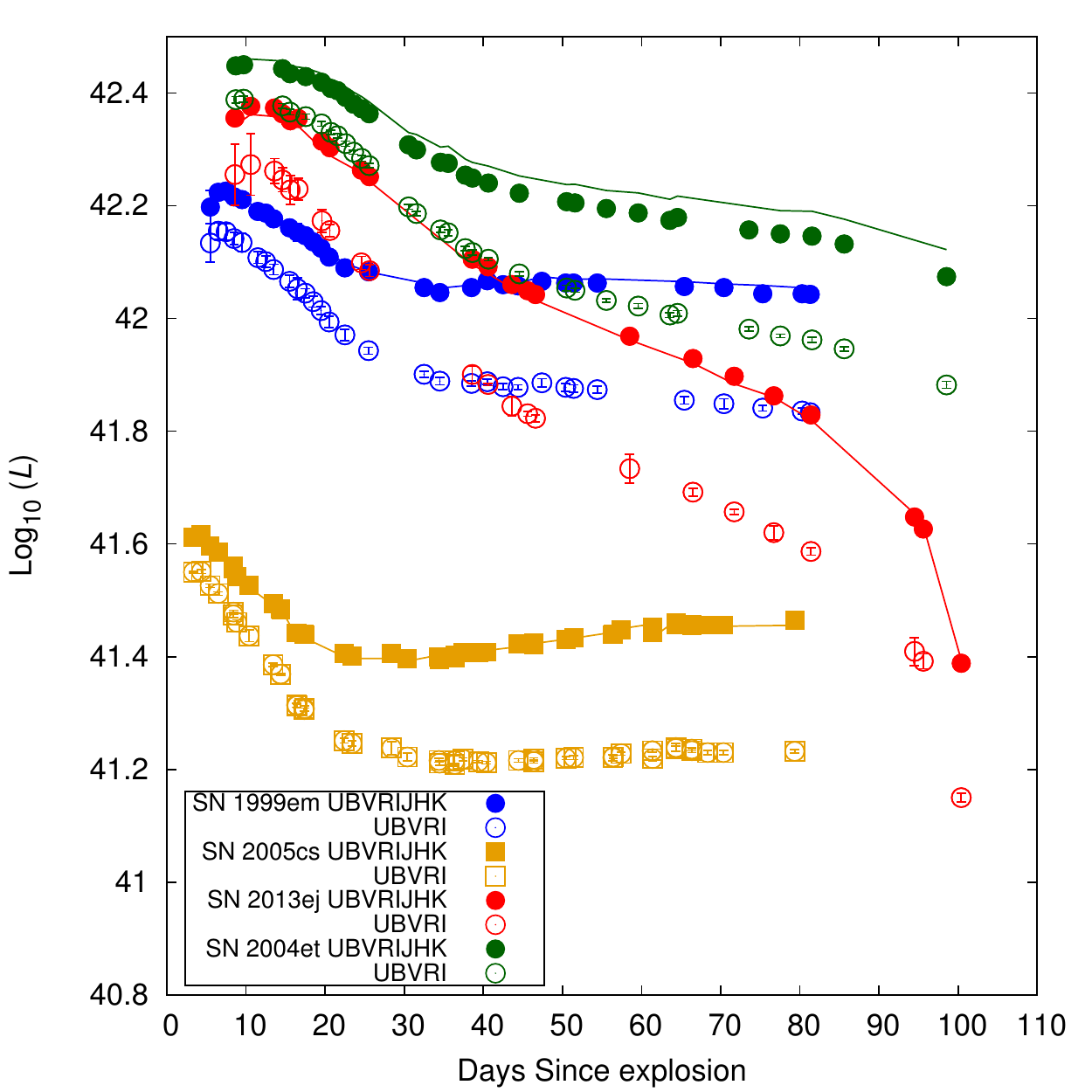}
\caption{Comparison of calibrated luminosity with measured luminosity. 
Solid lines represent $UtoK$ light curves while empty and solid circles 
represent $UBVRI$ and $UBVRIJHK$ light curves. While 1--2\% precision 
in the calibration is obtained for SN~2013ej, SN~2005cs, and SN~1999em, 
the outlier SN~2004et is also calibrated with about 6\% precision.}
\label{fig:bol2}
\end{center}
\end{figure}

\begin{figure*}
\begin{center}
\includegraphics[scale=0.9,keepaspectratio]{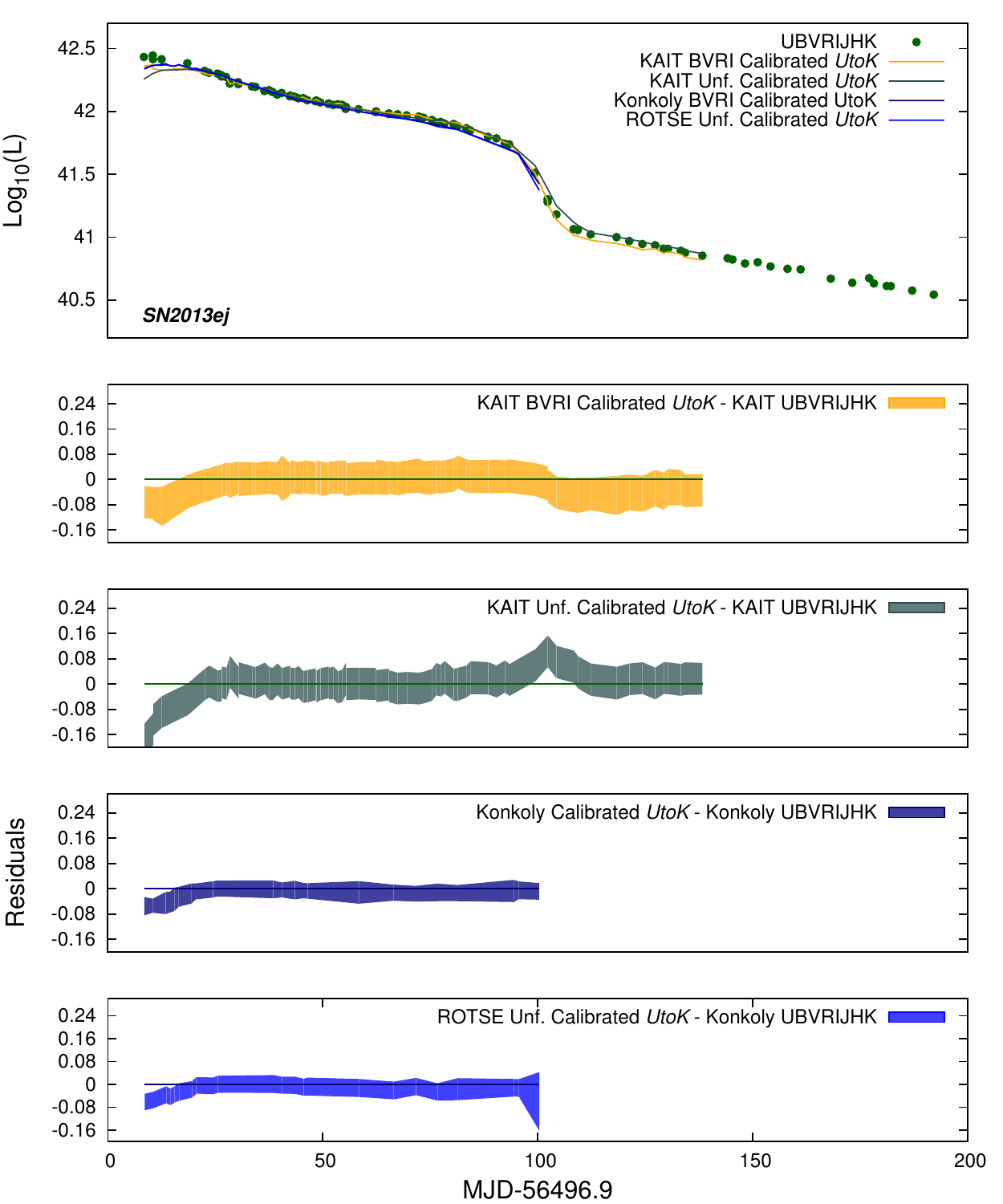}
\caption{$UtoK$ luminosity obtained from calibrating broadband 
and unfiltered photometry from KAIT, Konkoly, and ROTSE data for SN~2013ej. 
Data points $UBVRIJHK$ in the top panel are integrated U 
from $Swift$, KAIT $BVRI$, RATIR $JH$, and an
estimated $K$ flux (see text). Shaded regions in the residual 
plots indicate $1\sigma$ uncertainty from the RMS of the residuals. 
For KAIT data, we have added systematic uncertainty based on the 
differences of KAIT $UBVRIJHK$ and $UtoK$ values derived from 
$UBVRI$ luminosity using Eq. \ref{eq:calib}, which is derived 
using Konkoly data. The highest offset at very early times is a 
result of the high UV contribution to the total flux.}
\label{fig:bolfit}
\end{center}
\end{figure*}

The ratios of the $UBVRIJHK$ and calibrated $UtoK$ 
luminosities are shown in Figure \ref{fig:bolresidual}, while the 
$UBVRIJHK$ light curves are overlaid by calibrated $UtoK$ light curves 
using Eq. \ref{eq:calib} in Figure \ref{fig:bol2}. We can now 
combine the two-fold calibration: (1) ROTSE to $UBVRI$ using the 
fit shown in Figure \ref{fig:rotseubvri} (fit parameters are given in 
Table \ref{tab:bvrifit}), which gives $UtoI$, and (2) $UtoI$ 
obtained in the first step (equivalent to $UBVRI$) to $UtoK$ 
using Eq. \ref{eq:calib}. This yields the luminosity of SNe~IIP 
that have $B-V$ and unfiltered photometry. The relative 
uncertainty from this procedure is about 6\% or less. 
We perform the same analysis for KAIT unfiltered photometry. 
Figure \ref{fig:bolfit} shows the final calibrated $UtoK$ light 
curves from ROTSE and KAIT unfiltered photometry for SN~2013ej. 
Lower panels show the 
$1\sigma$ uncertainty from the calibration. Although the 
calibration was established by limiting to data before +102~d, it appears 
to provide reasonable estimates for the bolometric luminosity even during 
the early nebular phase (see Fig. \ref{fig:bolfit}).

\section{Spectroscopy} \label{spectro}
\subsection{Key spectral features} \label{keyspec}

  We present 17 optical spectra of SN~2013ej from the HET, Kast, and DEIMOS 
spectrographs in Figure~\ref{fig:sn13ej_spec}. All of the 
spectra are corrected for the recession of the host galaxy using 
$z = 0.002192$ (NED/IPAC Extragalactic Database\footnote{https://ned.ipac.caltech.edu/}). 
This is consistent with that determined by the Supernova Identification software
(SNID)\footnote{http://people.lam.fr/blondin.stephane/software/snid/}\citep{blondin07} 
with a fitted redshift of 0.002. 
Early-time spectra at +8~d, +9~d, and +11~d are primarily blue continuum, with a few 
P~Cygni profiles of neutral H Balmer lines and He~I lines. The opacity for all other
ions is too low to be conspicuously observed as spectral features at these early phases. 
H~I lines (H$\alpha$ $\lambda6562.85$, H$\beta$ $\lambda4861.36$, and H$\gamma$ $\lambda4340.49$) 
are very broad at early times. The strength of emission component of H~I lines decreases with time. 
At +11~d, O~I $\lambda7775$ is glimpsed. A week later at 
+19~d, several strong absorption signatures of SNe~II appear. 

Interestingly, the 
absorption line to the blue of H$\alpha$ is unusually strong. This feature, which we identify 
as Si~II $\lambda6355$ (see Section \ref{lineid}), subsequently becomes 
stronger until +19~d in our sample. It appears as
a small absorption 
notch at +44~d and disappears by +48~d. \citet{V14} 
showed this feature to get stronger than H$\alpha$ until +21~d in their dataset, and to
become weaker than H$\alpha$ by +23~d. 
The Si~II identification was also favored by \citet{V14}, \citet{bose15}, and \citet{Huang15}.
Si~II has seemed to occur much later in other SNe~IIP. While this strong early appearance of 
Si~II has not been observed previously, it may have been marginally detected 
at +10~d and +12~d and was not observed after +25~d in SN 2006bp \citep{Quimby2007}. 
Si~II is comparatively much stronger than in SN 2006bp at similar epochs. 
For SN 2006bp, the Si~II velocity 
profile evolves faster than that of H$\alpha$ before +25~d, 
whereas for SN~2013ej, 
it is more smooth and evolves more slowly than the H$\alpha$ velocity. All these factors make 
SN~2013ej exhibit unusual and strong early Si~II. 

As the ejecta expand, the subsequent spectral evolution of SN~2013ej shows typical 
SN~IIP singly ionized lines of Ca~II, Fe~II, Ti~II, Sc~II, and Ba~II. The He~I $\lambda$5876 
line gradually gets weaker until +15~d and is not seen at +19~d. This is evidence of the 
temperature decreasing below the critical temperature of excitation. More iron-group 
elements start to appear, corresponding to the commencement of the plateau 
phase where the photosphere penetrates deeper into the envelope. The 
same disappearance of He~I at around +16~d was also seen in SN~1999em 
\citep{leo02} and recently in SN~2012aw \citep{Bose13}. The Na~I~D lines are 
considerably stronger than the lines of other neutral elements, presumably coming from 
non-LTE effects \citep{Hatano99}. The Na~I~D feature is observed after +19~d and is probably 
blended with He~I at +15~d. We do not observe any narrow lines of Na~I.
No obvious evidence of high-velocity features (HVFs) is seen in our spectra. 
These observations may indicate negligible interaction of the ejecta 
with the circumstellar material (CSM). After +19~d, the
Ca~II near-IR triplet can be dissociated to at least a doublet at 8520~\AA\ and 
a singlet at 8662~\AA; however, the profile is well blended before +15~d, and we 
adopt this as a single Ca~II near-IR profile to determine the change in ion velocity with time. 

\subsection{Spectral Homogeneity in the UV}

\begin{figure}
\begin{center}
\includegraphics[scale=0.6,keepaspectratio]{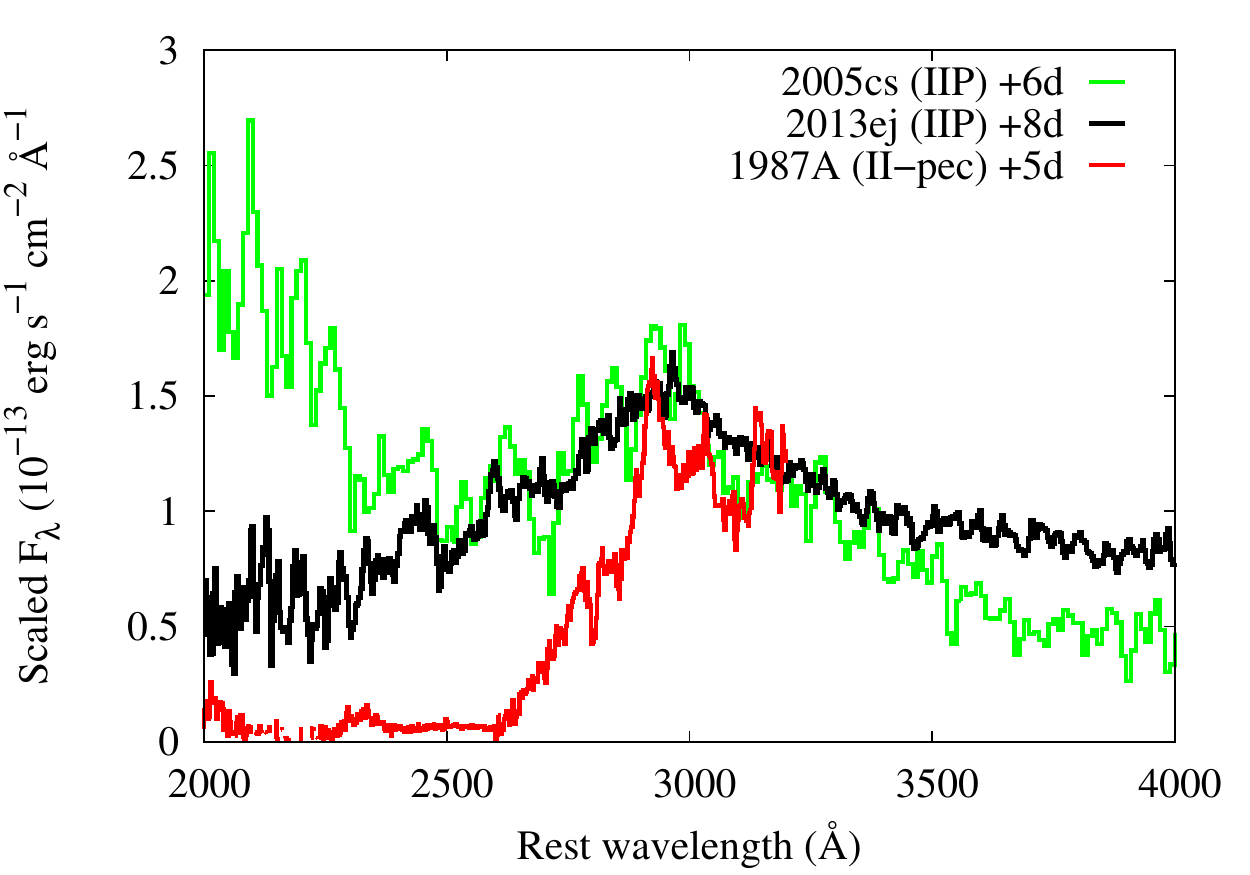}
\includegraphics[scale=0.6,keepaspectratio]{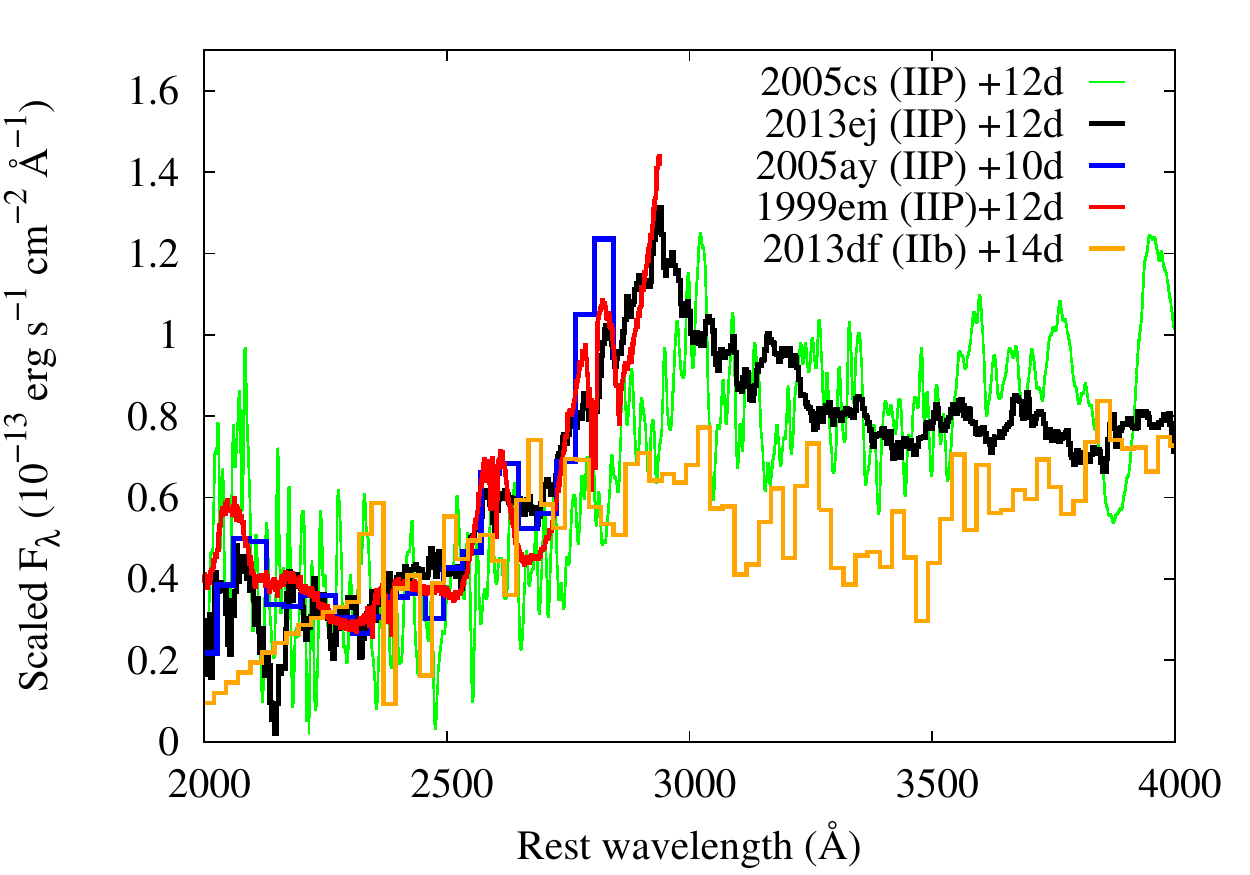}
\caption{Evidence of spectral homogeneity of SNe~IIP in the UV at early times. {\it Top panel}: SN~2013ej UV 
spectrum compared with the atypical SN 1987A, which shows a sharp UV cutoff, while the excess flux
of SN~2005cs below 2500~\AA\ is suspected to be coming from a different source. {\it Bottom panel}: Homogeneous 
SN~IIP UV sample at $\sim 12$~d. Also shown for comparison is a SN~IIb spectrum, which is clearly 
distinct from the rest.}
\label{fig:uvsp2}
\end{center}
\end{figure}

SNe~IIP are known to exhibit a remarkable homogeneity in their
UV spectra, as first pointed out by \citet{galyam08}. They found that
the early-phase UV spectra (2000--3000~\AA) of 
SNe 1999em,  2005ay, and 2005cs are very similar, both in the shape
of the continuum as well as in the visible spectral features. 
In comparison, \citet{benami15} recently pointed out that SNe~IIb, which are 
thought to have thinner H-rich envelopes than regular SNe~IIP, 
display relatively strong diversity in their UV spectra. 

The paucity of well-observed SNe~IIP having early-time UV spectra
impedes an in-depth study of this homogeneity verses diversity issue at present. 
It is therefore important to increase the size of the early-time UV sample. SN~2013ej is a valuable
addition to this sample because of its relative proximity, which enabled $Swift$ to
obtain near-UV spectra with its UVOT/UGRISM instrument (see Fig.~\ref{fig:uvsp}). 
Figure~\ref{fig:uvsp2} compares the +8~d and the +11~d spectra 
to those of other SNe~II taken at similar phases. All of these spectra
are corrected for interstellar extinction and scaled to match the fluxes 
in the region 2500--3000~\AA. 

Figure~\ref{fig:uvsp2} reveals that SN~2013ej nicely fits into the framework
of the UV spectral homogeneity of SNe~IIP, at least around 
10--12 days after explosion. We find that the similarity is not evident 
for spectra taken at $\sim 1$ week after explosion (Fig.~\ref{fig:uvsp2}, top panel)
in our sample.
Both SN~1987A and SN~2005cs showed some differences with respect to the
spectrum of SN~2013ej at this phase, although the rise of the UV flux in the
SN~2005cs spectrum below 2500~\AA\ may not be real. Close inspection of the 
UVOT/UGRISM frames revealed that this spectrum was contaminated 
by emission from the zeroth order of a nearby
source. Moreover, SN~1987A, which shows a sharp cutoff in the UV flux below
3000~\AA, was not a typical SN~IIP, as it had a blue supergiant progenitor. 
Nevertheless, the spectra taken around $11 \pm 1$ days after explosion confirms the
observed similarity nicely (Fig.~\ref{fig:uvsp2}, bottom panel). Figure \ref{fig:uvsp2} also
illustrates a SN~IIb UV spectrum at a similar epoch; it differs significantly 
from the SN~IIP sample. 

\begin{figure*}
\begin{center}
\includegraphics[width=1\textwidth]{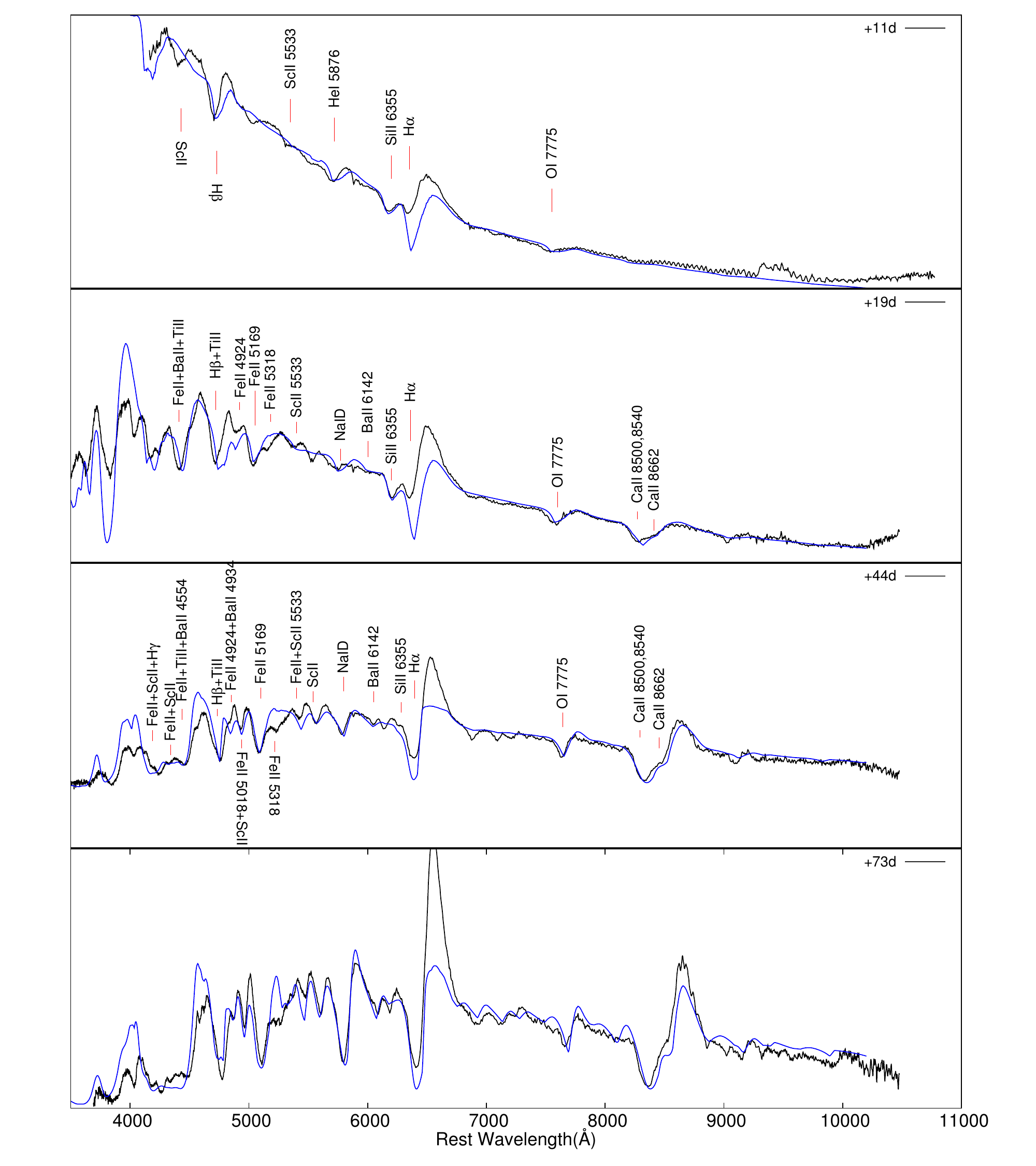}
\caption{Example Syn++ fits of SN~2013ej spectra are shown in blue while data are in black. 
The fits mostly reproduce the observed features. The inability to accurately 
reproduce the H~I line profile is perhaps a limitation of the model being 
purely scattering based and not accounting for the emission from recombination cascades and the NLTE effects. SN~2013ej exhibits most previously identified 
SN~IIP spectral features.}
\label{fig:syn++_804+812}
\end{center}
\end{figure*}

\subsection{Line Identification and Spectrum Modelling} \label{lineid}
Line identifications of most of the features in Section \ref{keyspec}
were first driven by the study of \cite{Hatano99} on ion signatures 
of SN spectra. Additional study and confirmation was performed 
by Syn++ \citep{Thomas11} modelling of a few of the optical spectra 
as shown in Figure~\ref{fig:syn++_804+812}. While the synthetic modelling produces most of the ionic signatures, the most obvious H$\alpha$ profile is not reproduced. 
This reflects the limitation of a purely scattering code: the emission is underestimated because it does not account 
for the emission due to recombination cascades. For H$\alpha$, there may also be a significant effect from 
non-thermodynamical equilibrium (NLTE) and time varying effects. The absorption 
notch blueward of the H$\alpha$ line in the +11~d and +19~d spectra 
is fitted with the Si~II line, and we have obtained the fit as shown 
in Figure \ref{fig:syn++_804+812}. One could argue that this feature is an 
HVF of H$\alpha$, but then we would expect to also see HVFs of
other Balmer lines. Taking an HVF input with such a high velocity, 
we were unable to reproduce a decent overall fit. Because of the lack 
of an HVF for other Balmer lines and no HVFs seen in the near-IR spectra, 
as reported at similar epochs by \cite{V14}, the HVF hypothesis is disfavored. Clearly, the SiII identification hypothesis will be settled with higher confidence only from more realistic modelling.
While \cite{bose15} also identified the blueward notch as a Si~II feature,
they have incorporated HVFs for H~I lines, albeit blended with the photospheric component, 
accounting for the broad Balmer lines beyond +42~d in their sample.
After +15~d, lines of intermediate-mass elements and iron-group elements 
start to appear. The s-process products Ba~II and Sc~II are also seen from
+19~d until the last spectra in the plateau at +94~d.

\subsection{Velocity Evolution} \label{velo}
While the average ejecta velocity is a direct tracer of kinematic properties, 
the photospheric velocity not only provides compositional clues 
but also traces the size of the photosphere, thereby aiding 
distance measurements (e.g., EPM). The photospheric velocities 
at early spectral epochs are estimated by fitting the 
He~I $\lambda 5876$ feature. After +19~d, the Fe~II $\lambda 5169$
line is most indicative of the photospheric velocity, 
since the minimum of the absorption profile tends to form 
near the photosphere \citep{branch03}.

\begin{figure*}
\begin{center}
\includegraphics[width=1\textwidth]{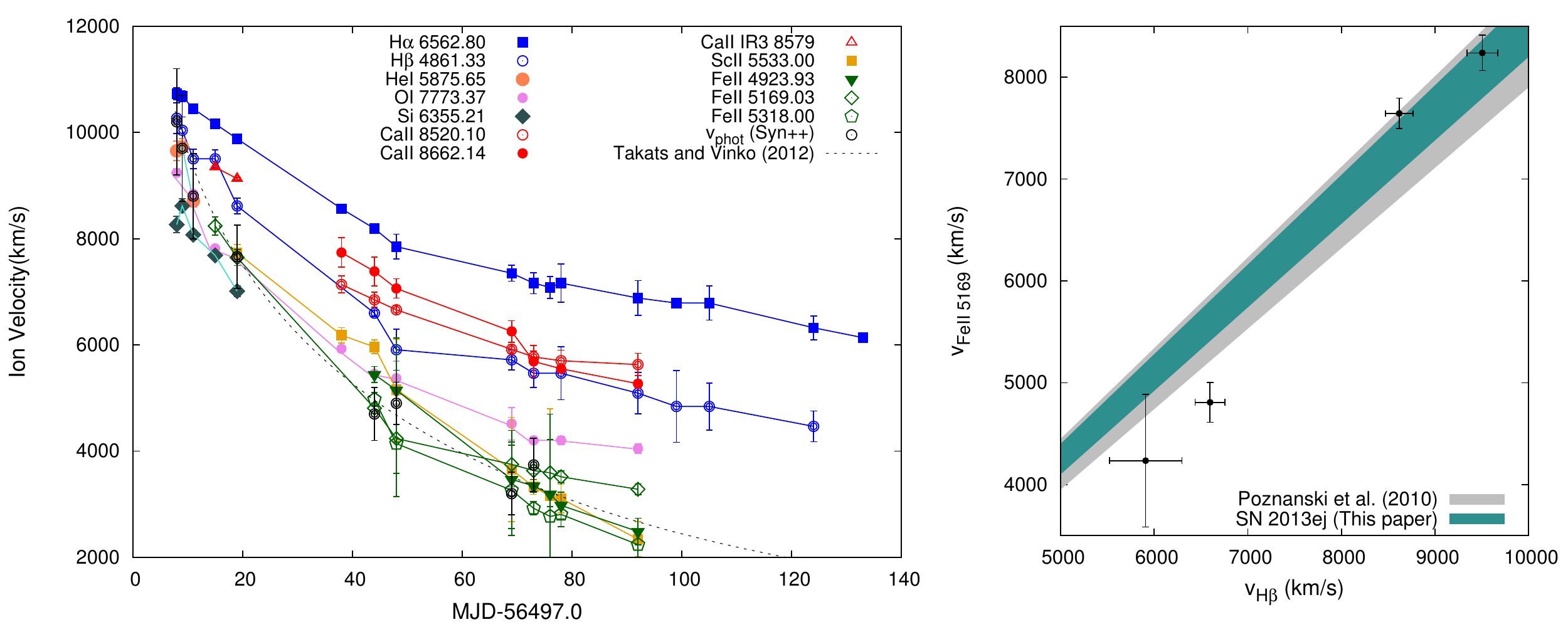}
\caption{$Left~Panel:$ SN 2013ej velocity evolution of strong ions. 
Empty black circles are the photospheric velocity derived from Syn++ fits. 
The dashed line is a fit to the photospheric velocity using the method 
of \cite{vinko12} (see text). $Right~Panel:$ Demonstration of correlation 
of $v_{\rm Fe~II}$ with $v_{{\rm H}\beta}$, as suggested by~\cite{poznanski10}. 
The shaded regions indicate the $1\sigma$ region of the correlation: 
gray, \cite{poznanski10}; navy, SN~2013ej (this paper).}
\label{fig:ion_vel}
\end{center}
\end{figure*}

The velocity evolution of some of the strongest ions is presented in 
Figure~\ref{fig:ion_vel}. Each line absorption feature is fitted by a 
Gaussian profile, and the minimum is converted to velocity 
using the relativistic Doppler equation. The H$\alpha$ line 
is decelerating more slowly than H$\beta$ and other metallic 
ions as expected, but the H~I lines show a flat velocity profile, 
which was also pointed out by \cite{bose15}. 
\cite{poznanski10} demonstrated a correlation of velocity of the 
Fe~II $\lambda5169$ line ($v_{\rm Fe~II}$) with that of the H${\beta}$ 
line ($v_{{\rm H}\beta}$) using 28 optical spectra of 13 SNe~IIP covering 
5--40 days after explosion. \cite{T12} have extended the validity 
of this relation to phases beyond 40 days. We have examined this 
behavior by taking four optical spectra of SN~2013ej from 15--48 days 
after explosion. Analysis of Fe~II is not justified before +15~d in our sample. 
We find that the velocities from SN~2013ej spectra are consistent with 
this correlation, as can be seen in the right panel of Figure
\ref{fig:ion_vel}. We obtain a linear relation of 
$v_{\rm Fe~II}~=~0.85\pm0.03~v_{{\rm H}\beta}$ for SN 2013ej, in agreement with 
$v_{\rm Fe~II}~=~0.84\pm0.05~v_{{\rm H}\beta}$ as obtained by \cite{poznanski10}.
 
\begin{table}
\begin{center}
\caption{Photospheric velocities of SN~2013ej determined by Syn++ fitting.
Phases are rounded to the nearest day since explosion}
\label{tab:synvel}
\begin{tabular}{lccc}
\hline
MJD & Phase & $v_{\rm phot}$ & Uncertainty \\
\hline
    & (days) & (km~s$^{-1}$) & (km~s$^{-1}$) \\
\hline 
56505.5 & +8 & 10200 & 1000 \\
56506.5 & +9 & 9700  & 1000 \\
56508.5 & +11 & 8800 &  800 \\
56516.5 & +19 & 7660 &  600 \\
56541.5 & +44 & 4700 &  500 \\
56545.5 & +48 & 4900  & 400 \\
56566.5 & +69 & 3200  & 400 \\
56570.5 & +73 & 3740 & 500 \\
\hline
\end{tabular}
\end{center}
\end{table}

In order to refine the photospheric velocity using more than one feature, 
we first approximate the velocity by estimating from the absorption 
minimum of the He~I $\lambda5876$ line for the two earliest spectra and the
Fe~II $\lambda5169$ line for the later spectra. We then performed 
synthetic spectral modelling with Syn++ and extracted photospheric 
velocities from the model. Velocities obtained from the fits are 
given in Table \ref{tab:synvel}.

\section{Distance Determination} \label{dist}
Recently, the distance to the SN~2013ej host, M74, was subjected to 
a number of measurements, from the value of $D \approx 7 \pm 2$ Mpc 
\citep{sharina96, vinko04, vandyk06} to $D \approx 9.5 \pm 0.5$ Mpc
\citep{zasov96, oliva10}; see Table \ref{tab:m74dist} for a summary.
Here we revisit this issue by inferring
the distance to SN~2013ej via EPM.
Modern versions of EPM have been applied for 
various samples of SNe~IIP (\citealt{hamuy01, leo02, dessart08, jones09,
vinko12, bose14, takats14}). We present the application of the
version presented by \cite{vinko12} by combining the data from 
two SNe that occurred in the same host galaxy, claiming the uncertainties
of EPM can be reduced and the reliability of the derived distance improved.
Thus, we take the advantage of having the necessary data for
both SN~2013ej (this paper) and SN~2002ap \citep{vinko04}, although
the latter object is a broad-lined SN~Ic for which the application
of EPM may not be fully justified. 
Despite the complications arising in modeling the atmospheres of such
stripped-envelope (SE) CC~SNe, we show
below that the combination of the two datasets results in surprisingly
consistent results, and the inferred distance is in very good agreement
with recently published independent estimates.

Following the procedure described by \cite{vinko12}, the basic equation
for EPM is
\begin{equation}
t ~=~ {D \times \left ( {\theta \over v_{\rm phot}} \right ) + t_0},
\label{eq:epma}
\end{equation}
where $t$ is the time, $D$ is the distance, $\theta$ is the angular radius of the photosphere, 
$v_{\rm phot}$ is the velocity of the photosphere at $t$, and $t_0$ is the moment of shock breakout. 
We estimate $\theta$ from the bolometric light curve by using
\begin{equation}  
\theta ~=~ {1 \over \zeta(T)} \sqrt{{f_{\rm bol}} \over {\sigma T_{\rm eff}^4} },
\label{eq:epmb}
\end{equation}
where $\zeta(T)$ is the dilution factor describing the alteration of the pure blackbody flux 
in a scattering-dominated SN atmosphere as a function of temperature \citep{E96, dessart05} 
and $f_{\rm bol}$ is the apparent bolometric flux. 
For SN~2013ej, we used the dilution factors determined by \cite{dessart05}, which are valid 
for H-rich SNe~IIP, but not for the H-free SE SN~2002ap. Since the atmospheres of such SE~SNe 
are much less known, we set $\zeta = 1$ as a first approximation. Note that the usage 
of $\zeta = 1$ worked surprisingly well when calculating the distance to the Type IIb 
SN~2011dh \citep{vinko12}. Since the ejecta of the Type Ic SN~2002ap contained practically no H, 
unlike the Type IIb SN~2011dh, the dilution of the blackbody flux due to Thompson scattering on
free electrons might be even less strong than in the case of Type II SNe~2013ej or 2011dh. 
Thus, setting $\zeta \approx 1$ may be a physically realistic approximation for SN~2002ap,
although its full justification would involve the computation of an NLTE model atmosphere
for SN~2002ap which is beyond the scope of this paper. Nevertheless, we estimate the 
probable amount of the systematic error of the distance introduced by the assumption 
of $\zeta = 1$ below.

The estimates of $\theta$ were based on the bolometric light curve of SN~2013ej, as described 
in Section \ref{bolo}. Moreover, we applied the $f_{\rm bol}$ fluxes of SN~2002ap similarly, 
after combining the optical light curves from \cite{F03}, \cite{P03}, and \cite{vinko04} with the 
near-IR measurements by \cite{Y03}. In the latter case, the UV contribution was estimated by 
assuming zero flux at 3000~\AA\ and a simple linear SED between 3000~\AA\ and the $U$ band. 
This approximation was justified by the shape of the spectra of SN~2002ap as they declined 
below 4500~\AA\ toward the blue (e.g., \citealt{vinko04}). 

The application of Eq.\ref{eq:epma} and \ref{eq:epmb} requires $v_{\rm phot}$ and $T_{\rm eff}$ values at 
several epochs, typically during the first 30--50 days after explosion. These can also 
be estimated directly from the observations. In case of SNe~Ic like SN~2002ap, the 
applicability of EPM is limited to no longer than a few weeks.

For SN~2013ej, values of $T_{\rm eff}$ were obtained in Section \ref{temp}. 
For SN~2002ap, we applied the reddening value of $E(B-V) = 0.09$ 
mag \citep{vinko04} and the relation
\begin{equation}
T_{\rm eff} ~=~ -0.122 (B-V) ~+~ 3.875
\end{equation} 
based on the SN~Ic-BL models by \cite{mazzali00}.  
  
In order to increase the sampling of the
velocity curve of SN~2013ej, we fit the velocity curve for 
SNe~IIP derived by \cite{T12} to the velocities obtained in Table 
\ref{tab:synvel}. This method involves velocity modelling as a power-law expansion of phase,
given a model velocity at some epoch. This model velocity is generally derived from synthetic modelling of the observed spectra 
or by direct measurement from the absorption profile of lines like Fe~II $\lambda5169$. 
See \cite{T12} for
a more detailed discussion of this kind of velocity measurement in SN~IIP  atmospheres.
The result of this fitting was applied in the procedure of EPM.

For SN~2002ap, we adopted the velocities based on the 
Si~II $\lambda6355$ feature as given by \cite{vinko04}.

The derived quantities needed for EPM are shown in Table~\ref{tab:epm}. 
The moments of the explosion were set to be $t_0$ = MJD~$56496.9$ (2013 July 23.9 UT), as derived in Section \ref{early}) 
and $t_0$ = MJD~$52302.0$ (2002 Jan 28.0 UT) for SNe~2013ej and 2002ap, respectively.

\begin{table}
\caption{Physical Quantities Derived for EPM.}
\label{tab:epm}
\begin{tabular}{rccc}
\hline
time & $\theta$ & $\theta / v_{\rm phot}$ & Uncertainty \\
\hline
(days) & ($10^8$ km Mpc$^{-1}$) & (day Mpc$^{-1}$) & (day Mpc$^{-1}$) \\
\hline
SN~2013ej \\
\hline
8.60 &9.44 &1.08 &0.12 \\
10.60 &10.10 &1.24 &0.12 \\
13.60 &11.45 &1.52 &0.17 \\
14.60 &12.78 &1.74 &0.19 \\
15.60 &12.98 &1.81 &0.20 \\
16.60 &13.30 &1.90 &0.20 \\
19.60 &13.93 &2.13 &0.22 \\
20.60 &15.31 &2.39 &0.24 \\
24.60 &16.39 &2.77 &0.28 \\
25.60 &15.80 &2.72 &0.28 \\
\hline
SN~2002ap \\
\hline
4.89 &11.96 &0.44 &0.34 \\
6.48 &12.34 &0.67 &0.30 \\
7.48 &13.00 &0.78 &0.32 \\
9.87 &14.70 &1.06 &0.35 \\
10.87 &15.17 &1.17 &0.37 \\
11.27 &15.53 &1.19 &0.38 \\
12.87 &16.15 &1.55 &0.40 \\
13.47 &16.11 &1.51 &0.39 \\
13.86 &16.10 &1.69 &0.39 \\
\hline
\hline
\end{tabular}
\end{table}

\begin{figure} 
\begin{center}
\plotone{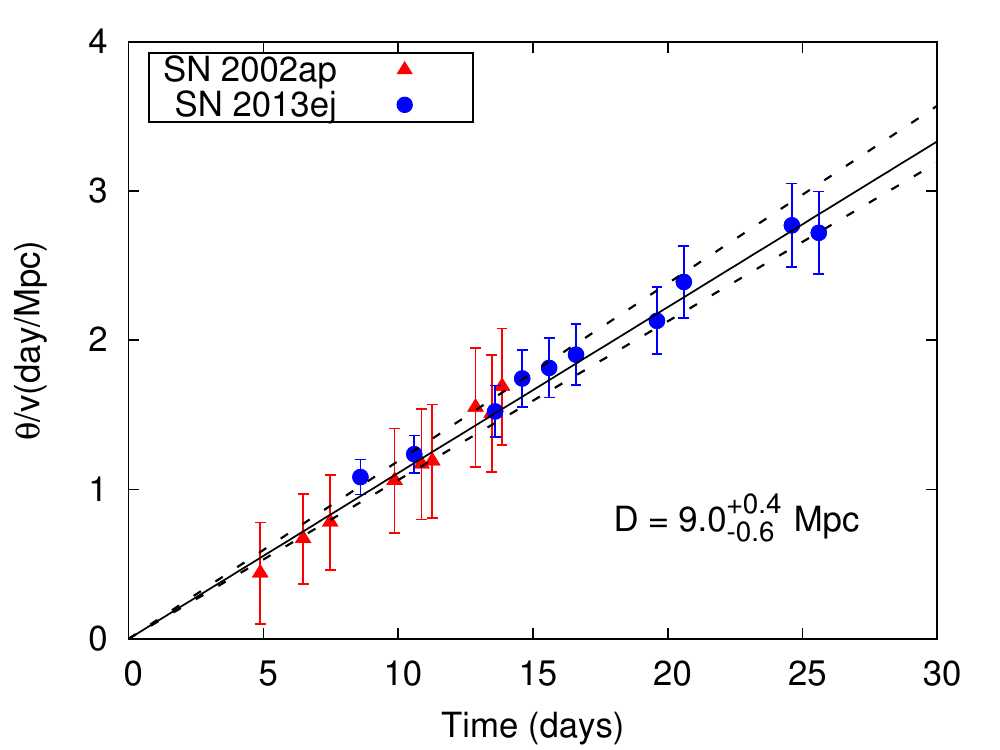}
\caption{Distance measurement of M74 using SN 2013ej and SN 2002ap. The black solid line shows
the final result that yields a distance of 9.0 Mpc, while the dotted lines mark its uncertainty.}
\label{fig:epm}
\end{center}
\end{figure}

The fit of Eq. \ref{eq:epma} to the data in Table~\ref{tab:epm} was performed assuming 
$\theta/v$ as the independent
variable, with either keeping $t_0$ fixed at the values given above or 
letting it float. 
The first fit resulted in $D = 8.86\pm0.21$ Mpc, while the second one gave $D 
= 9.09\pm0.30$ Mpc 
with $\Delta t_{0} = -0.59\pm0.47$ being consistent with our estimated $t_{0}$.
Alternatively, choosing $t$ as the
independent variable, one may get $D = 8.93\pm0.10$ Mpc and $D = 9.25\pm0.30$ Mpc 
with $\Delta t_0= 0.09\pm0.48$ days for fixed and floating $t_{0}$, respectively. 
The weighted average of these four values gives $D = 8.96\pm0.08$ Mpc. 
Accounting for any systematic effect that could have been introduced from our 
derived $t_{0}$ in Section \ref{early}, we take lower and upper bounds 
for $t_0$ as $-1.3$~d (obtained from floating index, see Section \ref{early}) 
and +0.9~d (Lulin detection epoch) from the derived $t_{0}$. Fitting with adjusted 
lower and upper bounds of $t_0$, we get 0.35 Mpc higher distance and 0.60 Mpc 
lower distance, respectively. We adopt these offsets as the 
systematic uncertainty and add in quadrature with the statistical uncertainty 
obtained above. Thus, $9.0_{-0.6}^{+0.4}$ Mpc is  adopted as the final
distance estimate of M74 from EPM using two SNe. This derived distance can be 
found from the inverse of the slope of the line shown in Figure~\ref{fig:epm}.

In order to test the effect of choosing $\zeta = 1$ artificially for SN~2002ap, we repeated
the fitting process described above after applying the dilution factors of \cite{dessart05} to the SN~2002ap data as well. This is clearly an overestimate of the effect of electron scattering 
(i.e. an underestimate of the $\zeta$ values)
in a Type Ic SN atmosphere, which may be somewhat less scattering-dominated than
a H-rich Type IIP atmosphere. However the amount of the systematic error introduced by such a strong
dilution might be useful for constraining the real uncertainty of the distance due to
the approximate dilution factors. Assigning the $\zeta(T)$ values from \cite{dessart05}
to SN~2002ap move those data (plotted with red triangles in Fig. \ref{fig:epm} by about $\sim 1\sigma$ 
upward, reducing the consistency between the two datasets. Performing the same
fitting process as described above, we get $D = 8.3\pm0.6$ Mpc, i.e. less than
$2 \sigma$ difference from the previous distance estimate from $\zeta = 1$. Since this test 
uses potentially underestimated values of $\zeta$ for SN~2002ap, we conclude that the uncertainty caused
by the inaccurate knowledge of the dilution factors for SN~2002ap probably does not exceed
the $\pm 0.5$ Mpc uncertainty estimated above.

 Using an independent dataset for SN~2013ej only, \citet{richmond14} applied the standard 
version of EPM to get $D = 9.1$ $\pm 0.4$ Mpc. From the values given in their Table~5, the uncertainty of that distance appears to be around $\sim 0.8$ Mpc instead of $\pm 0.4$ Mpc as noted. In either case, this is in very good agreement with our result.
Furthermore, we estimated the distance using the bolometric calibration 
for ROTSE unfiltered fluxes derived in Section \ref{bolo}. 
For this case, we include only data points beyond +15~d for SN~2013ej, as earlier 
data would include significant flux below the $U$ band that is not 
included in the calibration procedure. We get an EPM distance of $9.7\pm0.6$ Mpc, 
using the calibrated fluxes 
from ROTSE for SN~2013ej, in agreement with the previous derivation. 
Note that we have not combined SN~2002ap values in this case. 
Allowing upper and lower bounds to $t_0$ as before, we get the final 
distance estimate from calibration of ROTSE data to be $9.7_{-0.7}^{+0.9}$ Mpc, which is consistent 
with our preferred $9.0^{+0.4}_{-0.6}$ Mpc result.

\begin{table}
\caption{Recent Distance Estimates for M74.}
\label{tab:m74dist}
\begin{tabular}{lccl}
\hline
\hline
Method & $D$ & Reference \\
 & (Mpc) & \\
\hline
T-F & $9.68\pm1.63$ & \cite{tully88} \\   
BBSG & $7.31\pm1.23$ & \cite{sharina96} \\
Disk gravitational & $9.4$& \cite{zasov96} \\
~~~~~~~stability \\
Light echo & $7.2$& \cite{vandyk06} \\
SCM & $9.91\pm1.2$ & \cite{oliva10} \\
~~~~~~~(SN 2003gd) \\
EPM (SN 2002ap) & $6.7$ & \cite{vinko04} \\
TRGB & $10.2\pm0.6$ & \cite{jang14} \\   
{\bf EPM} & ${\bf 9.0_{-0.6}^{+0.4}}$ & {\bf present paper} \\
{\bf EPM: ROTSE}    &    ${\bf 9.7_{-0.7}^{+0.9}}$ & {\bf present paper} \\
~~~~~~~{\bf calibrated} \\
\hline
\end{tabular}
\end{table}

\section{Explosion Properties} \label{explo}
\subsection{Ni Mass} \label{nimass}

 The end of the plateau phase is believed to indicate the full recombination of 
hydrogen when the ionization front, and thus the photosphere, reaches the bottom 
of the hydrogen envelope in the ejecta. After this epoch, the light curve 
proceeds into a nebular phase. The subsequent luminosity is driven by the 
radioactive decay of elements that were produced during the explosion, so 
the light curve shows a characteristic exponential decay of flux output. 
This suggests that the gamma-rays and positrons from radioactive decay of $^{56}$Co 
thermalize in the ejecta. Here, we first assume the full trapping of gamma-rays and positrons in the 
ejecta. As the mass of freshly synthesized Ni should be proportional to the 
tail luminosity, we use the findings from the literature and use scaling to 
determine the nickel mass ($M_{\rm Ni}$) for SN 2013ej. 

\cite{Bose13} derived 
the $M_{\rm Ni}$ for SN~2012aw using the $UBVRI$ light-curve tail luminosity. 
As we will see below that the decline rate changes after +183d, we linearly fit the 
$UBVRI$ light curve from +120~d to +183~d and extrapolate to find the luminosity at 240~d to make a direct comparison with their result for SN2012aw. 
The luminosity, $L$(240~d) for SN~2013ej, is estimated to be $1.32\pm0.05\times10^{40}$ erg~s$^{-1}$, 
while that for SN~2012aw was found to be $4.53\pm0.11\times10^{40}$ erg~s$^{-1}$. 
The ratio is calculated to be $0.29\pm0.02$. Noting $M_{\rm Ni}$ for SN~2012aw to 
be $0.058\pm0.002$ M$_\odot$ \citep{Bose13}, we calculate for SN~2013ej, 
$M_{\rm Ni} = 0.017\pm0.001$ M$_\odot$. Alternatively, we use the 
method of \citet{hamuy03} to calculate the Ni mass from the tail 
luminosity ($L_{t}$), using the equation
\begin{equation}
 M_{\rm Ni}~=~7.866 \times 10^{-44} \, L_{t} \, {\rm exp}\Big[{{(t_{t}-t_{0})/(1+z)-6.1}\over 111.26}\Big]~ {\rm M}_\odot.
\end{equation}
We calculated $L_{t}$ at 20 epochs from +120~d to +183~d using the late-time 
$V$-band magnitude. We adopt the same bolometric correction of 0.26 mag
from \citet{hamuy03}. The weighted mean tail luminosity is calculated to 
be $5.82\pm0.26\times10^{40}$ erg~s$^{-1}$, corresponding to +157~d. The $M_{\rm Ni}$ value is 
then calculated to be $0.018\pm0.002$ M$_\odot$. Following the same procedure, 
but using the bolometric light curve derived in Section \ref{bolo}, 
we get $M_{\rm Ni}$ to be $0.019\pm0.003$ M$_\odot$. We take the weighted mean of the 
above three results as our best estimate of the synthesized radioactive material. 
This yields $M_{\rm Ni} = 0.018\pm0.001$ M$_\odot$

\cite{hamuy03} used a large sample of SNe~IIP to study the correlation between 
the Ni mass and mid-plateau (+50~d) photospheric velocity. From the modelling as 
described in Section \ref{velo}, we have derived a photospheric velocity 
of 4500 km~s$^{-1}$ at +50~d. Our results for $M_{\rm Ni}$ and $v_{50}$ are consistent 
with the results of \cite{hamuy03}. 

\begin{figure}
\begin{center}
\plotone{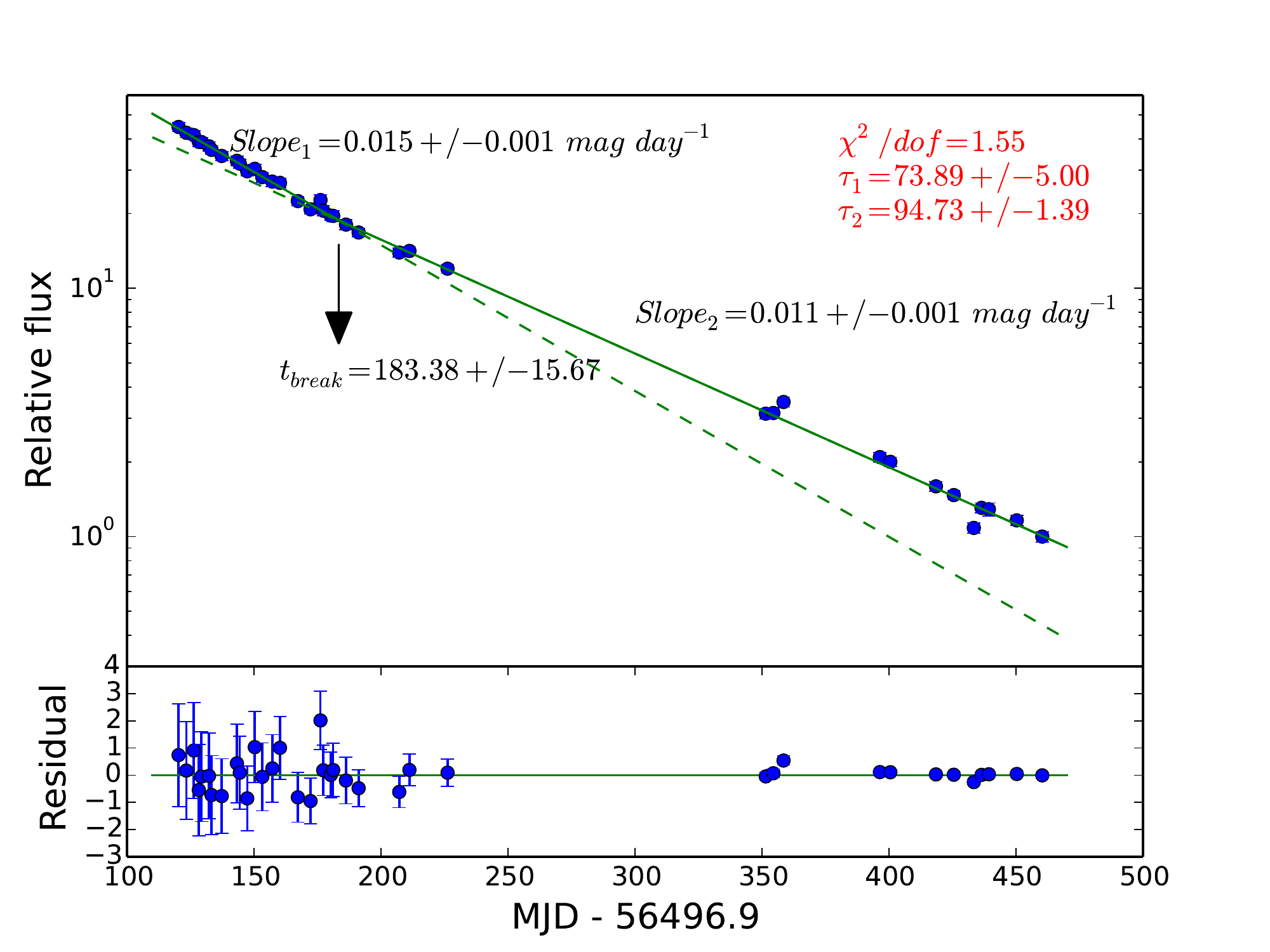}
\caption{Fit of the bolometric light curve of SN~2013ej beyond +120~d with a broken exponential law. The solid line is the fitted model while the dashed lines are the extrapolation of each exponential models in either direction. The crossover is at break time, $t_{br}$= $+183.4 \pm 15.7$ d. }
\label{fig:Lbolexp}
\end{center}
\end{figure}

We note, however, that from the late time data of SN~2013ej from the KAIT and Nickel telescopes, we observe two distinct slopes in the tail (see Fig. \ref{fig:Lbolexp}). To estimate the time of slope break, we fit the late time bolometric flux beyond +120~d with a broken exponential law of the form,
\begin{equation}
F(t)~=~S A e^{-\frac{t}{\tau_{1}}} \Big[1+e^{\alpha(t-t_{br})}\Big]^{\frac{1}{\alpha}(\frac{1}{\tau_{1}}-\frac{1}{\tau_{2}})}
\label{eq:expbreak}
\end{equation}
where, $\tau_{1}$ and $\tau_{2}$ are characterstic times for the first and second exponential profiles, $A$ is the initial flux, $t_{br}$ is the break time, $\alpha$ is the smoothing parameter and $S$ is the scaling factor, given by,
\begin{equation}
S~=~(1+e^{-\alpha t_{br}})^{\frac{1}{\alpha}(\frac{1}{\tau_{1}}-\frac{1}{\tau_{2}})}
\label{eq:scalefactor}
\end{equation}

Eq. \ref{eq:expbreak} has been analogously applied to study the radial profile of surface brightness from the disks of galaxies \citep [e.g.][]{mateos13}. The best fit parameters are found to be, $\tau_{1}=73.89\pm5.00$ days, $\tau_{2}=94.73\pm1.39$ days, $t_{br}=183.28\pm15.67$ and $\alpha=0.23\pm1.14$ with the $\chi^2/dof$ from the fit to be 1.55. The slopes before and after the break point are obtained to be, $Slope_{1}=0.015\pm0.001$ $\rm mag~day^{-1}$ and $Slope_{2}=0.011\pm0.001$ $\rm mag~day^{-1}$. While $Slope_{1}$ is found to be much steeper, $Slope_{2}$ is closer to the $^{56}{\rm Co} \rightarrow~^{56}{\rm Fe}$ decay rate.

To see the effect of these two distinct decline behaviors on the initial Ni mass, we further fit the late-time bolometric light curve of SN~2013ej with the simple model 
described in \citet{vinko04} and \citet{valenti2008}. This
model assumes an optically thin ejecta heated by the 
radioactive decay of $^{56}$Ni and $^{56}$Co. The decay energy
is emitted in the form of gamma-rays and positrons, which
may be partially trapped in the ejecta, thermalize and
emerge again as low-energy (mostly optical or near-infrared) 
photons. The deposition function for the gamma-rays 
at a given epoch $t$ can be expressed as 
\begin{equation}
D_{\gamma} ~=~ 1 - e^{-\tau_\gamma} ~=~ 1 - exp[-({\frac{T_0(\gamma)}{t})^2}],
\label{eq:Dgamma}
\end{equation}
where $\tau_\gamma$ is the optical depth for gamma-rays in the whole ejecta.
The timescale of the gamma-ray optical depth decrease \citep{wjc2015} is 
\begin{equation}
T_0(\gamma) = \sqrt{C \kappa_\gamma M_{ej}^2 / E_{kin}},
\label{eq:T0gamma}
\end{equation}
where $\kappa_\gamma$ is the gamma-ray opacity, $M_{ej}$ is the ejecta mass and $C$ is a constant depending on
the density distribution in the ejecta. For simplicity, we assumed a constant
density ejecta, which implies $C = 9 / 40 \pi$. 
The deposition function for positrons, $D_{+}$, takes the same form, 
except for the opacity. For the gamma-ray opacity, we 
adopted $\kappa_\gamma = 0.027$ cm$^2$~g$^{-1}$ and for positrons, we set $\kappa_{+} = 7$ cm$^2$~g$^{-1}$
 \citep[e.g.][]{Colgate80, valenti2008}. 

With these definitions, the late-time bolometric luminosity can be expressed
as 
\begin{eqnarray}
& L_{bol} ~=~ M_{\rm Ni} [ (S_{\rm Ni}(t) + 0.92 S_{\rm Co}(t)) D_\gamma \nonumber\\
&~~~~~~~~~+~ (0.03 + 0.05*D_\gamma) S_{\rm Co}(t) D_{+} ],
\label{eq:Lbol}
\end{eqnarray}
where $M_{\rm Ni}$ is the initial mass of the radioactive $^{56}$Ni synthesized during the explosion,
$S_{\rm Ni}$ and $S_{\rm Co}$ are the functions for the total energy input from the Ni- and Co-decay,
respectively (see \citet{szalai16,branchwheel16} for further discussion). This equation corrects for the typographical error in the expression given by \citet{valenti2008},
and accounts for the partial trapping of both gamma-rays and positrons via the deposition
functions given above. Since this light curve model assumes instantaneous release of the thermalized
deposited energy from radioactive decay, without considering any photon diffusion unlike the model 
of \citet{arnett1980}, it is applicable only when the ejecta is almost fully transparent in 
the optical, i.e. during the nebular phase. 

\begin{figure}
\begin{center}
\plotone{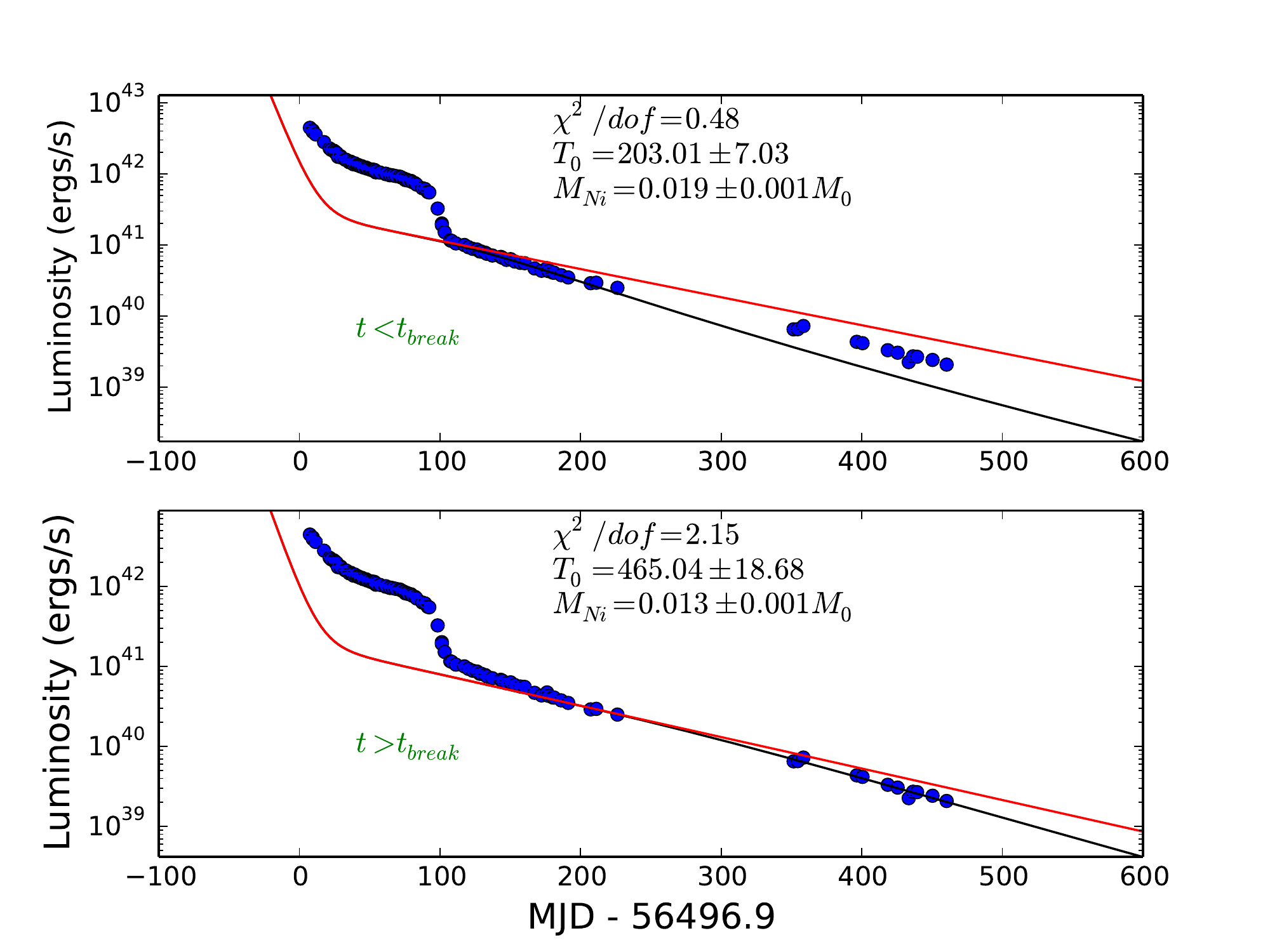}
\caption{Fit of the radioactive decay model of Eq. \ref{eq:Lbol} to the bolometric light curve tail of SN~2013ej. Pre-break time and post break time lightcurves are fitted separately. The upper panel shows the fit taking data between +120~d and the break point at +183~d. The lower panel shows the same fit taking data beyond the break time. The leakage
of both the gamma-rays and positrons were taken into account in the model (see text). The red line represents full trapping of gamma-rays and positrons on both panels.}
\label{fig:Lbolfit}
\end{center}
\end{figure}

The fit of Eq. \ref{eq:Lbol} to the bolometric lightcurve is plotted in Fig. \ref{fig:Lbolfit}.
We found that before +183~d, SN~2013ej exhibited a steeper decline in the bolometric
light curve than the rate of the $^{56}$Co decay, as also found by \cite{Huang15}
and \cite{bose15}; however, our extended photometry, revealing a shallower decay rate, 
has an implication for both the gamma-ray opacities and the estimated Ni-mass. 
From the fit beyond the break point of +183~d, the values inferred for the 
initial nickel mass and the gamma-opacity timescale are 
$M_{\rm Ni} = 0.013 \pm 0.001$ $M_\odot$ and $T_0(\gamma) = 465 \pm 18$ days. This timescale is
significantly longer than the value of $\sim 173$ day found by \cite{bose15} from the
data between +100 and +200 days. Our longer $T_0(\gamma)$ suggests that the light curve
of SN~2013ej was probably not yet settled on the radioactive tail before +183 days. The $Ni$ mass derived 
from the fit before break time, $M_{\rm Ni} = 0.019 \pm 0.001$ $M_\odot$ is consistent with our calculations above using methods of \cite{Bose13} and \cite{hamuy03}, and with previous independent estimates by Huang et al. (2015) and Bose et al.(2015), 
both assuming full gamma-ray and positron trapping in about the same time window, but all those estimates are 
probably overestimates and the lower value derived by restricting the analysis beyond +183~d is more likely to be correct.

\subsection{Explosion Physical Parameters}

To determine accurate explosion properties, a detailed hydrodynamic study is required. 
Recently, \citet{Huang15} have done such a study while \citet{bose15} use a 
semi-analytic approach following \citet{arnettfu89}.  In order to make an 
approximate estimate, we use the approach of \citet{LN85}. 
Even though there are complications concerning radioactive heating effects, 
inclusion of higher explosion energies in the models, simple 
physical assumptions will still be useful to compare the explosion 
parameters with those of the more elaborate studies.

From plateau length ($\Delta t$), the mid-plateau absolute $V$ magnitude $M_{V}$, 
and the corresponding expansion velocity, \citet{LN85} derive the explosion energy, 
ejected mass, and pre-SN radius. For the plateau midpoint of SN 2013ej, 
we use $(t_{\rm peak}+t_{\rm p})/2$; where $t_{\rm peak}$ is the  epoch of peak brightness 
in $V$, estimated to be +15~d using Gaussian Process regression of $V$ band data until +30d, 
and $t_{\rm p}$ is the epoch at which the plateau ends. 
It is nontrivial to precisely locate the end of the plateau. 
To determine this epoch, we again perform a Gaussian Process regression on the bolometric lightcurve from +80~d 
to +130~d, and determine the point of inflection from the obtained fit to be +109~d. 
We consider this to be the end of the plateau, $t_{\rm p}$. 
We therefore {\ estimate a plateau of length $94\pm7$ with +62~d 
as the midpoint}. The value of $v_{\rm ph}$  at +62~d is found to be $3800\pm500$ km~s$^{-1}$ from 
the fit described in Section \ref{velo}, while $M_{V}$ is determined to 
be $-16.47\pm0.04$ mag from linear interpolation of $V$ magnitudes from +50~d to +70~d. 
Using these values, we derive an explosion energy of $0.9\pm0.3 \times10^{51}$ ergs, 
and a pre-SN radius of $250\pm70$ R$_\odot$ based on \citet{LN85}. \citet{bose15} find 
an explosion energy of $2.3\times10^{51}$ ergs and a 
pre-SN radius of 450 R$_\odot$. \citet{Huang15} obtain values ranging 
over $0.7~-~2.1 \times10^{51}$ ergs and 230~--~600 R$_\odot$, respectively.  
Given the unstated
uncertainty in \citet{bose15}, and the range from \citet{Huang15}, 
our calculations 
appear to be consistent with theirs.  Our explosion energy and pre-SN
radius are also in agreement with measurements with the SN IIP sample
studies of \citet{hamuy03} and \citet{Nadezhin03}, both of which use the same
\citet{LN85} relations we used.

	  We further use these relations to determine the ejecta mass, $M_{ej}$, 
for SN~2013ej to be $13.8\pm4.2$ M$_\odot$.  \citet{hamuy03} and \citet{Nadezhin03} 
obtain $M_{ej}$ in the range 14--56 M$_\odot$ and 10--30 M$_\odot$, repsectively, which 
encompasses our result. As for the explosion energy and pre-SN radius, 
the ejecta mass is highly sensitive to plateau length; better knowledge 
of the plateau length can yield more accurate results.  For SN~2013ej
specifically, \citet{Huang15} and \citet{bose15} find ejecta masses 
of 10.6 M$_\odot$ and $12\pm3$ M$_\odot$, respectively. \citet{nagyvinko16} have 
derived an ejecta mass of 10.6 M$_\odot$ from light curve modeling 
using a two-component model incorporating a dense inner core and an extended low mass envelope. These 
three estimates are all consistent with our measurement based on \citet{LN85} relations. 
If we assume a remnant mass of 1.4 M$_\odot$, our
measurement for the final pre-explosion progenitor mass is $15.2\pm4.2$ M$_\odot$.

	    Other measurements have been performed of SNe IIP progenitor mass
in general or SN~2013ej specifically.  Care is required in comparing 
progenitor masses, however, as some correspond to initial pre-explosion 
masses or $ZAMS$ masses rather than the final pre-explosion progenitor mass we calculated above. 
\citet{Fraser14}, for instance, found the $ZAMS$ mass to
be in the 8--15.5 M$_\odot$ range for SN 2013ej, consistent with an M-type 
supergiant, from the archival HST images.  Using X-ray observations, 
\citet{sayan15} derived a $ZAMS$ mass of 14 M$_\odot$, accounting for 
the derived steady
mass loss of $3\times10^{-6}$ M$_\odot$ $yr^{-1}$ over the last 400 years.  Given the 
uncertainties, these measurements are consistent with our final progenitor 
mass of $15.2\pm4.2$ M$_\odot$.  For the general population of SNe IIP, \citet{smartt09b} 
obtain a $ZAMS$ mass range of 8--17 M$_\odot$. Use of nebular phase modelling \citep[e.g.][]{jerkstrand14} can also independently provide tighter constraints on the $ZAMS$ masses of these events. However, studies involving hydrodynamical modelling \citep[e.g.][]{UtrobinChugai09, UtrobinChugai13} and stellar evolutionary models \citep[e.g.][]{smartt09b}, along with nebular spectra modelling, have shown conflicts in the derived initial mass of the progenitor stars.

\begin{table*}
\caption{Calculated Physical Parameters of SN 2013ej.}
\label{tab:param}
\begin{tabular}{lccccc}
\hline
\hline
Parameter & \cite{bose15} & \cite{Huang15} & \cite{Fraser14}\tablenotemark{a} & \cite{V14} & {\bf This paper} \\
\hline
Explosion Energy ($10^{51}$ ergs)  & 2.3    & 0.7--2.1 & -- &  --  &  {$\bf 0.9\pm0.3$ }  \\
Progenitor Mass (M$_\odot$)  & $14.0\pm3.0$  & 12--13 & 8--15.5 & -- & {$\bf 15\pm4.2$} \\
Pre-SN Radius (R$_\odot$) & $450\pm112$  &  230--600   &  --    &  400--600 & {$\bf 250\pm70$}  \\
$M_{Ni}$ (M$_\odot$)  &   $0.019\pm0.002$ & $0.02\pm0.01$  & --   &  -- & {$\bf 0.013\pm0.001$}   \\
Plateau Duration (Days)   &  $\sim{85}$   & $\sim{50}$  & -- & -- & {$\bf 94\pm7$ } \\
Distance Assumed (Mpc)  & $9.57\pm0.7$ & $9.6\pm0.7$ & $9.1\pm1.0$ & 9.1 & - \\
Distance Measured (Mpc) & -  & - & - & - &  {$\bf 9.0_{-0.6}^{+0.4}$} \\
\hline
\hline
\end{tabular}
\tablenotetext{1}{Mass quoted is the $ZAMS$ mass of the progenitor, elsewhere it is the final progenitor mass immediately before explosion.}
\end{table*}

\section{Discussion and Conclusions} \label{discuss}
We present extensive photometry of the nearby SN~2013ej at UV, 
optical, and near-IR wavelengths. We also discuss well-sampled UV and 
optical spectroscopy from +8~d to +135~d after explosion. 
SN 2013ej looks kinematically similar to other normal SNe~IIP, but 
it also exhibits some unique features compared to a broader sample of 
SN~IIP, such as a steep plateau, early appearance of strong Si~II $\lambda 6355$, 
and a flat H$\alpha$ velocity profile. Such features hint 
at an intermediate class between SNe~IIL and IIP, 
or probably a continuum in the distribution of these CC~SNe. 
From a large sample of SNe~II, \citet{Anderson14} did not find 
any evidence of bimodality in the distributions of many photometric 
properties they studied based on the $V$ band, thereby suggesting a 
continuum in the properties of SNe~II. 

SNe~IIP show a wide 
range of plateau duration. Even the  typical SNe~IIP that exhibit 
flat plateaus, like SN~2006bp (\textless73~d, \citealt{Quimby2007}), 
SN~2013ab ($\sim80$~d, \citealt{bose15b}), and SN~2003hn ($\sim75$~d, 
\citealt{Bersten11}), have shorter plateau duration compared 
to that of SN~2013ej. This indicates that the envelope mass of 
SN~2013ej is not atypical. The amount of decrease of luminosity from the 
end of the plateau to the 
radioactive phase may be related to production of Ni in the ejecta, 
but the value of $M_{\rm Ni}$ calculated for SN 2013ej is also not 
atypical of SNe~IIP (e.g., \citealt{hamuy03, Bersten11, Anderson14}).

 Applying the $t^2$ model to the early-time data, we estimated 
the shock breakout epoch of SN~2013ej to be MJD $56496.9\pm0.3$ days; 
however, the validity of a $t^2$  model in the context of SNe~IIP at 
very early time remains to be studied thoroughly. 
The late time light curve of SN~2013ej shows a broken decline behavior in all $BVRI$ bands. While $M_{Ni}=0.019\pm0.001$ M$_\odot$, derived from the bolometric lightcurve before the break point at +183d, is consistent with the previous studies  \citep [e.g.][]{bose15, Huang15}, this is perhaps overestimated by ~50\%. Beyond the break point, the slope is shallower, close to the expected rate of decay from $^{56}$Co of 0.01 mag day$^{-1}$, resulting in $M_{Ni}=0.013\pm0.001$ M$_\odot$. The characterstic time scale of trapping beyond the break point is much longer than that found earlier, possibly suggesting that SN 2013ej had not completely  transitioned to the nebular phase before +183d, or there was some excess flux from CSM interaction with ejecta or other source before that time.

Collecting multiband photometry from $U$ through $K$ for a few 
well-observed SNe from the literature, 
we establish a calibration relation between 
the $UBVRI$ pseudo-bolometric flux and the $UBVRIJHK$ bolometric flux, 
which may reach 2\% precision. By performing a composite calibration, 
we showed that ROTSE or KAIT unfiltered measurements together 
with $B-V$ information may also be sufficient to derive the bolometric 
luminosity with high precision for SN IIP. We also present a pseudo-bolometric 
$BVRI$ calibration using a linear relation for unfiltered photometry. 
Given the position of SN~2013ej in relation to more typical SNe~IIP 
and more rapidly declining SNe~IIL, it would be interesting to explore the 
consistency of the bolometric calibration described here with a broader 
range of SNe II.

Even though SN~2013ej is mostly normal spectroscopically, the strong early 
appearance and subsequent evolution of the Si~II $\lambda 6355$ line is 
rather unusual. Fe~II might also have appeared somewhat early. 
The velocity evolution of weak and strong ions resembles usual 
SN~IIP behavior, 
but the flat H~I velocity profile is consistently high among 
the SN~IIP population. 
The correlation of $v_{\rm Fe~II}$ and $v_{{\rm H}\beta}$ as observed 
in a better-sampled study is also justified by SN 2013ej. One could expect somewhat 
different behavior with two components (photospheric and high velocity) of H~I lines, 
but we do not see any HVFs in our spectra. SN~2013ej adds a valuable UV spectrum to the early-time UV sample, supporting the spectral homogeneity around +10~d after explosion.
It is clear from the available sample that such homogeneity is not justified at earlier epochs. 
This also signifies the need of early time data.

By performing an EPM analysis of SN 2013ej, in combination with SN 2002ap, 
we estimated the distance to the host galaxy 
M74 to be $9.0_{-0.6}^{+0.4}$ Mpc. Using the 
calibrated bolometric flux for unfiltered ROTSE data for SN~2013ej, we obtain 
a distance of $9.7_{-0.7}^{+0.9}$ Mpc, consistent with the previous derivation. 
Various physical parameters derived here and the findings from other 
studies of SN~2013ej are listed in Table \ref{tab:param}. Generally, 
the values derived here from simple approximation models are consistent 
with other findings in the literature. Since we took the peak as the advent 
of plateau, we note that the expelled mass is likely overestimated while 
the radius could be underestimated.

\section{Acknowledgement}

This research was supported by NASA grant NNX10A196H (P.I. Kehoe), NSF 
grant AST-1109801 (P.I. Wheeler), and Hungarian OTKA grant NN 107637 (P.I. Vinko). 
J.M.S. is supported by an NSF Astronomy and Astrophysics Postdoctoral 
Fellowship under award AST-1302771. T.S. is funded by OTKA Postdoctoral 
Fellowship award PD 112325. K.S. has been supported by the Lend\"{u}let-2009 
program of the Hungarian Academy of Sciences and ESA 
PECS Contract No. 4000110889/14/NL/NDe. We also acknowledge support from the 
Hungarian Research Grants  OTKA K-109276 and OTKA K-113117, as well as 
the Lend\"{u}let-2009 and Lend\"{u}let-2012 Program (LP2012-31) of the 
Hungarian Academy of Sciences. K.T. was supported by CONICYT through the 
FONDECYT grant 3150473 and by the Ministry of Economy, Development, 
and Tourism's Millennium Science Initiative through grant IC12009, awarded 
to the Millennium Institute of Astrophysics, MAS. L.M. is supported by the 
J\'anos Bolyai Research Scholarship of the Hungarian Academy of Sciences.
A.V.F.'s group at U.C. Berkeley is supported by Gary \& Cynthia 
Bengier, the Richard \& Rhoda Goldman Fund, the Christopher R. 
Redlich Fund, the TABASGO Foundation, and NSF grant AST-1211916. 
Research at Lick Observatory is partially supported by a generous 
gift from Google. KAIT and its ongoing operation were made possible 
by donations from Sun Microsystems, Inc., the Hewlett-Packard 
Company, AutoScope Corporation, Lick Observatory, the NSF, the 
University of California, the Sylvia \& Jim Katzman Foundation, and 
the TABASGO Foundation.

HET is a joint project of the University of Texas at Austin, the Pennsylvania 
State University, Stanford University, Ludwig-Maximilians-Universitat Munchen, 
and Georg-August-University Gottingen. The HET is named in honor of its principal 
benefactors, William P. Hobby and Robert E. Eberly. The Marcario Low Resolution 
Spectrograph is named for Mike Marcario of High Lonesome Optics who fabricated 
several optics for the instrument but died before its completion. The LRS is a joint 
project of the HET partnership and the Instituto de Astronomica de la 
Universidad Nacional Autonova de Mexico. The ROTSE-IIIb telescope is owned and 
supported by Southern Methodist University. We thank the staff at McDonald 
Observatory and Lick Observatory for their excellent work during the observations. We also thank
U.C. Berkeley undergraduate students 
Minkyu Kim, Kevin Hayakawa, Haejung Kim, Heechan Yuk, Andrew Bigley, Goni Halevy, 
Samantha Cargill, Sahana Kumar, Kenia Pina, Kiera Fuller, 
Chadwick Casper, James Bradley, Philip Lu, Erin Leonard, Stephen Taylor, 
Jenifer Gross, Daniel Cohen, Michael Hyland, Kyle Blanchard, and Gary Li for their 
effort in taking Lick/Nickel data. We gratefully acknowledge Jon Mauerhan and Brad Tucker 
for helping obtain some of our optical spectra. We also thank the anonymous referee for many 
helpful discussions during the submission process of this paper.

We thank the RATIR project team and
the staff of the Observatorio Astronomico Nacional on
Sierra San Pedro Martir. RATIR is a collaboration between
the University of California, the Universidad Nacional
Autonoma de Mexico, NASA Goddard Space Flight
Center, and Arizona State University, benefiting from the
loan of an H2RG detector and hardware and software support
from Teledyne Scientific and Imaging. RATIR, the
automation of the Harold L. Johnson Telescope of the
Observatorio Astron´omico Nacional on Sierra San Pedro
Martir, and the operation of both are funded through NASA
grants NNX09AH71G, NNX09AT02G, NNX10AI27G, and
NNX12AE66G, CONACyT grants INFR-2009-01-122785
and CB-2008-101958, UNAM PAPIIT grants IN113810 and IG100414,
and a UCMEXUS-CONACyT grant.

\newpage
\newpage
\appendix
\begin{center}
\begin{longtable}{lc}
\caption{Rotse-IIIb Unfilterred photometry of SN~2013ej. Photometric 
Uncertainties are given inside parenthesis.}\label{tab:rotselc}\\

\hline
\hline
MJD & ROTSE magnitude\\ 
\hline
56498.38 & 13.32 (0.04) \\
56498.41 & 13.22 (0.02) \\
56504.36 & 12.30 (0.02) \\
56504.37 & 12.08 (0.05) \\
56505.32 & 12.30 (0.01) \\
56506.40 & 12.23 (0.01) \\
56506.32 & 12.24 (0.01) \\
56507.38 & 12.22 (0.01) \\
56507.29 & 12.21 (0.02) \\
56508.39 & 12.20 (0.01) \\
56508.32 & 12.21 (0.01) \\
56510.40 & 12.16 (0.02) \\
56510.31 & 12.19 (0.01) \\
56511.27 & 12.20 (0.02) \\
56512.27 & 12.21 (0.01) \\
56513.38 & 12.17 (0.02) \\
56513.31 & 12.20 (0.01) \\
56516.25 & 12.24 (0.02) \\
56517.32 & 12.21 (0.01) \\
56518.33 & 12.23 (0.01) \\
56520.41 & 12.27 (0.01) \\
56521.36 & 12.28 (0.01) \\
56521.28 & 12.27 (0.01) \\
56522.36 & 12.30 (0.01) \\
56523.35 & 12.36 (0.01) \\
56523.30 & 12.31 (0.02) \\
56524.37 & 12.38 (0.03) \\
56525.34 & 12.39 (0.02) \\
56526.25 & 12.32 (0.04) \\
56527.32 & 12.54 (0.03) \\
56530.24 & 12.49 (0.03) \\
56531.18 & 12.48 (0.07) \\
56533.28 & 12.58 (0.02) \\
56534.34 & 12.59 (0.02) \\
56534.26 & 12.60 (0.01) \\
56535.31 & 12.61 (0.02) \\
56536.25 & 12.63 (0.03) \\
56537.31 & 12.66 (0.01) \\
56537.24 & 12.64 (0.01) \\
56538.19 & 12.66 (0.01) \\
56539.32 & 12.70 (0.01) \\
56539.22 & 12.69 (0.01) \\
56540.31 & 12.72 (0.02) \\
56540.23 & 12.71 (0.01) \\
56541.23 & 12.75 (0.01) \\
56542.21 & 12.76 (0.01) \\
56543.31 & 12.77 (0.01) \\
56543.23 & 12.78 (0.01) \\
56549.36 & 12.88 (0.01) \\
56549.23 & 12.86 (0.01) \\
56563.21 & 13.07 (0.01) \\
56563.16 & 13.04 (0.01) \\
56565.28 & 13.11 (0.02) \\
56565.20 & 13.07 (0.01) \\
56567.24 & 13.11 (0.01) \\
56568.23 & 13.14 (0.01) \\
56568.17 & 13.14 (0.01) \\
56569.24 & 13.14 (0.01) \\
56569.11 & 13.09 (0.02) \\
56570.26 & 13.19 (0.02) \\
56570.19 & 13.17 (0.02) \\
56571.27 & 13.18 (0.02) \\
56571.19 & 13.19 (0.02) \\
56572.24 & 13.20 (0.02) \\
56572.21 & 13.21 (0.02) \\
56573.16 & 13.22 (0.02) \\
56574.27 & 13.25 (0.02) \\
56575.22 & 13.26 (0.02) \\
56575.16 & 13.24 (0.02) \\
56576.20 & 13.27 (0.02) \\
56576.18 & 13.27 (0.02) \\
56577.37 & 13.27 (0.02) \\
56578.12 & 13.30 (0.02) \\
56586.38 & 13.59 (0.04) \\
56586.34 & 13.54 (0.03) \\
56587.18 & 13.55 (0.03) \\
56587.18 & 13.59 (0.02) \\
56588.18 & 13.61 (0.02) \\
56588.17 & 13.61 (0.02) \\
56589.19 & 13.63 (0.02) \\
56589.15 & 13.62 (0.01) \\
56590.24 & 13.68 (0.01) \\
56590.16 & 13.66 (0.01) \\
56591.19 & 13.72 (0.01) \\
56591.15 & 13.72 (0.01) \\
56592.13 & 13.77 (0.01) \\
56593.17 & 13.88 (0.03) \\
56593.15 & 13.82 (0.03) \\
56594.27 & 13.96 (0.04) \\
56599.16 & 14.83 (0.02) \\
56602.19 & 15.24 (0.03) \\
56602.27 & 15.23 (0.03) \\
56603.27 & 15.38 (0.04) \\
56605.33 & 15.44 (0.05) \\
56605.31 & 15.44 (0.04) \\
56606.17 & 15.47 (0.04) \\
56606.21 & 15.44 (0.04) \\
56607.22 & 15.55 (0.06) \\
56608.12 & 15.65 (0.07) \\
56608.13 & 15.51 (0.04) \\
56616.09 & 15.55 (0.08) \\
56616.09 & 15.80 (0.07) \\
56617.13 & 15.58 (0.08) \\
56617.20 & 15.73 (0.04) \\
56618.17 & 15.56 (0.13) \\
56627.27 & 15.74 (0.09) \\
56627.13 & 15.72 (0.15) \\
56628.22 & 15.70 (0.24) \\
56630.21 & 15.78 (0.11) \\
56630.23 & 15.95 (0.07) \\
56631.16 & 15.85 (0.07) \\
56631.23 & 16.00 (0.07) \\
56650.20 & 16.17 (0.12) \\
56651.19 & 16.09 (0.08) \\
56651.16 & 16.23 (0.11) \\
56653.21 & 16.52 (0.14) \\
56670.16 & 17.04 (0.50) \\
56672.15 & 16.99 (0.34) \\
56672.13 & 16.57 (0.15) \\
56673.13 & 16.30 (0.07) \\
56673.15 & 16.74 (0.14) \\
56674.12 & 16.63 (0.12) \\
56675.13 & 16.98 (0.22) \\
56675.18 & 16.68 (0.11) \\
56676.14 & 16.88 (0.19) \\
56676.16 & 17.03 (0.19) \\
56677.11 & 16.58 (0.19) \\
56678.12 & 16.51 (0.12) \\
56678.11 & 16.96 (0.16) \\
56679.12 & 16.87 (0.14) \\
56679.17 & 16.58 (0.07) \\
56682.21 & 17.00 (0.21) \\
56683.17 & 16.61 (0.11) \\
56687.15 & 16.78 (0.15) \\
\hline
\hline
\end{longtable}
\end{center}

\begin{center}
\begin{table}[ht]
\caption{{\it BVRI} photometry of SN~2013ej from Konkoly Observatory.
Magnitudes are in the Vega-system, and uncertainties are given inside parentheses.}
\label{tab:konkolyphot}
\begin{tabular}{lcccc}
\hline
\hline
MJD & $B$ & $V$ & $R$ & $I$ \\
   & (mag) & (mag) & (mag) & (mag) \\
\hline
56505.5  &  12.64    (0.05)  &  12.57    (0.02)  &  12.49    (0.02)  &  12.43    (0.02)  \\
56507.5  &  12.61    (0.01)  &  12.52    (0.01)  &  12.39    (0.01)  &  12.36    (0.01)  \\
56510.5  &  12.66    (0.07)  &  12.49    (0.07)  &  12.35    (0.02)  &  12.29    (0.01)  \\
56511.5  &  12.75    (0.08)  &  12.51    (0.04)  &  12.31    (0.01)  &  12.26    (0.01)  \\
56512.5  &  12.78    (0.10)  &  12.52    (0.03)  &  12.31    (0.01)  &  12.24    (0.02)  \\
56513.5  &  12.78    (0.06)  &  12.48    (0.04)  &  12.27    (0.04)  &  12.21    (0.01)  \\
56516.5  &  12.99    (0.08)  &  12.53    (0.02)  &  12.31    (0.01)  &  12.20    (0.02)  \\
56517.5  &  13.05    (0.04)  &  12.54    (0.02)  &  12.30    (0.02)  &  12.19    (0.01)  \\
56521.5  &  13.38    (0.06)  &  12.61    (0.02)  &  12.32    (0.01)  &  12.17    (0.01)  \\
56522.5  &  13.39    (0.05)  &  12.64    (0.01)  &  12.34    (0.01)  &  12.19    (0.02)  \\
56535.5  &  14.15    (0.06)  &  13.06    (0.01)  &  12.61    (0.03)  &  12.42    (0.04)  \\
56537.5  &  14.22    (0.01)  &  13.11    (0.01)  &  12.65    (0.01)  &  12.43    (0.01)  \\
56540.5  &  14.42    (0.07)  &  13.19    (0.07)  &  12.70    (0.01)  &  12.50    (0.01)  \\
56542.5  &  14.46    (0.03)  &  13.23    (0.01)  &  12.74    (0.01)  &  12.51    (0.01)  \\
56543.5  &  14.49    (0.02)  &  13.26    (0.01)  &  12.76    (0.01)  &  12.52    (0.01)  \\
56555.4  &  14.89    (0.12)  &  13.48    (0.01)  &  12.95    (0.06)  &  12.66    (0.02)  \\
56563.4  &  15.01    (0.02)  &  13.59    (0.03)  &  13.03    (0.01)  &  12.77    (0.01)  \\
56568.6  &  15.09    (0.01)  &  13.69    (0.02)  &  13.11    (0.01)  &  12.84    (0.01)  \\
56573.6  &  15.29    (0.10)  &  13.77    (0.01)  &  13.19    (0.02)  &  12.91    (0.01)  \\
56578.3  &  15.34    (0.06)  &  13.88    (0.01)  &  13.27    (0.01)  &  12.98    (0.01)  \\
56591.4  &  15.93    (0.04)  &  14.37    (0.01)  &  13.67    (0.06)  &  13.40    (0.01)  \\
56592.5  &  15.94    (0.05)  &  14.41    (0.01)  &  13.73    (0.03)  &  13.43    (0.01)  \\
56597.3  &  16.65    (0.03)  &  15.10    (0.01)  &  14.28    (0.02)  &  13.99    (0.01)  \\
56604.4  &  17.65    (0.04)  &  16.24    (0.01)  &  15.25    (0.02)  &  14.92    (0.02)  \\
56628.2  &  17.95    (0.07)  &  16.68    (0.01)  &  15.63    (0.02)  &  15.36    (0.01)  \\
56629.2  &  17.96    (0.06)  &  16.70    (0.01)  &  15.68    (0.03)  &  15.43    (0.01)  \\
\hline 
\hline
\end{tabular}
\end{table}
\end{center}

\begin{center}
\begin{table}[ht]
\caption{Sloan {\it g'r'i'z'} photometry of SN~2013ej from Baja Observatory, Hungary. Magnitudes are in the AB-system, and uncertainties are given inside parentheses.}
\label{tab:bajaphot}
\begin{tabular}{lcccc}
\hline
\hline
MJD & $g'$ & $r'$ & $i'$ & $z'$ \\
    & (mag) & (mag) & (mag) & (mag) \\
\hline
56504.08 & 12.54 (0.06)  & 12.61 (0.03)  & 12.74 (0.09)  & 13.01 (0.17) \\
56505.07 & 12.49 (0.06)  & 12.58 (0.04)  & 12.63 (0.09)  & 12.94 (0.23) \\
56510.08 & 12.56 (0.10)  & 12.44 (0.02)  & 12.56 (0.02)  & 12.68 (0.05) \\
56512.09 & 12.60 (0.09)  & 12.42 (0.02)  & 12.55 (0.03)  & 12.66 (0.05) \\
56513.06 & 12.54 (0.04)  & 12.40 (0.02)  & 12.51 (0.04)  & 12.62 (0.06) \\
56516.06 & 12.64 (0.04)  & 12.39 (0.02)  & 12.49 (0.03)  & 12.61 (0.05) \\
56520.06 & 12.79 (0.04)  & 12.41 (0.02)  & 12.50 (0.02)  & 12.59 (0.05) \\
56521.06 & 12.84 (0.04)  & 12.43 (0.02)  & 12.51 (0.03)  & 12.61 (0.05) \\
56522.08 & 12.91 (0.04)  & 12.45 (0.02)  & 12.51 (0.05)  & 12.63 (0.13) \\
56535.00 & 13.48 (0.10)  & 12.78 (0.04)  & 12.90 (0.13)  & 12.76 (0.07) \\
56536.05 & 13.52 (0.05)  & 12.75 (0.02)  & 12.75 (0.03)  & 12.75 (0.04) \\
56539.07 & 13.61 (0.05)  & 12.81 (0.02)  & 12.83 (0.02)  & 12.79 (0.06) \\
56539.97 & 13.63 (0.03)  & 12.83 (0.02)  & 12.83 (0.02)  & 12.77 (0.05) \\
56541.92 & 13.69 (0.05)  & 12.87 (0.02)  & 12.87 (0.03)  & 12.80 (0.06) \\
56543.10 & 13.75 (0.03)  & 12.88 (0.03)  & 12.92 (0.06)  & 12.85 (0.19) \\
56543.95 & 13.75 (0.03)  & 12.91 (0.02)  & 12.91 (0.02)  & 12.83 (0.04) \\
56552.88 & 13.97 (0.08)  & 13.05 (0.04)  & 13.02 (0.05)  & 12.92 (0.06) \\
56558.91 & 14.06 (0.06)  & 13.14 (0.04)  & 13.13 (0.04)  & 12.97 (0.06) \\
56559.90 & 14.04 (0.08)  & 13.14 (0.03)  & 13.12 (0.03)  & 12.99 (0.08) \\
56568.88 & 14.25 (0.03)  & 13.25 (0.03)  & 13.26 (0.04)  & 13.09 (0.06) \\
56575.98 & 14.37 (0.05)  & 13.41 (0.04)  & 13.47 (0.05)  & 13.19 (0.07) \\
56576.95 & 14.61 (0.09)  & 13.44 (0.03)  & 13.44 (0.04)  & 13.26 (0.06) \\
56578.88 & 14.45 (0.08)  & 13.50 (0.04)  & 13.51 (0.03)  & 13.29 (0.06) \\
56586.87 & 14.73 (0.12)  & 13.65 (0.04)  & 13.67 (0.05)  & 13.41 (0.10) \\
56590.95 & 14.93 (0.06)  & 13.79 (0.02)  & 13.82 (0.03)  & 13.53 (0.05) \\
56592.96 & 15.25 (0.13)  & 13.95 (0.04)  & 14.00 (0.03)  & 13.65 (0.04) \\
56594.88 & 15.36 (0.03)  & 14.14 (0.02)  & 14.19 (0.02)  & 13.79 (0.04) \\
56596.03 & 15.78 (0.13)  & 14.37 (0.06)  & 14.45 (0.04)  & 14.02 (0.10) \\
56598.01 & 15.98 (0.06)  & 14.67 (0.03)  & 14.84 (0.05)  & 14.19 (0.11) \\
56603.90 & 16.88 (0.06)  & 15.39 (0.02)  & 15.54 (0.04)  & 14.86 (0.07) \\
56612.81 & 16.88 (0.34)  & 15.70 (0.14)  & 15.73 (0.10)  & 15.22 (0.15) \\
56626.97 & 17.21 (0.10)  & 15.81 (0.04)  & 16.04 (0.06)  & 15.42 (0.11) \\
56627.89 & 17.19 (0.06)  & 15.78 (0.03)  & 16.02 (0.04)  & 15.40 (0.07) \\
56628.75 & 17.36 (0.08)  & 15.87 (0.04)  & 16.03 (0.04)  & 15.45 (0.08) \\
56629.90 & 17.38 (0.08)  & 15.83 (0.04)  & 16.07 (0.04)  & 15.40 (0.07) \\
56636.81 & 17.28 (0.11)  & 15.94 (0.05)  & 16.21 (0.06)  & 15.48 (0.08) \\
56650.78 & 17.64 (0.10)  & 16.11 (0.03)  & 16.47 (0.06)  & 15.82 (0.12) \\
56660.84 & 17.95 (0.13)  & 16.32 (0.05)  & 16.78 (0.08)  & 15.88 (0.14) \\
56663.78 & 17.86 (0.12)  & 16.39 (0.06)  & 16.78 (0.09)  & 15.99 (0.12) \\
\hline
\hline
\end{tabular}
\end{table}
\end{center}

\clearpage
\footnotesize
\setlength\LTleft{30pt}
\setlength\LTright{30pt}
\begin{longtable*}{@{\extracolsep{\fill}}lccccc@{}}
\caption{KAIT Unfiltered and KAIT + Nickel BVRI photometry of SN~2013ej. Magnitudes 
are in Vega system and uncertainties are given inside parenthesis.}\label{tab:kaitlc} \\
\hline
\hline
MJD & Open CCD & $B$ & $V$ & $R$ & $I$ \\
    & (mag) & (mag) & (mag) & (mag) & (mag) \\
\hline
56498.45  &  13.30 (0.10) & \ldots &    \ldots  &  \ldots  &  \ldots  \\
56499.44  &  12.95 (0.04) & \ldots &   \ldots        &   \ldots        &   \ldots        \\
56504.39  &  12.29 (0.06) &  12.65  (0.06)   &  12.68  (0.03)   &   12.53  (0.03)  &   12.52  (0.03) \\
56507.51  &  12.23 (0.02) &  12.62  (0.03)   &  12.54  (0.03)   &   12.38  (0.03)  &   12.37  (0.04) \\
56509.52  &  12.24 (0.03) &  12.65   (0.10)  &  12.57  (0.05)   &   12.38  (0.02) &   12.37  (0.03)  \\
56515.51  &  12.21 (0.04) &  12.91  (0.04)   &  12.51  (0.04)   &   12.26  (0.03)  &   12.17  (0.03) \\
56516.51  &  12.21 (0.02) &  \ldots &    \ldots       &    \ldots       &     \ldots      \\
56519.48  &  12.29 (0.02) &  13.22  (0.04)   &  12.61  (0.03)   &   12.32  (0.02) &   12.20  (0.02)\\
56520.51  &  12.33 (0.01) &  13.33  (0.04)   &  12.64  (0.02)  &   12.33  (0.02) &   12.23  (0.02)\\
56522.50  &  12.36 (0.02) &  13.38  (0.04)   &  12.64  (0.02)  &   12.30  (0.01) &   12.20  (0.02)\\
56523.45  &  12.41 (0.03) &  13.50  (0.03)   &  12.68  (0.02)  &   12.37  (0.02) &   12.23  (0.02)\\
56524.50  &  12.40 (0.03) &  13.54  (0.04)   &  12.70  (0.02)  &   12.34  (0.02) &   12.21  (0.03) \\
56525.34  &  12.45 (0.03) &  13.86  (0.09)   &  12.87  (0.04)   &   12.44  (0.03)  &   12.33  (0.02)\\
56527.38  &  12.52 (0.02) &  13.82  (0.04)   &  12.85  (0.03)   &   12.47  (0.04)  &   12.32  (0.04) \\
56530.49  &  12.57 (0.03) &  13.91  (0.04)   &  12.92  (0.02)  &   12.49  (0.02) &   12.32  (0.03) \\
56532.53  &  12.62 (0.05) & \ldots &    \ldots       &    \ldots       &   \ldots        \\
56533.43  &  12.67 (0.03) &  14.11  (0.04)   &  13.01  (0.02)  &   12.58  (0.02) &   12.40  (0.02)\\
56534.49  &  12.64 (0.03) &  14.06  (0.03)   &  13.00  (0.02)  &   12.55  (0.02) &   12.37  (0.02)\\
56536.40  &  12.72 (0.03) &  14.23  (0.03)   &  13.08  (0.02)  &   12.67  (0.02) &   12.46  (0.02)\\
56537.45  &  \ldots       &  14.25   (0.12)  &  13.10  (0.05)   &   12.62  (0.02) &   12.35  (0.21) \\
56539.41  &  12.77 (0.03) &  14.30  (0.03)   &  13.12  (0.02)  &   12.65  (0.02) &   12.46  (0.03) \\
56540.41  &  12.78 (0.02) &  14.33  (0.03)   &  13.15  (0.01)  &   12.67  (0.02) &   12.48  (0.02)\\
56541.36  &  12.84 (0.02) &  \ldots &    \ldots       &   \ldots        &    \ldots       \\
56542.42  &  12.82 (0.03) &  14.41  (0.03)   &  13.19  (0.02)  &   12.70  (0.02) &   12.48  (0.02)\\
56543.38  &  12.87 (0.04) &  14.48  (0.04)   &  13.22  (0.02)  &   12.75  (0.02) &   12.53  (0.03) \\
56545.48  &  12.86 (0.02) &  14.47  (0.04)   &  13.23  (0.02)  &   12.74  (0.02) &   12.54  (0.03) \\
56546.41  &  12.88 (0.03) &  14.54  (0.04)   &  13.30  (0.03)   &   12.79  (0.04)  &   12.56  (0.04) \\
56548.39  &  12.92 (0.01) &  14.59  (0.03)   &  13.31  (0.01)  &   12.82  (0.01) &   12.57  (0.02)\\
56549.40  &  12.92 (0.01) &  14.65  (0.03)   &  13.33  (0.01)  &   12.84  (0.01) &   12.61  (0.01)\\
56550.43  &  12.95 (0.03) &  14.60  (0.03)   &  13.32  (0.02)  &   12.82  (0.02) &   12.59  (0.02)\\
56551.45  &  12.97 (0.03) &  14.62  (0.03)   &  13.33  (0.02)  &   12.83  (0.02) &   12.60  (0.02)\\
56552.29  &  12.98 (0.03) &  14.79  (0.08)   &  13.44  (0.04)   &   12.91  (0.04)  &   12.70  (0.04) \\
56553.36  &  13.02 (0.03) & \ldots &     \ldots      &    \ldots       &    \ldots       \\
56559.37  &  13.08 (0.02) &  14.90  (0.04)   &  13.50  (0.03)   &   12.93  (0.04)  &   12.72  (0.04) \\
56562.31  &  13.13 (0.01) &  14.96  (0.03)   &  13.56  (0.01)  &   13.01  (0.01) &   12.77  (0.02)\\
56563.40  &  13.15 (0.03) &  \ldots &    \ldots       &    \ldots       &   \ldots        \\
56564.37  &  13.15 (0.04) &  14.93  (0.04)   &  13.58  (0.02)  &   13.01  (0.02) &   12.76  (0.02)\\
56566.49  &  13.14 (0.01) &  15.03  (0.05)   &  13.59  (0.02)  &   13.02  (0.01) &   12.78  (0.02)\\
56567.38  &  13.18 (0.02) & \ldots &      \ldots     &    \ldots       &     \ldots      \\
56569.29  &  13.21 (0.01) & \ldots &    \ldots       &   \ldots        &    \ldots       \\
56570.39  &  13.21 (0.02) &  15.06  (0.03)   &  13.65  (0.02)  &   13.07  (0.02) &   12.82  (0.02)\\
56571.31  &  13.22 (0.03) & \ldots &   \ldots        &   \ldots        &     \ldots      \\
56572.35  &  13.25 (0.02) &  15.14  (0.04)   &  13.70  (0.03)   &   13.11  (0.03)  &   12.86  (0.04) \\
56573.35  &  13.22 (0.04) & \ldots &      \ldots     &     \ldots      &    \ldots       \\
56574.35  &  13.27 (0.01) &  15.18  (0.05)   &  13.74  (0.02)  &   13.15  (0.01) &   12.89  (0.01)\\
56575.37  &  13.30 (0.03) & \ldots &     \ldots      &     \ldots      &     \ldots      \\
56576.33  &  13.33 (0.04) & \ldots &     \ldots      &    \ldots       &   \ldots        \\
56577.38  &  13.34 (0.03) &  15.27  (0.04)   &  13.77  (0.03)  &   13.16  (0.02) &   12.94  (0.03) \\
56579.33  &  13.36 (0.02) & \ldots &  \ldots         &   \ldots        &    \ldots       \\
56580.32  &  13.41 (0.03) &  15.38  (0.04)   &  13.87  (0.02)  &   13.25  (0.02) &   13.02  (0.02)\\
56581.24  &  13.44 (0.02) &  \ldots &   \ldots        &    \ldots       &    \ldots       \\
56587.32  &  13.61 (0.03) &  15.65  (0.05)   &  14.10  (0.02)  &   13.45  (0.02) &   13.20  (0.03) \\
56588.33  &  13.64 (0.03) &  \ldots &   \ldots        &    \ldots       &    \ldots       \\
56589.32  &  13.70 (0.04) & \ldots &   \ldots        &    \ldots       &    \ldots            \\
56590.32  &  13.72 (0.02) &  15.79  (0.05)   &  14.23  (0.02) &   13.56  (0.02) &   13.32  (0.02)\\
56591.31  &  13.74 (0.04) & \ldots &   \ldots        &    \ldots       &    \ldots            \\
56592.30  &  13.89 (0.03) & \ldots &   \ldots        &    \ldots       &    \ldots          \\
56596.29  &  14.30 (0.02) &  16.49  (0.06)   &  14.91  (0.02) &   14.12  (0.01) &   13.85  (0.02)\\
56597.27  &  14.45 (0.02) &  \ldots &   \ldots        &    \ldots       &    \ldots          \\
56598.28  &  14.63 (0.03) &  \ldots &   \ldots        &    \ldots       &    \ldots         \\
56599.26  &  14.83 (0.02) &  17.00  (0.08)   &  15.54  (0.03)   &   14.62  (0.03)  &   14.35  (0.03)  \\
56600.27  &  15.03 (0.02) &  \ldots &   \ldots        &    \ldots       &    \ldots           \\
56601.26  &  15.16 (0.02) &  \ldots &   \ldots        &    \ldots       &    \ldots          \\
56603.31  &  15.38 (0.05) & \ldots &   \ldots        &    \ldots       &    \ldots \\
56604.27  &  15.38 (0.03) &  \ldots &   \ldots        &    \ldots       &    \ldots         \\
56605.28  &  15.42 (0.03) &  \ldots &   \ldots        &    \ldots       &    \ldots           \\
56606.27  &  15.45 (0.02) &  17.60  (0.09)   &  16.36  (0.05)   &   15.27  (0.04)  &   14.95  (0.04) \\
56608.22  &  15.59 (0.06) &  \ldots &   \ldots        &    \ldots       &    \ldots           \\
56615.27  &  15.55 (0.03) &  17.70  (0.29)   &  16.55  (0.09)   &   15.39  (0.03)  &   15.06  (0.04) \\
56618.22  &  15.63 (0.02) &  17.86  (0.22)   &  16.57  (0.05)   &   15.45  (0.02) &   15.17  (0.03) \\
56619.26  &  15.64 (0.02) &  \ldots &   \ldots        &    \ldots       &    \ldots           \\
56620.25  &  15.65 (0.03) &  \ldots &   \ldots        &    \ldots       &    \ldots          \\
56621.28  &  15.68 (0.03) &  17.83  (0.18)   &  16.65  (0.05)   &   15.49  (0.03)  &   15.24  (0.04) \\
56622.23  &  15.72 (0.04) &  \ldots &   \ldots        &    \ldots       &    \ldots          \\
56624.23  &  15.72 (0.03) &  18.06  (0.22)   &  16.65  (0.06)   &   15.49  (0.03)  &   15.24  (0.04) \\
56625.25  &  15.74 (0.03) &  \ldots &   \ldots        &    \ldots       &    \ldots          \\
56626.21  &  15.72 (0.03) &  \ldots &   \ldots        &    \ldots       &    \ldots          \\
56627.23  &  15.77 (0.02) &  18.18  (0.25)   &  16.70  (0.05)   &   15.57  (0.03)  &   15.31  (0.03) \\
56628.25  &  15.78 (0.03) & \ldots &   \ldots        &    \ldots       &    \ldots           \\
56629.25  &  15.81 (0.02) &  \ldots &   \ldots        &    \ldots       &    \ldots            \\
56630.20  &  15.83 (0.02) &  18.13  (0.18)   &  16.72  (0.05)   &   15.61  (0.03)  &   15.37  (0.03) \\
56631.18  &  15.88 (0.03) &  \ldots &   \ldots        &    \ldots       &    \ldots           \\
56632.17  &  15.89 (0.04) &  \ldots &   \ldots        &    \ldots       &    \ldots           \\
56634.24  &  15.90 (0.02) & \ldots &   \ldots        &    \ldots       &    \ldots \\
56635.15  &  15.92 (0.03) &  \ldots &   \ldots        &    \ldots       &    \ldots           \\
56636.17  &  15.97 (0.04) & \ldots &   \ldots        &    \ldots       &    \ldots           \\
56641.20  &  15.98 (0.03) &  18.17  (0.30)   &  16.88  (0.08)   &   15.78  (0.03)  &   15.54  (0.04) \\
56643.20  &  16.02 (0.06) &  \ldots &   \ldots        &    \ldots       &    \ldots           \\
56645.24  &  16.03 (0.04) &  \ldots &   \ldots        &    \ldots       &    \ldots           \\
56647.23  &  16.06 (0.03) &  \ldots &   \ldots        &    \ldots       &    \ldots            \\
56648.21  &  16.08 (0.03) &  18.06  (0.20)   &  17.13  (0.08)   &   15.83  (0.03)  &   15.63  (0.03) \\
56649.20  &  16.10 (0.02) &  \ldots &   \ldots        &    \ldots       &    \ldots         \\
56650.18  &  16.11 (0.04) & \ldots &   \ldots        &    \ldots       &    \ldots          \\
56651.19  &  16.15 (0.02) &  18.53  (0.36)   &  17.05  (0.08)   &   15.91  (0.03)  &   15.68  (0.04) \\
56653.18  &  16.18 (0.03) &  \ldots &   \ldots        &    \ldots       &    \ldots          \\
56655.15  &  16.23 (0.02) &  18.23  (0.30)   &  17.12   (0.11)  &   15.98  (0.04)  &   15.80  (0.05) \\
56656.17  &  16.24 (0.04) &  \ldots &   \ldots        &    \ldots       &    \ldots           \\
56658.15  &  16.30 (0.04) &  18.19  (0.21)   &  17.15  (0.07)   &   16.01  (0.03)  &   15.80  (0.04) \\
56660.15  &  16.39 (0.06) & \ldots &   \ldots        &    \ldots       &    \ldots          \\
56662.17  &  16.32 (0.02) & \ldots &   \ldots        &    \ldots       &    \ldots           \\
56668.12  &  16.46 (0.05) &  \ldots &   \ldots        &    \ldots       &    \ldots           \\
56669.12  &  16.41 (0.04) & \ldots &   \ldots        &    \ldots       &    \ldots \\
56673.15  &  16.52 (0.05) &  \ldots &   \ldots        &    \ldots       &    \ldots           \\
56674.11  &  16.48 (0.04) &  18.06  (0.25)   &  17.47   (0.10)  &   16.22  (0.05)  &   15.99  (0.05) \\
56676.16  &  16.55 (0.05) & \ldots &   \ldots        &    \ldots       &    \ldots            \\
56677.13  &  16.54 (0.04) &  \ldots &   \ldots        &    \ldots       &    \ldots           \\
56679.13  &  16.56 (0.04) &  18.53  (0.41)   &  17.54   (0.11)  &   16.33  (0.05)  &   16.11  (0.06) \\
56682.12  &  16.62 (0.06) &  \ldots &   \ldots        &    \ldots       &    \ldots           \\
56684.13  &  16.61 (0.04) &  18.63  (0.30)   &  17.66   (0.14)  &   16.39  (0.05)  &   16.21  (0.05) \\
56708.12  &  16.87 (0.05) & \ldots &   \ldots        &    \ldots       &    \ldots \\
56710.13  &  16.94 (0.05) & \ldots &   \ldots        &    \ldots       &    \ldots \\
\hline
& & & Nickel Photometry &\\
\hline
56505.4258  &  \ldots   &   12.37    (0.02)  &  12.52    (0.01)  &  12.43    (0.01)  &  12.40    (0.01)  \\
56507.4883  &  \ldots   &   12.39    (0.01)  &  12.50    (0.01)  &  12.37    (0.01)  &  12.32    (0.01)  \\
56523.3945  &  \ldots   &   13.40    (0.03)  &  12.66    (0.01)  &  12.36    (0.01)  &  12.21    (0.01)  \\
56527.3398  &  \ldots   &   13.77    (0.04)  &  12.79    (0.01)  &  12.45    (0.01)  &  12.26    (0.01)  \\
56531.3125  &  \ldots   &   13.90    (0.02)  &  12.94    (0.01)  &  12.50    (0.01)  &  12.32    (0.01)  \\
56535.2969  &  \ldots   &   14.14    (0.02)  &  13.05    (0.01)  &  12.60    (0.01)  &  12.39    (0.01)  \\
56539.3242  &  \ldots   &   14.29    (0.02)  &  13.17    (0.01)  &  12.69    (0.01)  &  12.45    (0.01)  \\
56541.3008  &  \ldots   &   14.38    (0.02)  &  13.19    (0.01)  &  12.72    (0.01)  &  12.49    (0.01)  \\
56545.3008  &  \ldots   &   14.51    (0.02)  &  13.27    (0.01)  &  12.79    (0.01)  &  12.52    (0.01)  \\
56548.2695  &  \ldots   &   14.59    (0.01)  &  13.33    (0.01)  &  12.84    (0.01)  &  12.57    (0.01)  \\
56552.3516  &  \ldots   &   14.67    (0.02)  &  13.41    (0.01)  &  12.89    (0.01)  &  12.63    (0.01)  \\
56555.3203  &  \ldots   &   14.80    (0.02)  &  13.45    (0.01)  &  12.92    (0.01)  &  12.68    (0.02)  \\
56559.3086  &  \ldots   &   14.92    (0.01)  &  13.52    (0.01)  &  12.98    (0.01)  &  12.72    (0.01)  \\
56562.3555  &  \ldots   &   15.00    (0.01)  &  13.55    (0.01)  &  13.00    (0.01)  &  12.73    (0.01)  \\
56569.2930  &  \ldots   &   15.02    (0.03)  &  13.54    (0.04)  &  13.13    (0.04)  &  12.78    (0.01)  \\
56573.3047  &  \ldots   &   15.27    (0.01)  &  13.74    (0.01)  &  13.17    (0.01)  &  12.91    (0.01)  \\
56575.2656  &  \ldots   &   15.30    (0.01)  &  13.78    (0.01)  &  13.18    (0.01)  &  12.91    (0.01)  \\
56578.3086  &  \ldots   &   15.42    (0.01)  &  13.74    (0.01)  &  13.25    (0.01)  &  12.98    (0.01)  \\
56581.2812  &  \ldots   &   15.52    (0.01)  &  13.94    (0.01)  &  13.31    (0.01)  &  13.04    (0.01)  \\
56585.3867  &  \ldots   &   15.67    (0.01)  &  14.07    (0.01)  &  13.43    (0.02)  &  13.19    (0.03)  \\
56589.2656  &  \ldots   &   15.87    (0.01)  &  14.24    (0.01)  &  13.57    (0.01)  &  13.28    (0.01)  \\
56599.2617  &  \ldots   &   17.07    (0.02)  &  15.62    (0.01)  &  14.70    (0.01)  &  14.41    (0.01)  \\
56601.2734  &  \ldots   &   17.39    (0.01)  &  15.97    (0.01)  &  14.99    (0.01)  &  14.67    (0.01)  \\
56605.2695  &  \ldots   &   17.68    (0.01)  &  16.33    (0.01)  &  15.28    (0.01)  &  14.94    (0.05)  \\
56609.2656  &  \ldots   &   17.79    (0.02)  &  16.44    (0.01)  &  15.36    (0.01)  &  15.04    (0.01)  \\
56626.2031  &  \ldots   &   18.07    (0.01)  &  16.72    (0.01)  &  15.60    (0.04)  &  15.31    (0.08)  \\
56631.2148  &  \ldots   &   18.13    (0.03)  &  16.79    (0.01)  &  15.68    (0.01)  &  15.39    (0.01)  \\
56635.2266  &  \ldots   &   18.17    (0.03)  &  16.84    (0.01)  &  15.73    (0.01)  &  15.46    (0.01)  \\
56642.3359  &  \ldots   &   18.27    (0.16)  &  16.78    (0.06)  &  15.89    (0.03)  &  15.53    (0.03)  \\
56645.2578  &  \ldots   &   18.41    (0.07)  &  16.96    (0.02)  &  15.89    (0.01)  &  15.61    (0.01)  \\
56665.1875  &  \ldots   &   18.47    (0.04)  &  17.32    (0.03)  &  16.21    (0.02)  &  15.93    (0.03)  \\
56670.1992  &  \ldots   &   18.66    (0.05)  &  17.36    (0.02)  &  16.27    (0.01)  &  16.03    (0.01)  \\
56675.1914  &  \ldots   &   18.64    (0.04)  &  17.44    (0.03)  &  16.27    (0.06)  &  16.03    (0.09)  \\
56678.1719  &  \ldots   &   18.66    (0.02)  &  17.51    (0.02)  &  16.32    (0.04)  &  16.08    (0.08)  \\
56689.1250  &  \ldots   &   18.81    (0.03)  &  17.63    (0.02)  &  16.52    (0.01)  &  16.25    (0.01)  \\
56705.1211  &  \ldots   &   18.96    (0.08)  &  17.86    (0.03)  &  16.71    (0.02)  &  16.47    (0.04)  \\
56709.1250  &  \ldots   &   18.78    (0.06)  &  17.78    (0.04)  &  16.74    (0.02)  &  16.51    (0.03)  \\
56724.1367  &  \ldots   &   19.17    (0.12)  &  17.91    (0.04)  &  16.88    (0.02)  &  16.68    (0.03)  \\
56849.4336  &  \ldots   &   19.96    (0.19)  &  19.27    (0.08)  &  18.56    (0.04)  &  18.34    (0.08)  \\
56852.4062  &  \ldots   &   20.65    (0.34)  &  19.20    (0.08)  &  18.46    (0.05)  &  18.09    (0.06)  \\
56856.4141  &  \ldots   &   19.74    (0.15)  &  19.21    (0.11)  &  18.35    (0.06)  &  18.41    (0.08)  \\
56894.3438  &  \ldots   &   20.31    (0.07)  &  19.88    (0.04)  &  19.21    (0.03)  &  18.49    (0.07)  \\
56898.4609  &  \ldots   &   20.27    (0.05)  &  20.02    (0.05)  &  19.08    (0.05)  &  18.73    (0.08)  \\
56916.4062  &  \ldots   &   20.49    (0.10)  &  20.27    (0.07)  &  19.55    (0.04)  &  18.78    (0.06)  \\
56923.4453  &  \ldots   &   20.55    (0.06)  &  20.09    (0.06)  &  19.70    (0.05)  &  19.03    (0.08)  \\
56931.3242  &  \ldots   &   20.86    (0.11)  &  20.40    (0.07)  &  19.84    (0.05)  &  19.61    (0.15)  \\
56934.3633  &  \ldots   &   20.77    (0.14)  &  20.21    (0.07)  &  19.69    (0.07)  &  19.21    (0.09)  \\
56937.2617  &  \ldots   &   21.06    (0.38)  &  20.36    (0.20)  &  19.59    (0.10)  &  19.02    (0.10)  \\
56948.2344  &  \ldots   &   20.56    (0.11)  &  20.41    (0.09)  &  19.82    (0.06)  &  19.70    (0.13)  \\
56958.2734  &  \ldots   &   20.72    (0.10)  &  20.74    (0.10)  &  20.13    (0.07)  &  19.54    (0.12)  \\
\hline
\end{longtable*}

\begin{table*}
\begin{center} 
\caption{Optical and NIR Photometry of SN~2013ej from RATIR. Magnitudes are in AB--System and Photometric uncertainties are given inside  parenthesis}
\label{tab:ratirphot}
\begin{tabular}{lcccccc}
\hline
\hline
 MJD & $r$ & $i$ & $Z$ & $Y$ & $J$ & $H$ \\
     & (mag) & (mag) & (mag) & (mag) & (mag) & (mag)\\
\hline
56500.0  &  13.0048 (0.02)  &  13.2938 (0.02)  &   13.3672 (0.02)  &   13.3388 (0.02)  &   13.8606 (0.05)  &   14.1553 (0.07) \\
56502.0  &  12.7652 (0.02)  &  12.9600 (0.02)  &   12.8912 (0.02)  &   13.1330 (0.02)  &   13.4210 (0.05)  &   13.8766 (0.07) \\
56503.0  &  12.6862 (0.02)  &  12.8271 (0.02)  &   12.8809 (0.02)  &   13.0350 (0.02)  &   13.3518 (0.05)  &   13.6459 (0.07) \\
56504.0  &  12.5889 (0.02)  &  12.7816 (0.02)  &   12.8486 (0.02)  &   12.9499 (0.02)  &   13.3239 (0.05)  &   13.5421 (0.07) \\
56505.0  &  12.5562 (0.02)  &  12.7271 (0.02)  &   12.8202 (0.02)  &   12.9174 (0.02)  &   13.2535 (0.05)  &   13.4951 (0.07) \\
56506.0  &  12.4911 (0.02)  &  12.6626 (0.02)  &   12.7144 (0.02)  &   12.9752 (0.03)  &   13.2002 (0.05)  &   13.4347 (0.07) \\
56508.0  &  \ldots              &   \ldots             &     12.7077 (0.02)  &   12.8287 (0.03)  &   13.1325 (0.05)  &   13.3669 (0.07)  \\
56509.0  &  12.4615 (0.02)  &  12.6214 (0.02)  &   12.6969 (0.02)  &   12.8081 (0.03)  &   \ldots                     &    \ldots       \\
56510.0  &  12.4368 (0.02)  &  12.6400 (0.02)  &   12.6499 (0.02)  &   12.7811 (0.03)  &   13.0620 (0.05)  &   13.3554 (0.07)  \\
56511.0  &  12.4306 (0.02)  &  12.6635 (0.02)  &   12.6288 (0.02)  &   12.7200 (0.02)  &   13.0414 (0.05)  &   13.3042 (0.07)  \\
56512.0  &  \ldots              &  12.6110 (0.02)  &   12.5960 (0.02)  &   12.7521 (0.03)  &   12.9975 (0.05)  &   13.3020 (0.07)   \\
56517.0  &  12.4202 (0.02)  &  12.5405 (0.02)  &   12.5384 (0.02)  &   12.6508 (0.03)  &   12.8858 (0.05)  &   13.1608 (0.07)  \\
56521.0  &  \ldots              &  12.6113 (0.02)  &   12.5129 (0.02)  &   12.6143 (0.02)  &   12.8340 (0.05)  &   13.0919 (0.07)   \\
56523.0  &  12.5081 (0.02)  &  12.5448 (0.02)  &   12.6007 (0.02)  &   12.5571 (0.02)  &   12.8797 (0.05)  &   13.0437 (0.07)   \\
56541.0  &  12.8768 (0.02)  &  12.8752 (0.02)  &   12.7664 (0.02)  &   12.7870 (0.03)  &   \ldots                     &    \ldots         \\
56552.0  &  13.0548 (0.02)  &  13.0406 (0.02)  &   12.8758 (0.02)  &   13.0518 (0.03)  &   13.1447 (0.03)  &   13.3756 (0.04)   \\
56559.0  &  13.1300 (0.02)  &  13.1293 (0.02)  &   12.9241 (0.02)  &   13.1574 (0.03)  &   13.2173 (0.03)  &   13.4510 (0.04)   \\
56571.0  &  13.2975 (0.02)  &  13.3096 (0.02)  &   13.0648 (0.02)  &   13.2636 (0.03)  &   13.3389 (0.03)  &   13.6176 (0.04)   \\
56578.0  &  13.4078 (0.02)  &  13.4114 (0.02)  &   13.1695 (0.02)  &   13.5447 (0.03)  &   13.4835 (0.03)  &   13.7575 (0.04)   \\
56585.0  &  13.5672 (0.02)  &  13.6186 (0.02)  &   13.2659 (0.02)  &   13.6321 (0.02)  &   13.6367 (0.03)  &   13.9024 (0.04)   \\
56588.0  &  13.7055 (0.02)  &  13.7697 (0.02)  &   13.4106 (0.02)  &   \ldots              &   13.7585 (0.03)  &   14.0406 (0.04)   \\
56589.0  &  13.7021 (0.02)  &  13.8036 (0.02)  &   13.4152 (0.02)  &   13.8345 (0.02)  &   13.7986 (0.03)  &   14.1202 (0.04)   \\
56593.0  &  13.9689 (0.02)  &  14.0149 (0.02)  &   13.5742 (0.02)  &   13.9862 (0.02)  &   13.9705 (0.03)  &   14.1668 (0.05)   \\
56595.0  &  14.1545 (0.02)  &  14.2544 (0.02)  &   13.7627 (0.02)  &   14.2009 (0.02)  &   14.1914 (0.03)  &   14.4074 (0.05)   \\
56600.0  &  15.0154 (0.02)  &  15.2823 (0.02)  &   14.5393 (0.02)  &   15.0502 (0.02)  &   15.1220 (0.03)  &   15.2142 (0.05)   \\
56604.0  &  15.3536 (0.02)  &  15.5905 (0.02)  &   14.8402 (0.02)  &   15.4730 (0.02)  &   15.3862 (0.03)  &   15.5139 (0.03)   \\
56608.0  &  15.4652 (0.02)  &  15.6747 (0.03)  &   14.9609 (0.03)  &   15.6256 (0.04)  &   15.5659 (0.04)  &   15.6812 (0.04)   \\
56623.0  &  15.6803 (0.02)  &  15.9268 (0.02)  &   15.2347 (0.03)  &   15.8464 (0.04)  &   15.8741 (0.06)  &   15.9175 (0.14)   \\
\hline
\end{tabular}
\end{center}
\end{table*}

\end{document}